\documentclass{article}
\usepackage{enumitem}
\usepackage[top=1cm, bottom=2cm, left=2cm, right=2cm]{geometry}
\usepackage[numbers]{natbib}
\usepackage{graphicx}
\usepackage{xr}
\usepackage[table,xcdraw]{xcolor}
\usepackage{hyperref}
\usepackage{upquote}

\usepackage{amsmath}
\usepackage{authblk}
\usepackage{booktabs, makecell, array, tabularx}
\usepackage[T1]{fontenc}
\usepackage[utf8]{inputenc}
\usepackage{fancyhdr}
\usepackage{tcolorbox, tikz}
\usepackage{float}

\title{Automated Materials Data and Knowledge Extraction using Machine Learning and Large Language Models}

\author[1]{Aravindan Kamatchi Sundaram}
\author[1]{B. Pabitramohan Prusty}
\author[1,2,3,*]{Rohit Batra}

\affil[1]{Department of Metallurgical and Materials Engineering, Indian Institute of Technology Madras, Chennai 600036, India}
\affil[2]{Center for Atomistic Modelling and Materials Design, IIT Madras, Chennai 600036, India}
\affil[3]{Wadhwani School of
Data Science \& Artificial Intelligence, IIT Madras, Chennai 600036, India}
\affil[*]{Corresponding author: rbatra@smail.iitm.ac.in}

\begin{document}
\newcolumntype{C}{>{\centering\arraybackslash}X}
\maketitle

\begin{abstract}
Over the past decade, substantial efforts have been devoted to developing machine learning (ML) methods for extracting materials data and, where possible, materials knowledge directly from the scientific literature. This review traces the evolution of these methods from heuristic, rule-based pipelines through natural language processing (NLP) models and transformer-based encoders to, lastly, large language models (LLMs) and agentic systems capable of extracting materials information across multiple modalities, including text, tables, and figures. Beside construction of large, queryable materials databases and knowledge graphs that support downstream materials discovery, we discuss how these advances have enabled new capabilities, such as, generic question answering across materials science, AI co-scientists and chatbots for research assistance, and closed-loop optimization and discovery when integrated with autonomous laboratories. We critically assess the datasets generated using these approaches, their extraction accuracy and computational costs, and the respective strengths and limitations of different methodologies, and even provide practical guidelines for developing more reliable and scalable materials data extraction pipelines. Finally, we identify key open challenges, including the extraction of full materials structure-property-processing-performance tetrahedron at the sample level, the scarcity of reliable benchmark datasets for systematic evaluation, and the resolution of non-standard materials terminology across different document modalities.
\end{abstract}



\section{Introduction}
\label{sec:Introduction}

Materials science can be viewed as a continual effort to understand and exploit the relationships captured by the structure-processing-properties-performance tetrahedron. Experiments, whether computational or physical, probe these relationships across an enormous space of compositions, structures, processing conditions, and applications, with their findings subsequently reported in peer-reviewed journal articles, patents, theses, books and even supplementary information. However, the resulting materials knowledge is predominantly stored in unstructured or semi-structured forms---text, tables, figures, captions, and supporting files---distributed across an enormous and rapidly expanding literature corpus, estimated to contain of the order of $10^7$ publications \cite{ward2014making, venugopal2024matkg}. Only a small fraction of this knowledge is available in structured form through handbooks, curated databases, and archival resources. Thus, while on one hand, the materials literature contains a vast body of evidence, relationships and trends (either established or hidden) that can reveal how materials behave, how their properties depend on structure and processing, and what design principles govern their success in a certain application. On the other hand, this knowledge remains difficult to access systematically at scale.

The distinction between materials data and materials knowledge becomes important in this context. Materials data comprise of individual observations, e.g., a composition synthesized by a particular route, a measured property under specified conditions, or a characterization result reported in a figure or table. Materials knowledge is a more general and compact representation of such observations in the form of relationships, trends, empirical rules, and physical models that enable prediction. Examples include the Hall–Petch relation \cite{hall1951deformation}, Hume–Rothery rules \cite{HumeRothery1961}, and Vegard’s law \cite{vegard1921konstitution}, which allow researchers to generalize beyond individual observations and anticipate the behaviour of previously untested materials. A materials researcher develops such knowledge as part of their formal training and by absorbing the literature, comparing observations, and gradually forming an internal model of composition--processing--structure--property relationships \citep{raccuglia2016nn}. This process of converting numerous observations into a compact, predictive understanding may be viewed as a form of \emph{knowledge compression}. It is, however, inherently limited by the amount of literature an individual can access and integrate, and by the difficulty of identifying connections distributed across different studies and subfields.

Machine learning (ML) offers a complementary route to this compression by learning patterns directly from large collections of materials data rather than relying on explicitly formulated physical relations or individual researcher experience. Property-prediction models can learn composition--structure--property relationships, while coupling these models with high-throughput calculations, active learning, and optimization methods allow exploration of materials spaces in a more systematic and efficient manner than possible through efforts of a single human researcher. However, a major bottleneck in this approach is the availability of reliable, machine-readable data. Although the information required to train such ML models is often already present in the literature, but not in a form that can be readily accessed and processed. For example, synthesizing a new polymer or ceramic may require determining suitable precursors, their concentrations, reaction conditions, and, in some cases, a sequence of intermediate reactions by examining synthesis reports across many related materials. Consequently, the central bottleneck is not the absence of materials information, but the gap between the form in which the information exists in published literature and the form in which information is required as structured, machine-actionable data. Manual curation can, and have been, used to produce high-quality databases, but its limited to specific class of materials and use-cases and cannot scale with the volume and diversity of the literature \cite{olivetti2020data, kononova2021opportunities, trewartha2022matbert}.

This gap motivates the development of materials data (and, if possible, directly materials knowledge) extraction methods, particularly those based on natural language processing (NLP) and, more recently, transformer-based large language (LLM) models. The objective is to transform the heterogeneous content of scientific publications---text, tables, figures, and supplementary information---into structured records that can act as a knowledge base for answering materials-related queries or support downstream materials discovery ML pipelines.

\begin{figure}[t]
  \centering
  \includegraphics[width=\linewidth]{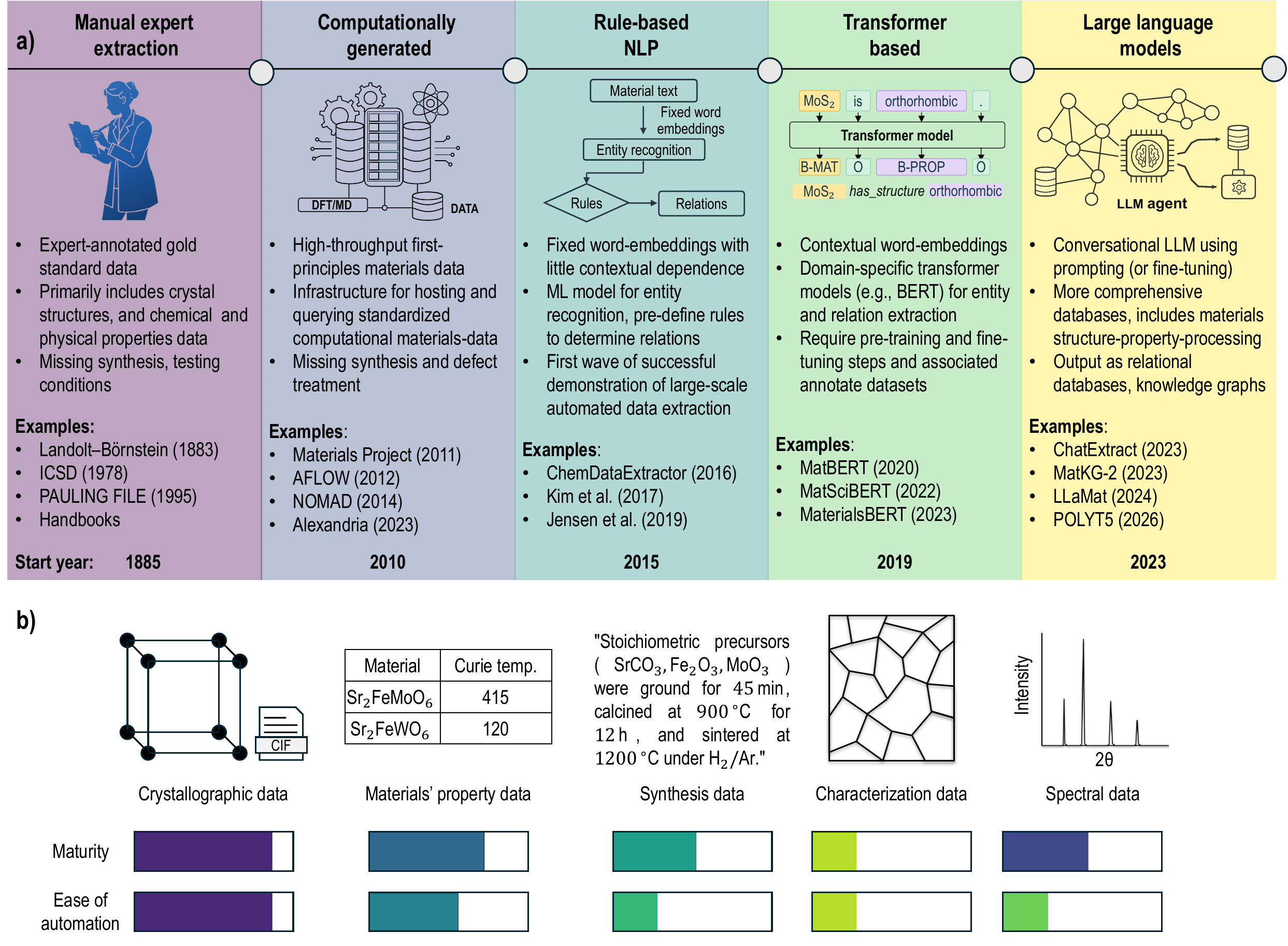}  
  \caption{a) The evolution of materials data extraction efforts. b) Different forms of materials data, and the level of maturity and ease of automation of each type.}
  \label{fig:timeline}
\end{figure}

This work tracks the evolution of materials data extraction as a progression from manual curation, to computational generation, to automated mining of the experimental literature (see Figure \ref{fig:timeline}). Early databases such as Landolt-B\"ornstein \cite{alma991024648149705706}, ICSD \citep{belsky2002icsd}, the Pauling File \citep{blokhin2018mpds} and PoLyInfo \citep{otsuka2011polyinfo}, established highly reliable, expert-curated and gold standard repositories of materials structures and properties. However, their construction is inherently slow and difficult to scale. High-throughput computational approaches provided an alternative by generating data rather than extracting it. Platforms such as the Materials Project \cite{jain2013materialsproject}, AFLOW \cite{curtarolo2012aflow}, OQMD \cite{saal2013oqmd}, NOMAD \cite{draxl2018nomad}, and Alexandria \cite{cavignac2026alexandria} enabled the systematic calculation of properties for millions of known and hypothetical structures, supporting materials screening beyond the experimentally characterised materials space. However, computational databases cannot readily capture several experimentally relevant aspects of materials, including synthesis and processing conditions, disorder, dopants, and device-level behaviour.

With the increasing maturity of AI/ML in other domains, rule-based NLP approaches such as ChemDataExtractor \cite{swain2016chemdataextractor} began to emerge around 2015 for materials data extraction. These methods typically followed a two-stage pipeline of \emph{entity recognition} and \emph{relation extraction}. During the first stage of entity recognition, an ML model identifies and classifies relevant terms present in text or tables into predefined categories such as \texttt{materials}, \texttt{synthesis methods}, \texttt{characterization methods}, or \texttt{properties}. Relation extraction then links these entities using predefined linguistic patterns and domain-specific rules to produce structured property and synthesis records \citep{swain2016chemdataextractor,kim2017synthesis,jensen2019zeolite}. These approaches provided the first demonstrations of large-scale automated extraction of materials data, but their dependence on hand-crafted rules limited their generalizability across different materials domain. Transformer-based encoders, such as MatBERT, marked the next era, where contextual representations of text enabled more robust entity and relation extraction, although these approaches required dedicated efforts for model development as well as task-specific labeled data \cite{trewartha2022matbert,shetty2023polymer}. The emergence of large language models (LLMs) around 2023 further shifted the paradigm towards general-purpose extraction. Prompting enabled structured extraction with little or no task-specific training, while fine-tuning enabled more reliable domain-specialized models from relatively small annotated datasets \citep{polak2024extracting,zheng2023chatgptmof,dagdelen2024structured}. Beyond improving the scale and flexibility of extraction, LLMs also expanded the target representation from flat material-property records to more comprehensive materials synthesis, processing, structure, and properties records, and to even knowledge graphs \citep{venugopal2024matkg2}. More recently, agentic frameworks have further enhanced the extraction quality by enabling dynamic coordination of specialized models and tools applicable on multiple modalities (figures, tables and text).  Knowledge graphs, on the other hand, provide a more efficient relational representation of materials data with explicit provenance, and the ability to achieve higher-level reasoning and hypothesis generation, bringing these models closer to the ultimate goal of materials knowledge extraction rather than mere data extraction. 

This review traces this progression in materials data extraction, highlighting the opportunities, limitations, and evolving solutions associated with each era. We thoroughly discuss the databases developed using these approaches, the accuracy and cost of extraction, and the diverse data modalities considered. Finally, we discuss key open challenges and potential opportunities to improve the quality and efficiency of materials data and knowledge extraction from the literature corpus.

\section{The forms of materials data and the difficulty of extracting them}


Before we discuss the different approaches adopted for data extraction, it is important to understand what are the different forms of materials data that are usually targeted and the associated challenges. Mainly materials information is reported in five broad categories: crystallographic structure files, physical property measurements, synthesis and processing procedures, characterization data in the form of images (e.g., micrographs and device performance curves), and spectral or diffraction data. These forms differ substantially in both their amenability to automated extraction and the maturity of existing extraction pipelines, and these two aspects are not necessarily correlated, as noted in Figure \ref{fig:timeline}b. Crystallographic data is the clearest case of both high maturity and high automatability. Its high automatability stems from being structured at the point of generation, with standardized CIF files \citep{hall1991cif} containing well-defined fields for unit-cell parameters, space group, and atomic coordinates. Thus, extraction simply amounts to parsing a structured file rather than interpreting information from free text. Further, its high maturity is reflected in dedicated repositories such as ICSD \cite{zagorac2019recent, belsky2002icsd}, CSD \cite{groom2016csd}, and COD \cite{grazulis2009cod}, which have accumulated and curated on the order of $10^5$–$10^6$ entries.


Property data is arguably the most sought-after category of materials data and has consequently attracted extensive development of ML-based extraction pipelines. Several large-scale efforts have produced text-mined databases (e.g., for
thermoelectrics~\citep{sierepeklis2022thermoelectric}, magnetic materials~\citep{court2018curie,kelly2025curieneel},
and general materials~\citep{weston2019matscholar}), placing this category at moderate-to-high maturity. Despite these efforts, property data still remains only moderately amenable to automated extraction because of four recurring challenges. First, the literature reported terminology is highly variable and lacks standardized forms. The same compound may be reported by its full name, abbreviation, or chemical formula (e.g., ``lithium cobalt oxide'', ``LCO'', and ``LiCoO$_2$'', or in case of doping as `5\% Mg-doped LiCoO$_2$'' or ``LiCo$_{0.95}$Mg$_{0.05}$O$_2$''). Property names may also be incomplete or require contextual interpretation---for example, `strength'' could refer to yield, ultimate, or compressive strength. Extraction therefore requires contextual disambiguation beyond than simply accurately extracting and matching to a fixed vocabulary \citep{duan2026information}. Second, extracting a complete record often requires resolving dependencies across different parts of a document, since information such as composition, processing conditions, and property values may be distributed across the abstract, methods, tables, and figure captions \citep{hira2024reconstructing}. Reconstructing a complete composition–structure–processing–property record thus requires linking information spread across these heterogeneous locations. Third, tables, despite their structured appearance, are difficult to parse reliably. Headers may be merged, transposed, or abbreviated, while essential qualifying information (e.g., symbol definitions) may appear only in captions or surrounding text rather than in the table itself \citep{yi2025matablegpt}. Lastly, even correctly extracted values can be semantically ambiguous as the full information may not be explicitly stated in the document but is rather implicit. A reported composition may represent the nominal target used during synthesis rather than the composition measured experimentally, while a property measured under specific testing conditions may be incorrectly tagged as an intrinsic property of the material. The combination of these challenges makes property-data extraction only moderately amenable to automation despite substantial progress in the field.

Synthesis and processing data remains low in both maturity and ease of automation. While dedicated synthesis databases exist (solid-state synthesis recipes \citep{kononova2019synthesis}, MOF synthesis records \citep{glasby2023digimof}), the absence of a structured standard for reporting this information, analogous to that of the CIF format, limits their maturity. In addition to the four challenges described for property extraction, the synthesis extraction entail additional challenges due to its inherent relational and sequential nature---precursors and their quantities, operations, and conditions involved must be correctly linked to one another to arrive at a correct synthesis record. These details are also often partly implicit, with conditions inherited from cited procedures or sometimes simply described as ``standard.'' This additional relational and sequential complexity makes synthesis data substantially harder to extract. Characterization images, such as micrographs and device performance curves, require visual and contextual reasoning, making them the least developed in both maturity and ease of automation. Few dedicated databases exist \cite{lupulescu2015asm}, while there is an additional need of multimodal reasoning to identify relevant figures, interpret their axes, legends, and scale, and link the visual information to the surrounding scientific text. Such capabilities are only beginning to be explored with vision-language models \citep{zheng2024gpt4v, odobesku2025nanominer}. Diffraction and other spectral data occupy a more unusual position. Diffraction data, in particular, has a long history of institutionalized curation---the ICDD Powder Diffraction File has served as a reference for phase identification since 1941 and now contains more than one million curated patterns \citep{gatesrector2019pdf}, while Rietveld refinement has been widely used to extract structural information from diffraction patterns \citep{rietveld1969}. However, this maturity has relied largely on manual expert curation, and efforts to extract diffraction data directly from literature figures remain relatively nascent \citep{mostafa2026eraf4xrd, dartsi2026automating, dong2022deepxrd, johansen2026decifer, choudhary2025diffractgpt}. Other spectral data, including Raman, FTIR, XPS, and NMR, are less mature, with existing resources typically limited to specific techniques or material classes, such as the RRUFF database of mineral Raman and infrared spectra \cite{lafuente2015power}. Overall, we consider this category moderately mature but still difficult to automate, particularly because data must be recovered from published figures, which requires the same multimodal reasoning challenges encountered in characterization-image extraction. Overall, this discussion forms the backdrop in which several efforts for materials data extraction are discussed next. Importantly, not all materials data types have achieved same level of success and each presents a unique set of challenges. 

\section{High level Pipeline}

\begin{figure}[H]
    \centering
    \includegraphics[width=0.95\linewidth]{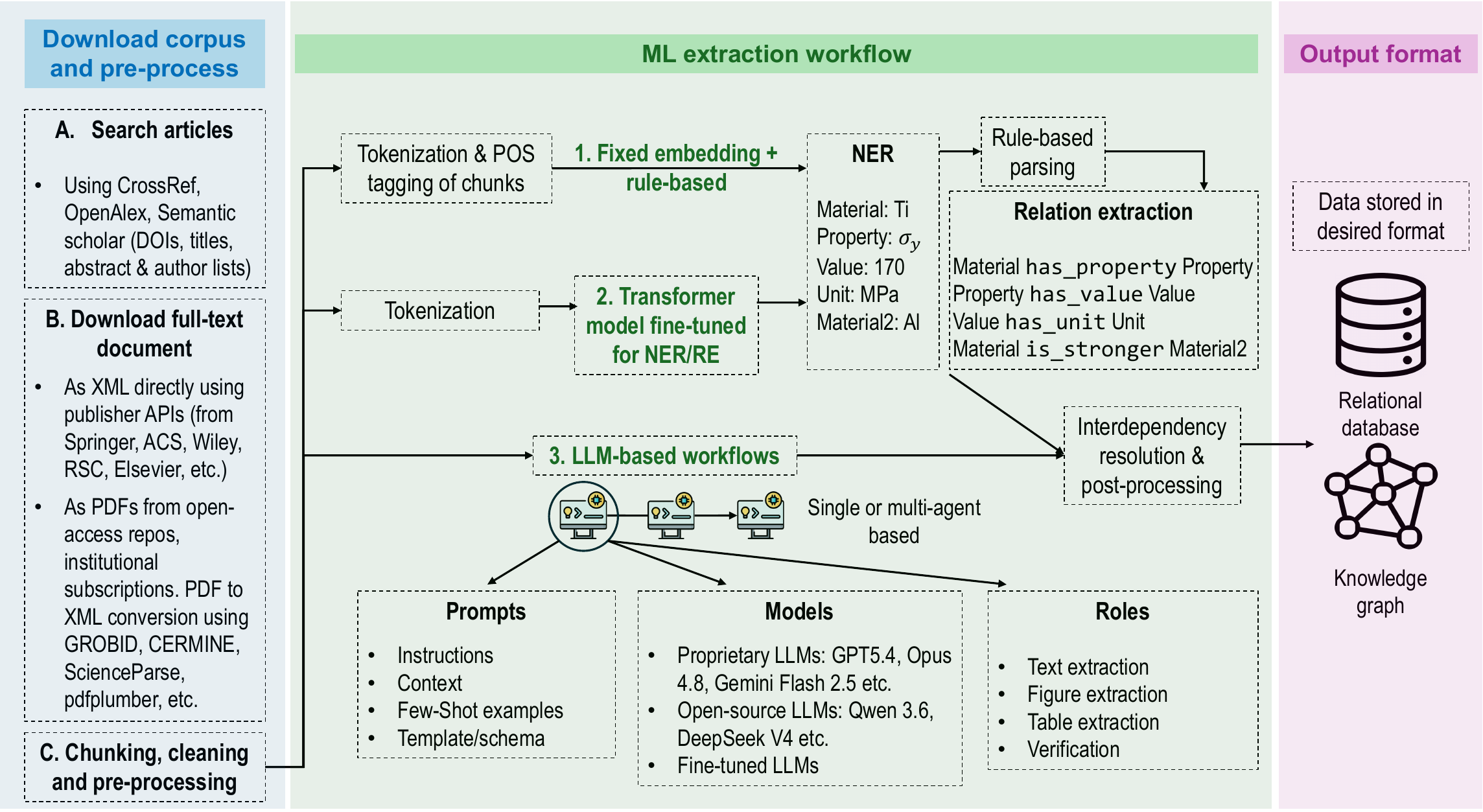}
    \caption{Overview of a general materials data extraction pipeline, from document identification, download, and preprocessing through entity recognition and relation extraction, dependency resolution, and post-processing, to the final storage of extracted data in a queryable, structured format.}
    \label{fig:pipeline}
\end{figure}


\label{sec:pipeline-overview}

Before surveying individual methods, it is helpful to see how the pieces fit
together. Almost every materials data extraction effort, whether built on
hand-written rules or the latest language models, follows the same high-level
skeleton, sketched in Figure~\ref{fig:pipeline}. One starts with acquiring and cleaning the source
documents, running one or more extraction workflows over the cleaned content, and then
reconciling and storing the results as a structured database or knowledge graph. Next
section walks through that skeleton at a high level, introducing the vocabulary and
the main design choices, while the later sections examine each stage in depth.

\subsection{Procuring and pre-processing the content}
\label{subsec:procure}

The pipeline begins not with an ML model but with the collection and cleaning of the literature corpus. Relevant articles are first identified by querying scholarly metadata services such as Web of Science, CrossRef, OpenAlex, arXiv, and Semantic Scholar with appropriate keywords or titles, yielding document identifiers (such as digital object identifiers (DOIs) for retrieval. Next, the actual document content (text, tables or figures) is obtained, ideally in a structured markup format such as XML accessed directly through publisher APIs (e.g., Elsevier, Springer Nature, Wiley, ACS, RSC), since XML preserves a document's logical structure (sections, tables, formulae) in a way that is difficult to recover from PDFs. When only PDFs are available, as is common for literature published before roughly the year 2000, they must be converted to structured text using layout-aware parsing tools such as GROBID \citep{grobid2008}, CERMINE \citep{tkaczyk2015cermine}, Science Parse \citep{scienceparse}, or pdfplumber \citep{pdfplumber}. A final pre-processing stage then strips the irrelevant markup, correctly re-encodes artifacts such as subscripts, superscripts, and units in chemical formulae and equations (e.g., ``$\mathrm{TiO_2}$'' is retained as a single meaningful word), and chunks the document into passages sized appropriately for the downstream pipeline. Optionally, this stage may further filter the corpus down to the most relevant passages such as the abstract or the experimental methods \cite{zheng2023chatgptmof,huo2019semisupervised} , or convert tables into structured representations suitable for extraction \cite{yi2025matablegpt,gupta2023discomat}.
This corpus procurement and pre-processing step is easy to underestimate, but errors introduced here, such as a truncated article or a garbled formula, propagate into every downstream record. For a more thorough treatment of corpus construction and cleaning, publisher access, text-and-data-mining licensing, and parsing tools, we refer the reader to dedicated reviews on this topic \cite{schilling2025text, kononova2021opportunities}.

\subsection{Extraction workflows}
\label{subsec:workflows}

Once cleaned, the content is passed to an extraction workflow. As shown in Figure~\ref{fig:pipeline}, three successive methodological generations of AI/ML frameworks have been developed for materials data extraction: rule-based (or pre-transformer) models, transformer-based models, and, most recently, LLMs. Below we briefly summarize each generation of model pipeline along with its advantages and limitations, while a more detailed review is provided in Section~\ref{subsec:extraction_pipeline}.
 
\paragraph{Rule-based (or pre-transformer) models.}
The oldest workflow treats extraction as classical NLP problem. Text is first tokenised, split into words and punctuation, and each token is tagged with its part of speech (noun, verb, adjective, and so on) in a step called part-of-speech (POS) tagging, typically using a probabilistic model such as linear-chain conditional random field (CRF). Named entity recognition (NER) follows next, wherein pre-defined materials-related categories (material, property, characterization, synthesis, etc.) are assigned to each token using another probabilistic model. Importantly, the features used for NER model training are hand-crafted and fixed, and can themselves include the POS tag. Hand-written grammatical rules then match patterns over these tagged tokens to identify entity relations, for example a rule matching the structure ``$\langle$\texttt{material}$\rangle$ has a $\langle$\texttt{property}$\rangle$ of $\langle$\texttt{value}$\rangle$ $\langle$\texttt{unit}$\rangle$'' will emit a corresponding data record capturing the material-property-value relation. Chemistry-aware toolkits have been developed with their own specialized tokenisers, dictionaries, and rules required for this. ChemDataExtractor \citep{swain2016chemdataextractor} is among the most prominent example of this generation and has been successfully used to auto-generate several materials property databases directly from the literature. The strength of this workflow is its precision (extracted records are mostly correct) and provenance (every extraction is traceable to an explicit rule). However, its weakness lies in the brittleness of the manually constructed rules and the fixed, context-independent nature of the underlying token representation. 
Another issue is the ability to fuse information from various modalities such as figures, tables and text.

\paragraph{Fine-tuned transformer models.}
The second workflow replaces fixed token features and hand-written rules with a learned language model, such as an encoder of the BERT family (bidirectional encoder representations from transformers) \cite{devlin2019bert}. Such a model reads a tokenised sentence and produces a context-dependent vector for each token, meaning the representation of a word reflects the sentence it appears in rather than a fixed dictionary entry. Separate lightweight task-specific ``heads'' are then trained on top of this encoder, one for NER and the other for relation extraction (RE), i.e., linking identified entities into tuples such as $\langle$\texttt{material}, \emph{has\_property}, \texttt{property}$\rangle$. Because the general BERT model is trained on english text (e.g., Wikipedia, BooksCorpus), it underperforms on scientific text containing chemical names and symbols. Thus, the BERT encoder is typically first adapted to the materials domain by pre-training on millions of materials paragraphs \cite{trewartha2022matbert,gupta2022matscibert}, before proceeding to another round of training for downstream NER and RE tasks. This is followed by applying the model to whole literature or selected materials corpus to construct databases of material entities and their relation \cite{shetty2023polymer,foppiano2023automatic,mullick2024matscire}.
The main advantage of this workflow is that it overcomes the brittleness of the rule-based models; fixed word features are replaced by context-dependent token embeddings, and fixed grammar rules are replaced by RE models learned from training examples. However, the limitations is the restricted size of the context window (around 512 tokens), the requirement of well-labeled training data, separate rounds of model training for each task, pre-determined entity or relation types, and the inability to handle multi-modal data beyond text.

\paragraph{LLM-based workflows.}
The most recent workflow uses LLMs, or very large generative models which given an instruction, produce structured output directly, often with little or no task-specific training. As highlighted in Figure \ref{fig:pipeline}, configuring an LLM-based workflow involves three knobs. The first is the prompt, the natural-language instruction given to the model, which typically combines task instructions, relevant context, a few worked examples (or few-shot prompting, which shows the model the desired input-output behaviour), and an output template such as a target JSON schema defining the target materials structure-property-processing relations that need to be extracted. The second is the choice of the model, which can be a proprietary LLM (e.g., Claude \cite{anthropic2024claude3}, ChatGPT \cite{achiam2023gpt, openai2022chatgpt}) accessed via an API, an open-source model (e.g., LLaMA \cite{touvron2023llama2}, Pixtral \cite{agrawal2024pixtral}, Qwen \cite{yang2025qwen3})  running locally, or a model that has been fine-tuned on domain examples to specialize its behaviour on materials literature. The last knob is the workflow architecture, which in a more advanced setup involves multiple cooperating models (or agents), each assigned a specific role, for instance, one agent extracts from text, another from figures, and a third verifies the extracted outputs. Various flavors of LLM-based extraction have been developed for materials, including prompt-based mining of synthesis parameters \cite{zheng2023chatgptmof}, conversational property extraction with follow-up verification \cite{polak2024extracting}, fine-tuned models for joint NER and RE \cite{dagdelen2024structured}, and multi-agent systems with coordinated specialized roles for domain-specific materials extraction \cite{roy2026comproscanner}.
The strength of this workflow is its generality and shallow learning curve, one capable model can handle many tasks (from NER, RE to more advanced hypothesis generation and question and answering) with no bespoke training. Its weaknesses are the risk of model hallucination, sensitivity to prompt wording, and most importantly the cost of running large models over an entire corpus, which to a certain extent can be mitigated using role-based multi-agent setup discussed later.

\subsection{The target data schema}
\label{subsec:schema}

As highlighted in  Figure \ref{fig:pipeline}, the next important aspect of the pipeline is the target data schema, which primarily consists of 
a small ontology defining the entity types to be recognized (\texttt{material}, \texttt{property}, \texttt{value}, \texttt{unit}, \texttt{condition}, \texttt{synthesis}, etc.) and the relation types (\emph{has\_property}, \emph{has\_value}, \emph{has\_unit}, etc. as shown in Figure \ref{fig:pipeline}) to be identified between the tagged entities. In other words, it is the target structured schema into which the unstructured materials literature is re-expressed. This schema is generally designed by a domain expert prior to the execution of the extraction workflow according to the anticipated downstream needs.
Importantly, different extraction workflows offer different degrees of schema flexibility. For a fine-tuned BERT model, the schema is fixed at training time and baked into the training labels, yielding consistent, well-typed output but making the model rigid: adding a new entity or relation type generally requires re-annotating training data and re-training the model. In contrast, for a LLM, the schema is supplied at inference time in the prompt, which makes the workflow far more flexible, since changing the target is as simple as updating the prompt. A further consideration in schema design is its granularity: rather than specifying an exact property name as a field (e.g., hardness), the field can instead be a generic placeholder (e.g., material property), shifting the burden of interpreting the specific property onto the model itself, a task better suited for an LLM-type model with broad background knowledge. A generic schema thus enables construction of large general-purpose materials databases, but its reliance on the LLM for property identification reduces its accuracy relative to a domain-specific target schema.

\subsection{Reconciliation and post-processing}
\label{subsec:postprocess}
Even after the extracted data has been re-structured into the target schema, it does not yet constitute a usable database. Achieving that requires the final critical step of \emph{interdependency resolution} and \emph{normalization}.
The interdependency resolution refers to reconnecting facts that were extracted in isolation but belong together, such as linking an abbreviation defined in the introduction text to its use in a table, associating a property value obtained with the specific sample processing and testing condition, or resolving a dopant label mentioned only in a table caption. This is the step that reconstructs a complete composition-processing-property record from fragments/chunks scattered across the document. The second is normalization, i.e., canonicalizing materials identifiers so that ``LiCoO$_2$'', ``lithium cobalt oxide'', and ``LCO'' collapse to one entity, and standardizing units and reference conditions so that extracted numbers are comparable across different articles. Normalization also makes it possible to identify and remove duplicate records from the same article. A link to the source article is then added to each deduplicated record for provenance, and the final result is stored either as a relational database of tuples or as a knowledge graph, whose nodes are entities and whose edges are the extracted relations \citep{bai2025kgfm}. Overall, this stage is what converts a pile of individually correct extractions into a coherent, queryable resource. Unfortunately, interdependency resolution remains among the less developed parts of the pipeline, and is a common source of difficulty in using extracted materials databases. With this high-level extraction anatomy in place, the following sections review the progress made in each part of pipeline in detail.

\subsection{Extraction accuracy}
Another important aspect of an extraction pipeline is its performance. Standard information-retrieval metrics such as precision (the fraction of extracted data points that are correct), and recall (the fraction of relevant data points in the source that are successfully extracted) are commonly used, often summarized by their harmonic mean, the $F_1$ score. Both precision and recall matters since a low recall limits the size of the resulting database, whereas low precision introduces incorrect values that can degrade usability of the extracted data. Arguably, low precision is more problematic because erroneous entries remain invisible without independent verification. Measuring precision and recall in practice is itself challenging because materials science lacks the broad, standardized benchmark datasets available in mainstream NLP. The diversity of materials subfields, terminology, and reporting conventions makes it difficult to construct general-purpose benchmark datasets. Consequently, most studies evaluate extraction pipelines against smaller, manually curated ground-truth datasets annotated by domain experts, since no automated proxy pipelines have reached the level to replace an expert judgment in determining whether an extracted value is genuinely correct \citep{foppiano2024mining}. This reliance on bespoke, expert-annotated evaluation sets makes it difficult to compare performance across multiple studies and is itself an open problem. Nonetheless, some of the state-of-the-art LLM-based pipelines typically achieve precision and recall in the high-80s to mid-90s range on materials extraction tasks \citep{polak2024extracting,yi2025matablegpt}. While these results enable useful databases to be constructed at scale, they remain insufficient to eliminate the need for human verification. Thus, further progress requires not only development of accurate materials data extraction pipelines but also more comprehensive standardized materials benchmark datasets and evaluation strategies. 

\section{ML-based extraction approaches}
\label{subsec:extraction_pipeline}

In this section, we examine each of the ML-based extraction workflow in detail, identifying the types and scale of databases developed to date (see Supporting Information), the advantages and limitations of each method, and the overall progress of the field.

\subsection{Pre-transformer approaches: Static word embeddings and rule-based parsing (2015--2021)}
\label{sec:pre_transformer}

The pre-transformer extraction era is dominated by rule-based NLP models, which combine fixed word features for NER with hand-written grammars for RE. As mentioned earlier, ChemDataExtractor is the dominant example of this type and has been used to build materials databases across several domains. These include, the largest battery database (at the time) comprising of 292,313 records (with 214,617 unique material-property relations)
spanning five properties (capacity, voltage, conductivity, Coulombic efficiency, energy) 
\citep{huang2020nerbattery}, a thermoelectric database built with an upgraded version of ChemDataExtractor incorporating a rule-based sentence-simplification step \cite{sierepeklis2022thermoelectric}, mined from 60,843 articles across the Elsevier, RSC, and Springer publishing groups to yield 22,805 records covering 10,641 unique chemical names at 82.25\% precision, with each record consisting of a compound with one of the five transport properties ($ZT$, $\kappa$, $S$, $\sigma$, power factor) and its measurement temperature. Beyond composition-property relational databases, ChemDataExtractor has also been used to mine synthesis details for perovskite solar devices from over 3,000 journal articles, producing records with 30 attributes (such as deposition method, rate, and duration, alongside composition data) to build the Perovskite Database
\citep{ref53}. Once such a database becomes available, the downstream task of ML property prediction can also be pursued. For instance, ChemDataExtractor has been integrated with ML and generative models to predict the magnetocaloric effect of Heusler alloy candidates
\citep{ref22}. Other rule-based pipelines have also been built for different classes of materials data. For instance, DigiMOF adapted ChemDataExtractor to the synthesis side of MOF discovery, scraping 43,281 journal articles identified through the Cambridge Structural Database MOF subset to extract 15,501 unique MOFs and over 52,680 associated properties spanning synthesis method, solvent, organic linker, metal precursor and topology \citep{glasby2023digimof}. Similarly, ZeoSyn combined table parsing, NLP-based text extraction, and regex-based text-table linkage, wherein relations were recovered from the underlying table-row structure and pattern matching rather than using a learned relation classifier, followed by a manual verification step to compile 23,961 zeolite synthesis routes (gel composition, conditions, precursors, organic structure-directing agent, and product information, including failed/negative outcomes)
\citep{ref5}. 

As mentioned earlier, the reliance on hand-written grammar rules is one of the major limitation of these methods. To give a specific example, the precision accuracy of ChemDataExtractor fell from 90\% to just 66\%, when the hand-written phrase-parsing rules tuned on organic-materials literature, was directly applied to extract magnetic phase-transition temperatures from inorganic chemistry and physics articles \citep{court2018curie}. Thus, some efforts have been made to learn these rules directly from the literature corpus.
For instance, a modified Snowball algorithm was used to auto-generate grammatical rules to extract quaternary relation tuples $\langle$\texttt{property specifier}, \texttt{chemical entity}, \texttt{value}, \texttt{unit}$\rangle$ from around 500 example sentences which were manually verified to contain such entity information. Once learned, these auto-generated grammatical rules were applied alongside conventional ChemDataExtractor rules to over 68,000 articles, raising precision to 82\% from a baseline of 73\%, and yielding an overall database of 39,822 records \citep{court2018curie}. 
Overall, these methods enabled literature-scale information extraction, but remained constrained by patterns anticipated in advance by the curator. Importantly, not all properties are amenable to such rules, making it difficult to capture them using predefined rules as they lack any consistent syntactic patterns of reporting. Others require new rules whenever the schema is extended, for example, adding cyclability to a battery database. Consequently, although these methods can achieve high precision for the patterns they are designed to capture, they typically suffer from low recall, as no fixed set of rules can anticipate every way in which a fact may be phrased.

Another limitation of these works was the fixed hand-crafted features used for word representation. To overcome this, the field shifted toward learning word embeddings directly from the literature corpus.
The idea is to represent each word (or token) as a dense vector, learned purely from the distributional statistics of a large unlabelled corpus, so that words (or tokens) used in similar contexts land near one another in vector space. Common algorithms for learning such static embeddings include Word2vec \cite{mikolov2013word2vec} and GloVe \cite{pennington2014glove}, among others \cite{bojanowski2017enriching}. These models are trained in a self-supervised manner: Word2vec learns to predict a word from its surrounding context words (or the context from the word), while GloVe fits the vectors to global word co-occurrence counts, so no manual labels are required. While general-purpose word-embedding tools trained on much larger newswire corpora exist, embeddings learned on a materials-specific corpus remain more useful for downstream tasks, since general-purpose models may not only miss materials-specific co-occurrences but also mis-tokenise chemical formulae and other technical words or symbols.

Once learned, these word embeddings can be fed through a sequence of BiLSTM layers and a CRF output layer to perform the NER task, though such training requires a corpus of expert-annotated sentences.
In an exemplary work, 800 abstracts were hand-annotated across seven entity types (inorganic material, symmetry/phase label, sample descriptor, property, application, synthesis method, and characterization method), motivated by the materials science processing-structure-property-performance tetrahedron  \cite{weston2019matscholar}. The resulting BiLSTM-CRF model, followed by a separately trained entity-normalisation classifier that merges synonymous mentions (``TiO$_2$'', ``titania'', ``titanium dioxide'') into a canonical formula, achieved an overall $F_1$ score of 0.87, and was then applied to 3.27 million abstracts, extracting over 80 million entities to build the publicly available Matscholar database. Similar approaches have been used to extract inorganic synthesis parameters \cite{kim2017synthesis, kononova2019synthesis}.

Beyond serving as input features, embeddings learned from unlabelled text can be queried directly to reveal material--property associations never stated explicitly in the literature. The defining demonstration is Tshitoyan et al.'s \emph{mat2vec} study \cite{tshitoyan2019word2vec}, which trained Word2Vec-style embeddings on roughly 3.3 million materials-science abstracts using no labels and no explicit chemistry rules. The learned vectors alone recovered structure resembling the periodic table, recovered chemical analogies by vector arithmetic (``ferromagnetic $-$ NiFe $+$ IrMn $\approx$ antiferromagnetic''), and ranked several compositions as promising thermoelectrics years before those materials were independently reported as such \cite{tshitoyan2019word2vec}. A similar approach has been applied within specific subdomain of polymers, wherein an unsupervised embedding model trained on half a million polymer papers demonstrated material--property associations without any labeled data
\cite{ref36}. Performance of aforementioned pre-transformer models is mainly limited by static embeddings, which assign a single vector per word irrespective of context, and by the high cost of expert annotated datasets required for every new entity schema \cite{Friedrich2020SOFCExp}. For a more comprehensive account of the pre-transformer era, we refer readers to other reviews \cite{olivetti2020data, kononova2021opportunities}, and also mention in the passing that graph-based methods have also been used to extract materials data from tables \cite{gupta2023discomat, hira2025matskraft}.
\subsection{The transformer era: Contextual embeddings, foundation models, prompting and fine-tuning}
\label{sec:transformer_intro}

The transformer architecture \cite{vaswani2017attention} uses self-attention mechanism to learn contextual embedding of every token, meaning a token's vector is determined by the sentence in which it appears. Thus, context-dependent terms such as ``conductivity" (electrical or thermal) or ``cell" (an electrochemical cell or a unit cell) can acquire different representations, in contrast to the fixed embeddings discussed in the previous section. Transformers can also be efficiently pre-trained on very large corpora using self-supervised learning, allowing the model to acquire syntactic patterns, domain-specific terminology, and broader semantic relationships from unlabelled text \cite{vaswani2017attention}. Crucially, unlike the BiLSTM systems discussed earlier, where the embedding layer is fixed during downstream NER training, transformer-based models can transfer the entire pre-trained network to a downstream task and further adapt its parameters through fine-tuning. Broadly three families of transformers architectures are available, encoder-only, encoder–decoder, and decoder-only, which differ primarily in whether they are designed to represent, transform, or generate text, respectively \cite{min2023recent, zhao2023surveyllm}.

Before discussing the use of transformer models for data extraction, it is useful to clarify some of the terminology used in this field. Models pretrained at large-scale on broad and diverse corpora are commonly referred to as \textit{foundation models}, i.e., general-purpose models that are not designed for a single task but instead provide a reusable starting point that can be adapted to a wide range of applications \cite{bommasani2021foundation}. Examples include BERT~\cite{devlin2019bert} and GPT-3~\cite{brown2020language}. Foundation models can be adapted to specific tasks in two broad ways, \textit{fine-tuning} and \textit{prompting}. In the former case, the model is further trained on task-specific labelled data, optionally following continued pretraining on domain-specific text. Parameter-efficient approaches (e.g., LoRA), reduce the fine-tuning cost by updating only a small subset of parameters while keeping most of the pretrained weights frozen \cite{howard2018ulmfit,hu2022lora}. In prompting, by contrast, the weights are left untouched and the task is specified through the input prompt itself, an ability that emerges only at large scale \cite{brown2020language}. Prompt-based task performance was particularly enabled by the emergence of large decoder-only models, commonly referred to as large language models (LLMs). In this article, we reserve the term LLM specifically for models with more than roughly a billion parameters (e.g., GPT and Claude families \cite{achiam2023gpt, anthropic2024claude3}), while using ``foundation model'' as the broader term encompassing models of the BERT family as well.

\subsection{Scientific BERT models}
\label{subsubsec:advanced_embeddings}

As highlighted in Figure \ref{prompt_engg}a, a typical workflow for materials-data extraction using transformer models such as BERT comprises of four stages: (1) pretraining on a large corpus, (2) fine-tuning for downstream tasks such as NER or RE, (3) evaluation on manually annotated test datasets, and (4) large-scale extraction using the validated models. On the pretraining front, the general-purpose BERT model was pretrained on web and book text and consequently underperforms on specialised scientific writing, whose vocabulary (e.g., chemical formulae, units, alloy and dopant notation, and mineral and compound names) and sentence structures differ substantially from everyday language \citep{beltagy2019scibert,trewartha2022matbert}.
Thus, a family of domain-adapted encoders pretrained on scientific corpora using the same self-supervised masked-language-modelling objective as BERT were developed.
For example, SciBERT was pretrained on 1.14 million biomedical and computer-science papers \cite{beltagy2019scibert}; MatBERT on roughly 50 million paragraphs drawn from about 2 million materials-science articles \cite{trewartha2022matbert}; and MatSciBERT on approximately 154,000 papers (285 million words) spanning inorganic glasses and ceramics, metallic glasses, alloys, and cement and concrete \cite{gupta2022matscibert}. Importantly, these domain-specific models consistently outperformed general-purpose BERT on materials-specific benchmarks. In an ablation study, domain-specific pretraining resulted in a 1-12\% improvement in NER performance over general-purpose BERT, with larger gains in data-scarce regime, precisely where annotated materials datasets are often limited \cite{trewartha2022matbert}. Another important aspect that has been investigated is the role of BERT model size where authors trained 770M-parameter ScholarBERT model on 221 billion tokens. Surprisingly, both the model size and the domain-specific pretraining were found to offer little advantage over substantially smaller domain-specific models \citep{hong2023diminishing}.


\subsubsection{Fine-tuning BERT models}
Once pretrained, the model is fine-tuned by adding a lightweight, task-specific ``head" to the encoder for downstream tasks such as NER or RE. NER is generally simpler of these tasks, as it requires identifying entity boundaries and types, whereas RE must additionally determine whether two entities are related and, if so, the type and direction of that relation. Several unique approaches employing different BERT-based architectures have therefore been developed for RE. The simplest treats RE as a classification problem, where candidate entity pairs are marked with special tokens in the input sentence, and the model predicts either a specific relation type or ``no relation" \cite{gupta2022matscibert, beltagy2019scibert}. 
A second approach performs NER and RE jointly using a single model, allowing information from the two tasks to interact during prediction. For example, DyGIE++ \citep{wadden-etal-2019-entity} constructs and scores candidate entity spans and the relations between them jointly, and has been applied to oxygen-reduction-reaction catalyst extraction using BERT-derived encoders \citep{ref41}. Pointer networks \citep{mullick2024matscire} instead sequentially decode entity boundaries and classify the relation between them within a single encoder-decoder model, offering an alternative joint architecture.
A third approach reformulates RE as an extractive question answering, in which questions such as ``What is the Curie temperature of $\langle$material$\rangle$?'' are used to identify the corresponding answer span within the input sentence. This strategy was adopted in the MagBERT \citep{ref23} for Curie-temperature extraction and BatteryBERT \cite{huang2022batterybert} for identifying material names corresponding to anode, cathode, or electrolyte.

A fourth approach avoids training an explicit RE model itself. Instead, entities identified by BERT or other NER models are linked using co-occurrence statistics, proximity, syntactic dependencies, or other domain-specific heuristics. For example, MatKG \citep{venugopal2024matkg} connects any two entities that co-occur within the same paper, with the edge weight given by the number of papers in which the pair co-occurs. Importantly, co-occurrence indicates only that two entities were discussed together; it does not establish a specific scientific relationship between them. Some other examples of rule-based RE include: MaterialsBERT \citep{shetty2023polymer} and the polymer solar-cell pipeline \citep{ref1} used proximity and role-assignment rules, SuperCon2 \citep{foppiano2023automatic} used ordering cues and token-distance constraints, and a stress--strain extraction pipeline \citep{ref9} used syntactic dependency distances. Such approaches avoid the need to construct separate relation-labeled datasets for model training, but inherit the same brittleness and limited generalizability of predefined rules discussed earlier.


Despite a variety of architectures adopted for RE, it remains substantially more difficult than NER. For example, the DyGIE++ catalyst-extraction model achieved an $F_1$ score of 0.82 for entities but only 0.66 for relations \citep{ref41}. Similarly, in SuperCon2 linking superconducting critical temperature $T_c$ to its measurement method reached only 0.64 $F_1$ score, well below the model's overall entity-recognition micro-average of 0.77 \citep{foppiano2023automatic}. The difficulty mainly arises because of co-occurrence of multiple relations involving materials, properties, values, or experimental conditions within the same sentence or across neighbouring sentences.

It is worth noting briefly the cost involved in building these BERT models. Pretraining is expensive mostly from a computational point of view as it requires processing very large corpora over many optimization steps. For example, SciBERT required approximately one week of training on an 8-core TPU v3 system \citep{beltagy2019scibert}. However, this cost is largely incurred only once and can be amortized across multiple downstream tasks. Fine-tuning is relatively inexpensive computationally but introduces the additional burden of expert annotation, particularly for RE, where annotators must label both entities and their relationships. Thus, extending a BERT pipeline to new extraction tasks often shifts the dominant cost from computation to data annotation.

\subsubsection{Databases extracted using BERT models}
BERT-based extraction pipelines have enabled the construction of large-scale materials databases spanning entities, properties, synthesis, and domain-specific knowledge.
MatKG \citep{venugopal2024matkg} extended the earlier Matscholar work from entity extraction to knowledge-graph construction by applying MatBERT to approximately 5 million papers, identifying 85 million entity mentions and consolidating them into $\sim$70,000 unique entities. Rather than performing explicit RE, MatKG connected entities that co-occurred within a paper, producing 5.4 million weighted edges, with edge weights corresponding to the number of papers containing each entity pair. Other large-scale resources include 
databases of computational methods and parameters, magnetic materials, and battery materials \citep{dieb2021matscie,ref23,huang2022batterybert}.

Other class of resources extracts explicit, typed relations between entities, particularly in polymer and battery applications. For example, MaterialsBERT, obtained by further pretraining on materials-science literature and fine-tuned for materials NER, was combined with heuristic relation-assignment rules to extract $\sim$300,000 material-property records from 130,000 abstracts, covering thermal, mechanical, electrical, and gas-transport properties \citep{shetty2023polymer}. A related polymer solar-cell pipeline used NER together with proximity and role-assignment rules to construct 2,585 unique donor-acceptor records, with associated power-conversion efficiency, fill factor, and voltage properties, from 3,307 papers \citep{ref1}. Another interesting aspect of this study is the degradation in the accuracy of extraction quality with increasing number of tuples in a relation---the $F_1$ score falls from 0.69 for two-entity tuples to 0.58 for three and 0.43 for the full quaternary $\langle$donor, acceptor, property, value$\rangle$ tuple. This highlights the challenge of extracting $N$-ary relations in materials databases, which are often required to link composition, processing, structure, and properties within a single record but cannot be reliably captured through pairwise extraction alone. This challenge was later formalized as a benchmark by PolyIE, the first full-text (rather than abstract-only) NER and relation-extraction resource for polymer literature \citep{ref37}. BatteryBERT instead used extractive question answering to assign materials to the roles of anode, cathode, or electrolyte, yielding 308,964 device-component records involving 14,052 unique materials \citep{huang2022batterybert}. The corresponding database also contains 231,345 records of properties (capacity, voltage, conductivity, Coulombic efficiency, and energy), although these relations were identified using a rule-based grammar parser rather than a trained model \citep{huang2022batterybert,huang2020nerbattery}.

BERT-based approaches have also resulted in extraction of synthesis and specialized materials-property records. A gold-nanoparticle pipeline identified 5,154 synthesis-relevant papers from a corpus of nearly 5 million publications and used MatBERT-based NER coupled with dependency-based linking to organize information (synthesis procedures, morphology, and size) from 7,608 experimental and 12,519 characterization paragraphs \citep{ref65}. A stress-strain database used SciBERT-based NER combined with grammar parsing and syntactic dependency information to construct
720,308 records linking materials, mechanical properties, and measured values
\cite{ref9}. For an oxygen-reduction-reaction catalyst database, joint entity-relation modeling was used to simultaneously identify catalyst-related entities and their relations \citep{ref41}. However, all the above databases lack comprehensive records that link composition, structure, processing, and properties within a single entry. Instead, they primarily capture individual relations, such as material–property, material–process, or device–component relations, but not the complete materials information.

\section{Large language models based extraction approaches}
Despite their success, BERT-based approaches impose several practical limitations for materials data extraction that general-purpose LLMs can substantially alleviate. First, pretraining and task-specific fine-tuning for NER, RE, and other tasks require dedicated datasets and expert annotation, in addition to substantial computational resources. For example, domain-specific encoders such as SciBERT and MatBERT required pretraining on 1.14 million scientific articles \citep{beltagy2019scibert} and 50 million paragraphs from $\sim$2 million materials-science articles \citep{trewartha2022matbert}, respectively, followed by task-specific fine-tuning. Although pretraining or fine-tuning an LLM can itself be computationally demanding, prompting a pretrained general-purpose LLM largely eliminates this burden for the end user, allowing the same frozen model to be applied across different materials domains without collecting task-specific annotations \citep{ekuma2025propertyextractor, polak2024extracting}. Second, BERT-based models are typically limited to a 512-token context window, preventing them from jointly processing information that is distributed across different parts of a paper. For example, a material composition reported in the experimental section may need to be linked to a property reported in a table or to processing conditions described elsewhere in the article (say, Supporting Information). Modern LLMs offer context windows extending to hundreds of thousands or even millions of tokens \citep{schilling2025text}, enabling substantially larger portions of a paper to be processed jointly, thereby preserving long-range contextual relationships that would otherwise be lost in sentence- or passage-level extraction. Third, multimodal LLMs overcome the text-only limitation of BERT-based models by enabling information extraction directly from figures, in addition to text and tables. This provides access to data often reported only graphically, such as diffraction patterns, microscopy images, isotherms, phase diagrams, and property plots. For instance, GPT-4V has been demonstrated to extract information from sorption isotherms, powder X-ray diffraction patterns, and thermogravimetric analysis curves in MOF literature \citep{zheng2024gpt4v}, while more recent systems have demonstrated automated composition-property extraction from scientific charts \citep{roy2026comproscanner, roy2026beyond}. Lastly, in contrast to BERT-based models that require a separate fine-tuned model for each task, a single general-purpose LLM can be used to perform a variety of tasks, including NER, RE, identifying relevant paragraphs, summarizing a synthesis procedure, and answering a free-form question, simply by changing the prompt. This flexibility substantially lowers the barrier to develop new extraction workflows for different materials sub-domains, and shifts the central effort from model training towards prompt and workflow design, motivating our next discussion on LLM prompting strategies.

\subsection{Prompt engineering}
\label{subsubsec:prompt_engineering}
\begin{figure}[h]
    \centering
    \includegraphics[width=1\linewidth]{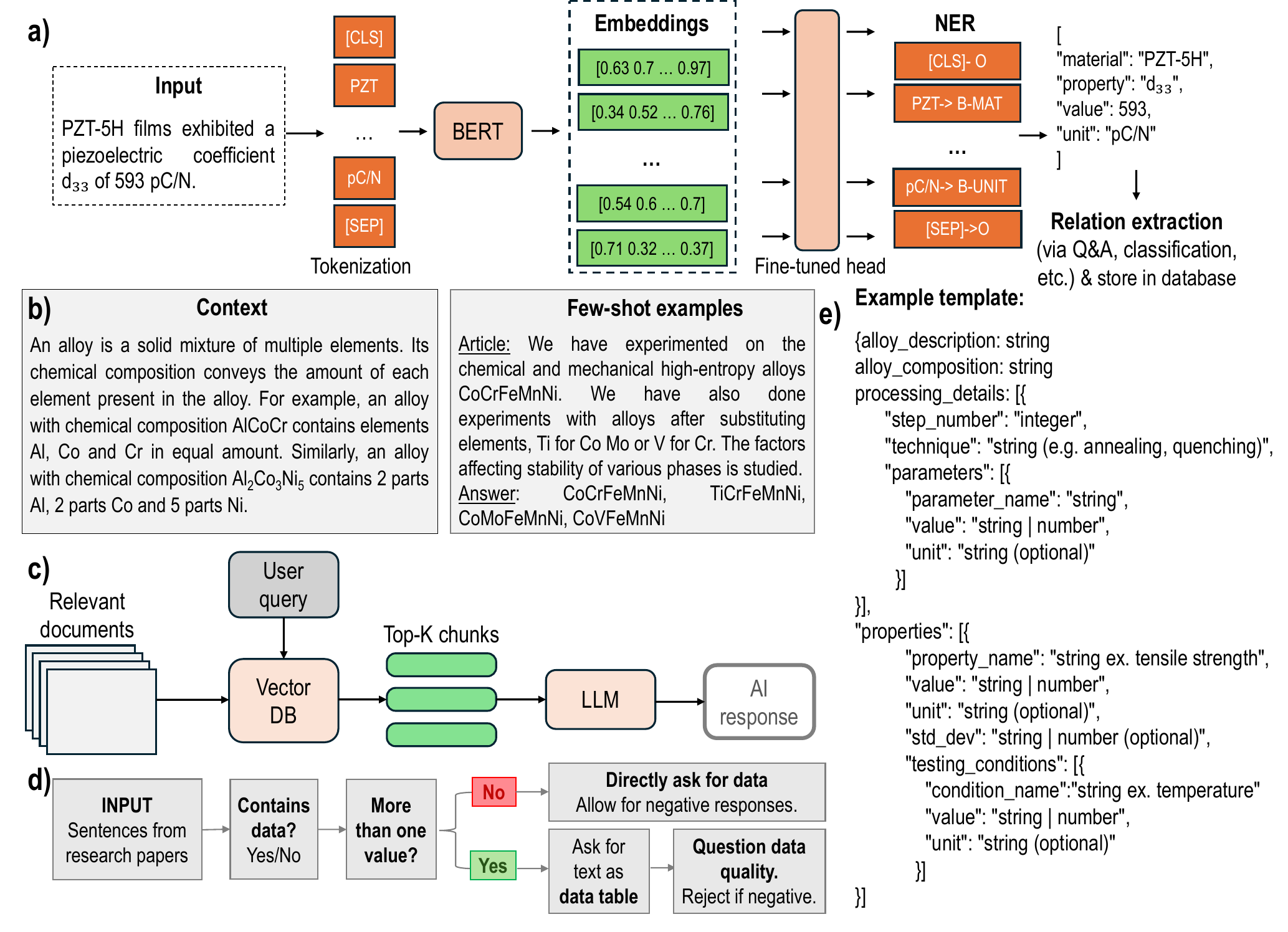}
    \caption{Transformer- and LLM-based approaches for materials data extraction: (a) Typical workflow for BERT-based extraction, (b) examples of prompt-engineering using contextual information and few-shot examples to extract alloy compositions from scientific text, (c) typical RAG workflow, (d) an example of a multi-step prompting workflow \cite{polak2024extracting}, and e) a sample template used for data extraction.}
    \label{prompt_engg}
\end{figure}

The performance of an LLM-based extraction pipeline depends not only on the underlying model type but also on the prompt, i.e., the natural-language instructions, context, examples, and output constraints provided at the inference time \cite{white2023pattern}. Because the model weights remain unchanged, the same LLM can be redirected to tasks such as NER, property extraction, classification, table parsing, and summarization simply by modifying the prompt \cite{lei2024perspective}. However, prompt quality strongly affects extraction reliability. For example, an underspecified query (\texttt{What is the fracture toughness of Al$_2$O$_3$?}) may produce a plausible but unsupported output, whereas explicitly instructing (\texttt{From the text below, extract the fracture toughness of Al$_2$O$_3$ and its testing conditions. Return JSON with keys `material', `testing condition', `value', and `unit'; return `null' if no value is reported.}) can substantially improve the output quality. This motivates the need of prompt engineering, namely the systematic design, evaluation, and refinement of prompts to improve accuracy, consistency, and reproducibility, for materials data extraction \citep{white2023pattern, schulhoff2024promptreport, chen2025unleashing}.

\subsubsection{Prompting strategies}
Effective prompt engineering for materials extraction requires domain knowledge because the user must anticipate the ambiguities that a generic prompt would miss---whether ``strength'' refers to yield or ultimate tensile strength, whether a composition is reported in atomic or weight percent, how to handle different ways of reporting compound names, non-standard abbreviations, and symbols specific to a subfield. Embedding such domain-specific technical context, together with a small set of curated annotations and worked examples, directly into the prompt has been shown to reduce hallucination and to outperform generic, domain-agnostic prompting strategies across chemistry and materials tasks~\cite{liu2025chemistry}. In practice, prompt development is an iterative process in which candidate prompts are evaluated against expert-annotated examples and refined according to observed errors \citep{zheng2023chatgptmof,kamatchi2026automated}.

Next, we discuss several prompting strategies that are particularly relevant to materials data extraction. \emph{In-context learning} (ICL) refers to the ability of an LLM to adapt to a task, such as property extraction, classification, or table parsing, using information provided in the prompt, without updating its parameters \citep{brown2020language,ekuma2025propertyextractor}. This includes zero-shot prompting, in which the model receives only a task description, for example, \texttt{identify sentences reporting a band gap and extract the material, value, and unit.} The ChatExtract pipeline is an example of zero-shot formulation \citep{polak2024extracting}. In contrast, in few-shot prompting, the prompt additionally contains a small number of expert-curated examples illustrating the desired input-output mapping (see Figure \ref{prompt_engg}b). Few-shot ICL has been used to extract polymer property data from full-text articles, with both the selection and number of in-prompt examples found to affect extraction accuracy and computational cost \citep{gupta2024polymer}. Similarly, the PropertyExtractor tool dynamically combines zero- and few-shot prompting based on the model's performance for a given property \citep{ekuma2025propertyextractor}.

Prompting can be further augmented with strategies that provide additional context or guide multi-step processing. Chain-of-thought (CoT) prompting forces the model to go through intermediate reasoning steps and can be useful when extraction requires resolving information across tables, compositions, or processing conditions \citep{wei2022cot, kamatchi2026automated}.
Lastly, and among the more popular approaches, retrieval-augmented generation (RAG) couples an LLM with an external retriever that fetches relevant passages, tables, or records from a corpus, typically indexed in a vector database, and inserts them into the prompt before generation, as shown in Figure \ref{prompt_engg}c \cite{lewis2020rag}. This helps to address two practical limitations of prompt-only approaches, one an LLM's fixed context window cannot hold an entire literature corpus, and, second, its parametric knowledge may be outdated or simply absent for a niche material. In the context of extraction, RAG has been used to dynamically select the most relevant few-shot examples for each target paragraph rather than using a large fixed set for an entire corpus \cite{kamatchi2026automated}, and in knowledge-assistant settings, it has been used to ground LLM responses according to information retrieved from experimental records and scientific literature, thereby increasing the depth and quality of response \cite{schilling2025text}. Beyond these core mechanisms, prompt engineering encompasses a broad range of design techniques, including assigning task-specific roles, providing domain context, and constraining outputs to predefined templates or schemas \citep{white2023pattern,schulhoff2024promptreport,chen2025unleashing}.

\subsubsection{Databases extracted using prompt engineering}
Prompt-engineered pipelines are among the most successful approaches that have produced large-scale databases with sufficient accuracy and consistency to support genuine materials discovery. A multicomponent-alloy dataset containing more than 100,000 composition-processing-property records was extracted using an LLM-based prompting workflow \citep{kamatchi2026automated}. The workflow combined system instructions, contextual definitions, expert-annotated few-shot examples, and formatting constraints within a chain-of-thought prompting framework, and achieved $F_1$ scores of $\sim$0.9 against an expert-curated reference dataset. The resulting database was further integrated with elemental sustainability indicators to screen alloy chemistries, representing one of the early applications of literature-scale extraction to sustainable alloy design. Similarly, ChatExtract wrapped a conversational model in a fixed sequence of prompts with verification steps (see Figure \ref{prompt_engg}d) and achieved an $F_1$ score of 0.9 on a manually curated benchmark \citep{polak2024extracting}. Applied to 4,029 papers, the pipeline extracted 10,269 raw yield-strength values for high-entropy alloys, together with a companion database of critical cooling rates for metallic glasses.


Other examples of materials databases constructed using LLM prompting include GPTArticleExtractor, which curated 7,123 thermoelectric compounds with Seebeck coefficients, power factors, and $ZT$ values from the literature \citep{itani2025thermoelectric}. At substantially larger scale, a pipeline combining a MaterialsBERT-based NER filter with similarity-based few-shot GPT-3.5 prompting was applied to 2.4 million full-text articles to jointly extract material, property, value, and condition fields. Relations were either inferred directly by GPT-3.5 from the prompt context or established using MaterialsBERT-based rule-driven entity linking, yielding more than one million records spanning 24 properties and 106,000 unique polymers \citep{gupta2024polymer}. In heterogeneous catalysis, CatMiner employed sequential prompting to first extract property values, then identify the associated catalysts, and finally retrieve their operating conditions. Applied to the literature on oxidative coupling of methane, the workflow produced 3,628 structured records from 1,029 papers \citep{walls2025catminer}.

Synthesis conditions, among the least structured yet most valuable classes of materials information, have also been targeted using prompt engineering. For example, a whole-document, context-accumulating workflow using a Mamba-Transformer hybrid LLM (Jamba 1.5 Large) and five sequential prompts extracted phase, crystallographic, and synthesis/processing information for 12,427 high-entropy alloys from 4,625 papers \citep{chizhevskiy2026hea}. In another alloy-focused study, LitXBench used a synthesis-centric prompt schema to identify each material by its synthesis process and represent the associated processing steps as a linked sequence, with composition, microstructure, and measurements attached to the resulting material. Importantly, this representation avoids conflating materials with identical compositions but different processing histories. Further, the extracted information is returned as executable Python objects, enabling programmatic validation and iterative correction. The benefit of this design was evident as the composition-centric pipeline (LeMat-Synth) achieved an $F_1$ score of only 0.58, compared with 0.82--0.88 for the LitXBench schema. The framework was further used to construct LitXAlloy, a benchmark containing 1,426 measurements from 19 alloy papers, enabling evaluation of the trade-off between extraction accuracy and LLM cost \citep{ref34}.

Beyond dataset construction, several studies have used LLM-extracted data for downstream model development and materials discovery. NEMAD, for example, first assembled a 67,573-entry magnetic-materials database from 40,000 candidate articles, capturing composition, magnetic phase-transition temperatures, structural information, and magnetic properties. This data were used to train classifiers distinguishing ferromagnetic, antiferromagnetic, and non-magnetic materials with 90\% accuracy, and regressors predicting Curie and N'eel temperatures with $R^2$ values of 0.87 and 0.83, respectively. These models were subsequently used to identify candidate high-temperature magnetic materials from the Materials Project \citep{itani2025nemad}. In electrocatalysis, a single-prompt workflow was used to construct a structure database of catalyst, applied potential, electrolyte, catalyst structure, and product involved in CO$_2$-reduction experiment. Applied to 2,302 papers, the extracted data were analyzed using decision-tree and random-forest models, which recovered known catalyst trends and indicated that copper catalysts require more negative potentials 
to produce multicarbon products \citep{farris2025accelerating}. Similarly, a GPT-4o-mini workflow extracted structured experimental records from 433 MOF studies on CO$_2$ capture, linking metal, ligand, solvent, synthesis method, pore properties, adsorption conditions, and CO$_2$ uptake. Statistical analysis of the resulting database identified combinations of metals, ligands, and MOF families associated with high CO$_2$ uptake \citep{mashhadimoslem2026co2}. In the perovskite domain, ChemDataExtractor2 followed by Llama 3.2 prompting was used to extract composition, crystal structure, and synthesis evidence for 206 unique halide and chalcogenide perovskites from over 1,000 papers. Combined with DFT-derived decomposition energies and band gaps, these data were used to train a positive-unlabelled learning classifier and screen 20,000 enumerated candidates for synthesizability \cite{desai2025perovskite}.

Other works have focused on benchmarking performance of different extraction strategies or RAG-based approaches to improve extraction quality, mainly revealing that raw model size alone is not the most direct predictor of materials extraction performance. A model-agnostic benchmark comparing commercial GPT-3, GPT-3.5/4, and free BART and DeBERTaV3 models found that GPT-4 outperformed GPT-3.5 only with more detailed prompting, while a fine-tuned GPT-3 model achieved still better performance at substantially lower cost. The free BART and DeBERTaV3 classifiers ranked lowest overall but still performed reasonably well, and were recommended as practical, low-cost alternatives. The workflow reached up to 90\% precision at 96\% recall for bulk-modulus extraction and was subsequently used to construct a metallic-glass critical cooling-rate database more than twice the size of previous human-curated databases \citep{polak2024flexible}. Similarly, a study of GPT-4 for extracting composition and properties from polymer-nanocomposite tables showed that the input modality can be as important as the model choice. While image-based input achieved 0.91 composition accuracy and 0.86 property-name $F_1$ score, structured text with captions achieved higher raw composition accuracy (0.95) when missing predictions were excluded. However, strict matching of property value, unit, and condition yielded only 0.42 $F_1$ score \citep{circi2024tables}. A comparison of GPT-4o, DeepSeek-V3, and Gemini Flash 2.0 across 265 single-atom catalysis papers further found that no single model consistently showed better score across different data fields. Thereby, routing each field to its best-performing model increased overall accuracy to 92.5\%, highlighting the importance of task-specific model selection \citep{wu2026empowering}.

RAG-based approaches provide another route to improving extraction performance. For layered sodium-ion battery cathodes, an advanced retrieval strategy based on recursive query clarification improved $F_1$ from 0.922 to 0.937 and accuracy from 85.6\% to 88.6\% relative to a naive-RAG baseline, extracting 945 data points from 312 publications \citep{ref28}. MOF-ChemUnity \citep{ref62} similarly combined RAG with knowledge-graph-grounded extraction to identify more than 70,000 material properties and 2,500 application mentions across over 15,000 MOF structures from approximately 10,000 publications. Its extraction achieved 94\% accuracy for entity matching, 89\% precision for properties, and 97.5\% for synthesis procedures.

Several recent studies have gone beyond conventional data extraction to extracting higher-level materials knowledge. For example, one study traced the evolution of scientific consensus on mechanisms of high-temperature superconductivity across 179,538 publications spanning 1950--2025. Zero-shot BERTopic clustering identified candidate mechanisms from the literature, which were subsequently refined by domain experts into nine categories. A few-shot GPT-5-mini model then scored the degree of support for each mechanism in individual papers, while citation-graph analysis tracked how these ideas evolved and propagated over time, producing a knowledge base of 5,446 papers with non-zero mechanism scores and 9,740 mechanism assignments \citep{ref51}. Another study used a three-tier hierarchical prompting workflow to extract structured shock-physics experimental records spanning 37 experimentally relevant fields (such as hardness, spall strength, sample thickness, and gas gun diameter). Rather than relying solely on direct extraction from text and tables (tier 1), the workflow additionally used physical relations (e.g., $G = \rho_0 c_s^2$) to populate fields not directly stated in the article (tier 2) and extracted information hidden in plots (tier 3), generating entries beyond what any single extraction pass could recover \citep{ref4}. LLM-based extraction has also been analyzed for systematic biases. SuperMat identified a first-entity default bias, whereby models tended to associate a property with the first-mentioned entity when multiple compounds and values co-occurred in a sentence. Fine-tuning with shuffled entity ordering was used to mitigate this bias \citep{foppiano2024mining}.

Other studies have transformed extracted materials knowledge into conversational and interactive systems. ChemPrompt extracted approximately 26,000 synthesis parameters for nearly 800 MOFs, achieving precision, recall, and $F_1$ scores of 90-99\%. These data were used to train a classifier for predicting crystallization outcomes and were subsequently repackaged into a MOF-specific chatbot for querying individual synthesis procedures \citep{zheng2023chatgptmof}. ChatMOF further integrated a GPT-4 conversational agent with MOF databases, text-mined synthesis data, a machine-learning property predictor, and a genetic-algorithm-based structure generator. The system supported natural-language queries, property prediction, and inverse design of MOF structures, achieving accuracies of 96.9\%, 95.7\%, and 87.5\% for search, property prediction, and structure generation, respectively \citep{kang2024chatmof}.

\subsubsection{Thumb-rules for better prompt engineering}

We provide a few practical rules of thumb and pedagogical examples for incorporating domain knowledge into prompts and improving extraction reliability. However, an important caveat to note is that using a larger or more capable model, as measured by standard benchmarks \cite{hendrycks2021mmlu}, often yields larger performance gains than modifying the prompt or pipeline design. The first thumb-rule is that prompts should assign the LLM an explicit role and a well-defined task rather than a broad, open-ended instruction. For example, \texttt{You are assisting a materials scientist in extracting coating thickness values from experimental papers on physical vapor deposition (PVD) coatings; extract only the values explicitly reported in the text} is preferable to a generic instruction such as \texttt{extract relevant information.} Second, prompts should provide explicit definitions when domain-specific terms have meanings that differ from common usage or vary across subfields. Although LLMs generally have broad materials science knowledge, they may not reliably capture such specialized semantics. For example, in coatings literature ``adhesion'' refers specifically to the bond strength between a coating and its substrate (commonly measured via pull-off or scratch testing), which is a narrower and more specific meaning rather than its broader everyday meaning. Terms with potentially ambiguous or subdomain-specific meanings should therefore be explicitly defined in the prompt to reduce semantic errors.

Third, the expected output should be defined using a rigorous data schema (see Figure \ref{prompt_engg}e rather than left to the model’s discretion. Beyond specifying the name of each field, its expected data type, unit, permissible range, and the value to return when no information is reported (e.g., \texttt{null} or \texttt{not reported}) should be defined explicitly. For example, a field for alloy composition should explicitly state that the values are expected as atomic fraction (at.\%) rather than weight fraction or as unspecified ``composition'', since papers routinely report using both conventions and an LLM given only the field name ``composition'' has no principled way to infer which one is intended. Understandably, defining such a schema is itself an iterative process that may require multiple refinements to eliminate ambiguities. Furthermore, each extracted data point should have one clearly defined destination field. For example, separate \texttt{heat\_treatment} and \texttt{post\_processing} fields could create uncertainty about where to place hot rolling performed after heat treatment, because of the ambiguity involved in understanding exactly what processes are covered under ``post-processing''. Replacing the latter with a more specific field such as \texttt{mechanical\_post\_processing} makes the intended scope of each field explicit and thus reduces inconsistent extraction.


Finally, rather than going for a full complex extraction as a single task, it should be decomposed into smaller, single-purpose subtasks, with separate verification steps where necessary. LLMs are generally more reliable on narrow, well-defined tasks than on prompts requiring multiple reasoning steps simultaneously. For example, it is preferable to first identify all distinct alloy compositions mentioned in a passage and then, in a separate step, extract properties (say, mechanical) associated with each composition, rather than performing both tasks in a single pass \cite{khot2023decomposedpromptingmodularapproach,li2025slm}. Few-shot examples should follow the same principle and cover diverse cases, including straightforward extractions, negative examples where the target quantity is absent and the model should learn to abstain, and difficult or borderline cases, such as values reported only in figure captions or compositions specified as nominal targets rather than measured values. The prompt should also explicitly state how such ambiguities are meant to be handled rather than leaving this to be inferred from the examples alone.


Another aspect that must be considered during prompt design is the associated cost–performance trade-offs. Verification-heavy pipelines that ask the model several follow-up questions for each candidate data point, as in ChatExtract \cite{polak2024extracting}, increase both the number of API calls and the resulting monetary cost and latency required to extract a single value \cite{polak2024extracting}. Similarly, chain-of-thought prompting can increase token usage, while RAG requires additional infrastructure, including an embedding model, vector database, and the retrieval pipeline. The number of few-shot examples also presents a trade-off between extraction performance and API cost \cite{gupta2024polymer}. And, lastly, there is the hidden cost of domain-expert time incurred during iterative prompt refinement, although it can substantially improve the extraction quality.

\subsection{Fine-tuning of generative LLMs}
\label{subsubsec:finetuning}
Prompting has two structural limitations for materials data extraction. First, many tasks are highly domain-intensive and thus require specialised knowledge, such as dopant notation, phase labels, unit conventions, and composition syntax, that if often poorly represented in general-purpose foundation models, limiting their performance even with carefully designed prompts. Second, encoding such domain knowledge through prompting can become increasingly costly and impractical. While performance of generic LLMs can be improved using many-shot prompting with hundreds or thousands of examples, this approach is constrained by context-window limits, increases per-call token costs, and can suffer from reduced performance as relevant information gets buried in long contexts \citep{agarwal2024manyshot,liu2024lostmiddle}. Fine-tuning, instead, transfers this domain-specific knowledge into the model parameters, improving the model's ability to handle specialized terminology and tasks while consistently adhering to fixed output schemas. These improvements can even be seen in zero-shot settings, allowing shorter deployment prompts and lower inference costs per extracted record, which becomes important when processing millions of paragraphs.


Two variants of fine-tuning are used in practice. First is the domain-adaptive (continued) pretraining that retains the self-supervised next-token objective but applies it to a large unlabeled in-domain corpus, shifting the model's tokenization and terminology toward materials text. Supervised fine-tuning, in contrast, trains the model on labeled prompt-completion pairs---for example, a passage and prompt paired with the desired JSON record---teaching both the extraction task and the target schema \citep{dagdelen2024structured}. As discussed earlier in the context of transformer (BERT) models, full fine-tuning of billion-parameter models can be computationally expensive, thereby motivating parameter-efficient fine-tuning methods such as LoRA \citep{hu2022lora}. 
QLoRA further reduces memory requirements by storing the base model in 4-bit precision, enabling adaptation of very large models on a single GPU \citep{dettmers2023qlora}. In materials data extraction, however, the primary bottleneck in full fine-tuning is often the availability of expert annotated datasets rather than the compute.

In an exemplary applications of fine-tuned LLMs, GPT-3 and Llama-2 models were fine-tuned on only a few hundred ($\sim$500) annotated prompt-completion pairs to perform joint NER and RE across three materials-chemistry schemas, linking dopants to host materials, cataloguing metal-organic frameworks, and extracting composition, phase, morphology, and application information \cite{dagdelen2024structured}. The study released both the annotations and extracted JSON records as reusable datasets (with the best dopant $F_1$ score of $\sim$0.82). Comparable fine-tuned pipelines have been used to construct structured solar-cell datasets for downstream property prediction \citep{xie2024solarcell} and for substitution mining in resource-criticality assessment \citep{duan2025substitution}. Fine-tuning has also been used to adapt GPT-3 for low-data materials-property prediction, including high-entropy-alloy phase classification and molecular solubility \citep{jablonka2024predictive}. More recently, an LLM-driven workflow was used to construct a superconductivity database containing 78,203 records for 19,058 unique compositions, which was then used to LoRA-fine-tune open-source models for superconductivity classification, $T_c$ prediction, and conditional generation of candidate compositions and space groups \citep{ref50}. As an example of domain-adapted materials foundation models, LLaMat was pretrained on $\sim$30B tokens of materials text and crystallographic data followed by instruction tuning, which reportedly outperformed substantially larger general-purpose commercial models on structured materials extraction tasks \citep{ahlawat2026llamat}. We however note that large commercial LLMs combined with prompt engineering often provide a simpler and more accessible approach to data extraction than fine-tuned models. Fine-tuning is more advantageous for highly specialized tasks, such as crystal structure generation \cite{choudhary2024atomgpt, choudhary2025diffractgpt}, where task-specific capabilities may be difficult to achieve through prompting alone. 

\subsection{Agentic systems}
\label{sec:agentic}
Before discussing agentic systems, it is important to recognize that materials-data extraction is often a multi-step process requiring information to be located, interpreted, and linked across different sections and modalities of a paper. For example, material composition may be reported in the Results section, synthesis conditions in the Experimental section or Supporting Information, and properties in tables or figures \cite{zheng2023chatgptmof, jiang2025applications}. Sample names and abbreviations may also be defined in one part of the article and used throughout the remainder of the text. Thus, establishing a complete composition-processing-structure-property record requires repeated retrieval of relevant information, resolution of cross-references and abbreviations, and association of information reported at different locations. Passing an entire article to a model in a single pass conflates these sub-decisions, whereas naively dividing the article into fixed-size passages to accommodate context-window limits destroys precisely these cross-section dependencies. Prompting or fine-tuning alone cannot address this need for dynamic, multi-step information retrieval.

A second limitation is that different extraction subtasks may require different capabilities. Text, figures, and tables have distinct representations and may be better handled by specialized language, vision, or table-based models rather than an expensive general-purpose LLM. Conversely, a large general-purpose LLM may provide high accuracy for complex reasoning tasks but be unnecessarily expensive for simpler operations such as unit conversion or abbreviation resolution. Thus, an effective extraction pipeline may involve dynamically selecting different models or tools depending on different subtasks, rather than relying on a single model throughout. Recent studies corroborate this, showing that decomposing materials-data extraction into specialized subtasks can improve both accuracy and efficiency \citep{odobesku2025nanominer,roy2026comproscanner}.

This motivates the need of agentic approach, in which extraction is treated not as a single model inference or a predefined sequence of steps, but as a dynamically coordinated, multi-step process. An LLM-based agent can be viewed as a goal-directed system in which an LLM determines which tools or APIs to invoke, interprets their outputs, and decides the next action based on the information obtained. An \emph{agentic system} extends this concept by coordinating one or more agents, tools, models, and external memory through a control algorithm to accomplish tasks that cannot be reliably completed by a single LLM call or a fixed sequence of operations \citep{abou2025agentic}. This shift from single-shot extraction to adaptive, coordinated, and multi-step processing via agentic systems represents the current frontier of materials knowledge extraction.

\begin{figure}[h]
    \centering
    \includegraphics[width=1\linewidth]{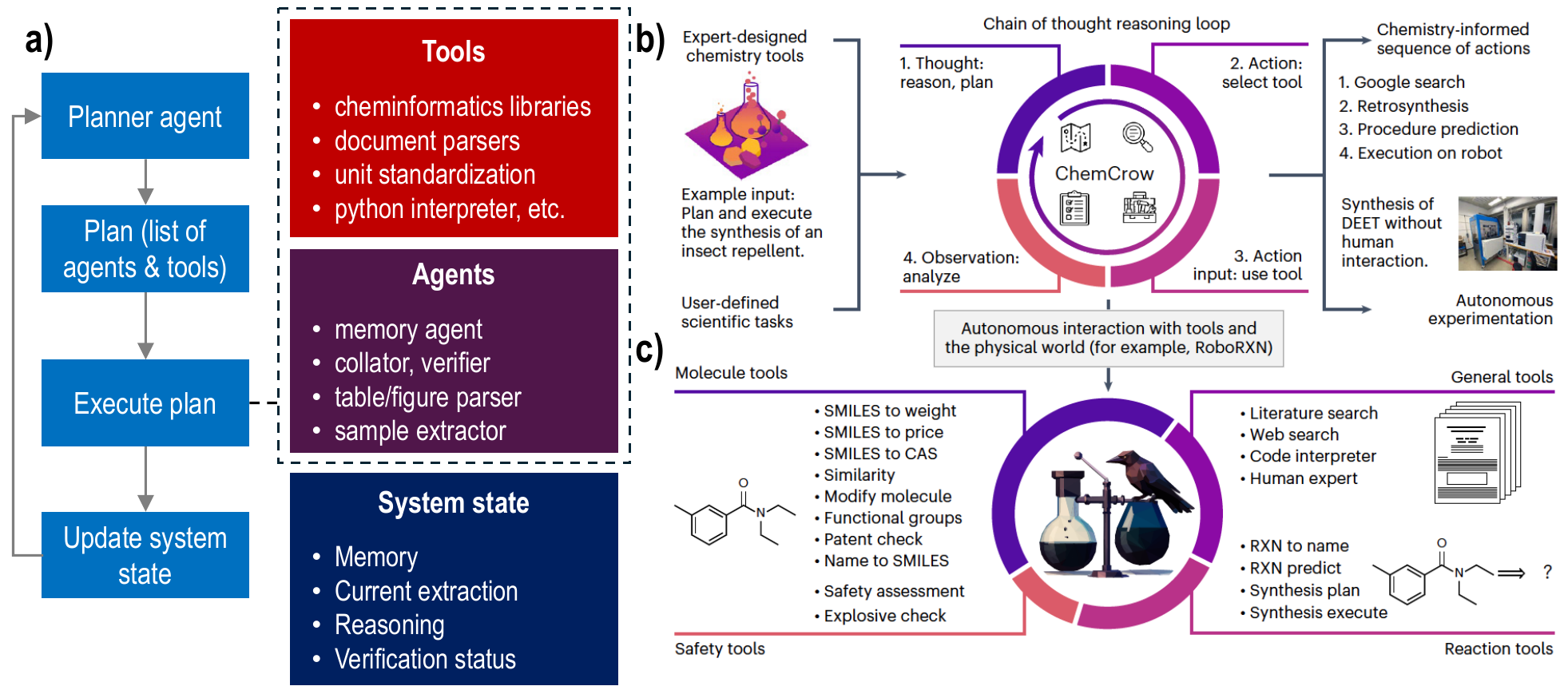}
    \caption{Agentic approaches for materials data extraction: (a) a sample agentic extraction workflow within the Plan-and-Solve framework supported by specialized tools, auxiliary agents, and memory; (b) and (c) a representative chemistry-focused autonomous agent architecture illustrating iterative reasoning, tool selection, observation, and autonomous execution using literature, molecule, reaction, and safety tools. Reproduced from Ref. \cite{bran2024chemcrow}, licensed under the Creative Commons Attribution 4.0 International License.}
    \label{agenticAI}
\end{figure}

\subsubsection{Architecture and design choices of agentic systems}
Agentic systems can be characterized by largely five design choices: the control algorithm that determines how LLM actions are sequenced, the degree of autonomy granted to them, the division of labour and specialized role among agents, the tools available to them, and the management of memory and state for coordination. Together, these choices determine how effectively and efficiently unstructured documents can be transformed into structured, validated, and useful materials records.

The first, and perhaps the most critical, design choice is the control algorithm that determines how an agent reasons and sequences its actions, and here the materials field has borrowed a well-defined toolbox from the general agent literature. ReAct-style interleaving of reasoning and tool calls is widely used in materials and chemistry applications, including nanoMINER and LLaMP \citep{yao2023react, odobesku2025nanominer, chiang2024llamp, mcnaughton2024cactus}. In ReAct, the agent proceeds one step at a time. At each step it first reasons about the query based on what it already knows, then issues a tool call action, such as a database query, a literature search, or a code execution. It then reads the returned observation/result and decides on its next step, repeating until the final answer is produced. Because each decision is conditioned on the result of the previous tool call, the agent can ground its reasoning in retrieved evidence, recover from failed tool calls, and adapt its plan as new information becomes available. Other control algorithms include Plan-and-Solve \citep{wang2023planandsolve}, ReWOO \citep{xu2023rewoo}, Reflexion \citep{shinn2023reflexion}, Self-Refine \citep{madaan2023selfrefine}, Tree-of-Thoughts \citep{yao2023tot,zhou2024lats}, and router architectures such as MRKL \citep{karpas2022mrkl}. These reasoning strategies are implemented within an \emph{agent harness}, which provides the execution infrastructure for message passing, tool invocation, state management, error handling, and context-window constraints. Examples include general frameworks such as AutoGen and LangChain/LangGraph, as well as custom harnesses developed for individual applications \citep{wu2023autogen,sumers2023coala,liu2023agentbench}.

Cutting across the choice of control algorithm is a distinct and, we argue, under-discussed design variable, degree of autonomy, i.e., how much of the action sequence is decided by the developer in advance versus by the agent at run time. At one extreme, the workflow is a fixed directed graph in which agents execute predefined steps, as in the LangGraph implementation of Ghosh et al. \cite{ghosh2026agentic} and role-based architectures such as ComProScanner \cite{roy2026comproscanner}. Intermediate architectures introduce conditional branches, allowing the agent to choose among a bounded set of actions according to the information obtained during execution. ChemGraph, for example, distinguishes fixed directed edges from conditional edges and allows further tool calls to be selected at run time \citep{pham2026chemgraph}. At the other extreme, agents can formulate and revise open-ended plans based on intermediate results, as demonstrated in systems such as Coscientist, AgentCAT, and CICERO \citep{boiko2023coscientist,yang2026agentcatllmagentextracting,ritchhart2026cicero}, and the six-role co-scientist architecture \citep{gottweis2025coscientist}. For materials data extraction, higher autonomy is particularly useful when the required information, its location, or the sequence of operations cannot be specified reliably in advance---for instance, when it is unknown \emph{a priori} which modality contains a particular data field, or when the target schema contains deeply nested or interdependent fields, as is often the case with materials synthesis.

The third design choice concerns how the extraction task is decomposed into well-scoped sub-tasks, each assigned to an agent with its own prompt, schema, and, often, its own backbone model. Specialization may be functional, with separate agents responsible for composition, properties, synthesis, formatting, or verification, or it may be modality-specific, with different models handling text, tables, figures, and captions. ComProScanner, for example, assigns separate agents to composition, property, synthesis, formatting, and screening \citep{roy2026comproscanner}, while other systems employ simpler task-based decompositions \citep{odobesku2025nanominer,chen2025chemeagle,ghosh2026agentic}. Since the LLM model can be selected according to the complexity and requirements of each subtask, such specialization can improve both extraction accuracy and computational efficiency of the agentic workflow. Importantly, verification can itself be assigned as a specialized role to a dedicated agent. ComProScanner uses a screening agent to check extracted records \citep{roy2026comproscanner}, while an ``LLM-Judge'' was employed to identify records that violate schema or physical constraints \citep{ghosh2026agentic}. Such extract-verify architectures provide an additional opportunity to detect and correct errors before records are incorporated into the final database.

The fourth design aspect is the external tools made available to agents to perform operations more reliably than error-prone token generation. These include cheminformatics and reaction libraries, simulation packages, database and property-prediction APIs, unit-conversion utilities, schema validators, and chemical-name normalization services, among others. ChemCrow (see Figure \ref{agenticAI}c), for example, provides an LLM reasoner with access to 18 expert-designed chemistry tools \citep{bran2024chemcrow}. CACTUS provides property-prediction and similarity-search functions \citep{mcnaughton2024cactus}, while ChemGraph exposes simulation capabilities through Python functions that wrap established libraries such as RDKit and the Atomic Simulation Environment, allowing quantities to be computed rather than recalled \citep{pham2026chemgraph}. ChatMOF, LLaMP and AGAPI similarly connect LLMs to materials databases and prediction models \citep{kang2024chatmof, chiang2024llamp, lee2026agapi}. Such tools are particularly valuable for deterministic operations such as unit conversion, composition checks, and schema validation. SciKGExtract, for example, combines schema-guided LLM extraction with deterministic cleaning, unit conversion, and chemical-name normalisation using resources such as PubChem \citep{sadruddin2026scikgextract}.

The final design choice is the explicit management of the agentic system's state and memory. Because information often needs to be accumulated across sections, documents, or multiple extraction steps, intermediate results such as partial records, resolved abbreviations, sample identities, and provenance need to be stored outside context window of a specific LLM agent, often in form of a structured JSON objects, relational records, or knowledge graphs. This temporary information can be selectively retrieved when needed \citep{chiang2024llamp,ghafarollahi2025sciagents}. This allows agents to revisit previously extracted information without repeatedly supplying the entire document. AgentCAT, for example, represents catalytic data as an explicit graph linking catalysts, reactions, conditions, and performance \citep{yang2026agentcatllmagentextracting}, while CICERO maintains experimental provenance throughout its workflow \citep{ritchhart2026cicero}. External memory therefore serves not only as persistent storage but also as a coordination layer through which specialized agents can share intermediate results and progressively construct a complete materials record efficiently. An example agentic workflow based on Plan-and-Solve framework and various associated tools, agents and memory state is shown in Figure \ref{agenticAI}a.

\subsubsection{Databases produced by agentic pipelines}
Beyond prompt engineering, some of the most comprehensive databases at scale have been produced using the agentic systems. We organize the discussion by placing them into three broad classes, i) multi-entity linked property databases built from text alone, ii) databases that additionally required fusing evidence across text, tables and figures, and, iii) databases that an agentic system not only mines from existing literature but also generates through its own actions. The first, and perhaps the most common, is the construction of comprehensive databases combining several aspects of materials (such as structure, property, composition, and even synthesis and processing history) into a single structured record, moving beyond the extraction of isolated material-property pairs as prevalent in the pre-LLM era approaches. For example, a thermoelectric database was constructed using a LangGraph-orchestrated, four-agent workflow with dynamic token allocation, a dedicated table agent, and an optional LLM-Judge to process roughly $10^{4}$ full-text articles \citep{ghosh2026agentic}. The resulting database contains 18,869 unit-normalized, temperature-resolved records covering Seebeck coefficient, electrical conductivity and resistivity, thermal conductivity, power factor, and $ZT$, together with structural attributes such as crystal class, space group, and doping strategy, making it larger than previously discussed databases based on both the rule-based and the earlier presented LLM pipeline. Another example is ComProScanner, which uses multi-agent extraction to link mechanical properties and phase information to processing histories in the piezoelectric and functional-ceramics literature, thereby connecting synthesis recipes with measured properties \citep{roy2026comproscanner}. Similarly, ALLMAT processed 1,531 high-entropy-alloy papers through a three-stage pipeline of labelling, extraction, and post-processing. Rather than processing paragraphs independently, it jointly considered synthesis, phase, and strength paragraphs with the title and abstract, allowing structural descriptors such as ``fine grains'' to be linked to the corresponding processing route and composition. The resulting database contains 5,393 records linking composition, yield strength, ultimate strength, strain, grain size, phase, and processing history, with $F_1$ scores ranging from 0.89 for grain size to 0.98 for yield strength \citep{ref32}. 

Other examples include AgentCAT, which constructs structured catalytic reaction data linking catalysts, reactions, operating conditions, and performance through a LangGraph-orchestrated pipeline of planning, schema-evolution, extraction, and review agents. The review agent enables closed-loop re-extraction by providing error hints, while the resulting reaction network is stored as a persistent, schema-evolving knowledge graph with traceability to the source PDFs \citep{yang2026agentcatllmagentextracting}. SciKGExtract, similarly, combines LLM-based extraction with deterministic processing to construct atomic-layer-deposition knowledge bases for ZnO and IGZO, using separate extraction, reflection, feedback, and orchestration agents to implement self-correction and a dedicated lookup tool to normalize compound names against PubChem \citep{sadruddin2026scikgextract}. Instead of multiple agents, some works employ a single prompted LLM with various tools. For example, the MOF synthesis database used a single prompted LLM augmented with deterministic utilities for text chunking, embedding-based retrieval, and tabulation \citep{zheng2023chatgptmof}. The MOF-SPK adopts a similar tool-augmented, single-model approach but at a larger scale, combining 133,737 CSD structures and computed pore properties with application, stability, and characterization information mined from more than 35,000 papers, and yielding over 100,000 MOF records \citep{ref63}. Likewise, an LLM-based text-mining workflow for lithium-metal batteries used specialized LLM roles for cell-name extraction, categorization, and value extraction across 29 entity types, and subsequently linked these records with independently extracted cyclability data using the MatGD graph-mining tool, producing 8,074 data points with 0.99 $F_1$ score for text-graph matching \citep{ref30}. A two-dimensional-materials database took a different route across about 50,000 papers, converting the entire literature corpus to plain text and extracting synthesis and performance data with a single fine-tuned model rather than a multi-agent pipeline \citep{ref66}. 
Collectively, these studies highlight the advanced capability of agentic system to construct multi-entity linked materials records.

A second class of databases contain information extracted across multiple modalities, particularly text, tables, figures (including molecular or reaction graphics). nanoMINER combines fine-tuned NER agents, a GPT-4o text extractor, and a YOLO detector with GPT-4o-based image interpretation to construct experiment-level records linking nanozyme kinetics ($K_m$ and $V_{max}$) with reaction conditions and particle structure \citep{odobesku2025nanominer}. Notably, the vision component provided information that could not be recovered from text alone, particularly substrate concentrations reported only in plots. OpenChemIE similarly links drawn molecular structures to textual labels and reconstructs R-groups from reaction schemes, achieving an $F_1$ score of 0.69 on substrate-scope diagrams \citep{fan2024openchemie}. ChemEAGLE further introduced an agentic planner that dynamically selects specialized agents for template parsing, R-group substitution, condition interpretation, and text alignment, improving end-to-end $F_1$ score from 0.39 to 0.76 \citep{chen2025chemeagle}. Similarly, multimodal extraction has been applied to additive-manufacturing and high-entropy-alloy literature, where process parameters and melt-pool geometry, or mechanical measurements, are reconstructed from graphical and textual evidence \citep{chandrasekhar2026materialdatabaseagentmultimodal}. Multimodal capabilities really help the agentic pipelines to overcome the issue of entity resolution at the document level by combining information distributed across the document.

The last and the more advanced class of datasets is those generated by agents through their own actions rather than mined directly from the published literature. CICERO provides a representative example, wherein near autonomous separations campaigns were used to produce provenance-complete, machine-readable records of hypotheses, executed experiments, and experimental outcomes for critical-material recovery from produced water and magnet leachates \citep{ritchhart2026cicero}. Importantly, it doesn't construct a static database of existing literature, but the database progressively expands as the agent formulates hypotheses, selects and executes experiments, observes their outcomes, and records the resulting evidence. The resulting provenance enables the accumulated experimental knowledge to be reused in subsequent research. ChemGraph provides a related example in computational chemistry, where structured-output modes convert the intermediate results of agentic computational workflows into machine-readable records \citep{pham2026chemgraph}.

\subsubsection{From extraction to autonomous discovery: Co-scientists and closed-loop agents.} 
Thus far, we have discussed agentic systems primarily from the perspective of data extraction. However, their role can extend well beyond just structuring existing data to extracting materials trends/knowledge, reasoning over it, proposing new hypothesis or experiments, executing them through robotic APIs, and incorporating the resulting measurements into the database to guide subsequent experimental cycles, thereby closing the discovery loop. In this paradigm, the agent evolves from a data extractor to an informed-data generator and, ultimately, an autonomous materials-discovery system. More specifically, in these systems, literature and documentation retrieval, materials informatics tools, code execution, simulation, and laboratory APIs are all orchestrated within a reasoning loop to allow autonomous materials discovery. A representative example is ChemCrow \citep{bran2024chemcrow}, which uses a GPT-4 agent wrapped in an iterative thought-action-observation (ReAct) loop (Figure \ref{agenticAI}b) to select at each step from a curated library of expert-designed chemistry tools (Figure \ref{agenticAI}c), including cheminformatics utilities, reaction predictors, literature and web search, and a code interpreter, and then communicates with hardware by issuing API calls to IBM's cloud-connected RoboRXN synthesis platform. It autonomously synthesized the insect repellent DEET and three thiourea organocatalysts, with a built-in safety mechanism to prevent potentially problematic executions. A-Lab \citep{szymanski2023alab} provides another example of this paradigm, using Mitsubishi and UR5e robotic arms to dispense and mix powders, load and unload furnaces, and transfer samples to an XRD diffractometer for automated ML-based phase analysis, executing recipes proposed by a synthesis-similarity-based model supported by a precursor-completion model and an XGBoost temperature regressor. Iterating this loop, A-Lab realized 36 of 57 targeted novel inorganic compounds.

Other examples of agent-driven autonomous systems include Coscientist \citep{boiko2023coscientist}, a GPT-4-driven system that integrates internet search, retrieval of instrument/API manuals, code execution, and robotic-experimentation APIs. It generates and executes control code for an Opentrons OT-2 liquid-handling robot and was demonstrated to autonomously optimize palladium-catalyzed Suzuki and Sonogashira cross-coupling reactions. ORGANA \citep{darvish2024organa} similarly plans and executes physical experiments while directly controlling laboratory robots using visual feedback, successfully demonstrating solubility screening, pH measurement, recrystallization, and a 19-step parallel electrochemical characterization of quinone derivatives for flow-battery electrolytes. MatPilot \citep{ni2024matpilot} drives a physical ceramic solid-state sintering platform, covering raw-material dispensing, ball milling, sintering and molding, with Digital Holographic Microscopy and dielectric measurements providing characterization feedback to its optimization models. CICERO \citep{ritchhart2026cicero} agentic workflow 
drives automated synthesis and characterization instrumentation to execute selective-precipitation recipes, recovering magnesium from oil-and-gas produced water, and neodymium, praseodymium and samarium from rare-earth magnet leachate, thereby reducing optimization times from months to days using low-cost commodity precipitants. 

Then there are partly autonomous labs where all tasks except the final experimental validation is done by the agentic system. For instance, the AI co-scientist \citep{gottweis2025coscientist} organizes specialized Generation, Reflection, Ranking, Evolution, Proximity, and Meta-review agents (built on Gemini) into a generate-debate-evolve tournament, in which agents iteratively debate and critique candidate drugs to converge on stronger ones. Applied to acute myeloid leukemia, the system proposed drug-repurposing candidates that were subsequently validated in vitro by human laboratories. Similarly, the Virtual Lab \citep{swanson2024virtuallab}, mimics an academic setting, with a principal-investigator agent coordinating with domain-expert agents (an immunologist, an ML specialist, a computational biologist) to address a computational nanobody-design problem. The resulting 92 candidates were subsequently synthesized and experimentally validated by human researchers. 

Another class of agentic systems close the discovery loop in silico by  evaluating candidate designs against a simulator or a surrogate model rather than a physical experiment. LLMatDesign \citep{jia2024llmatdesign} used a propose-modify-evaluate-reflect strategy to iteratively edit crystal compositions and query surrogate property-prediction models, with its best-performing model (GPT-4o) reaching a target band gap in an average of 10.8 modifications versus 27.4 for random sampling. dZiner \citep{ansari2024dziner} performed several inverse-design tasks, such as improving CO$_2$ adsorption in MOFs, lowering the critical micelle concentration of surfactants, and designing small-molecule binders for a drug target, guided by natural-language design rules together with docking simulations or property-prediction surrogate models. Similar agentic systems have been built in the field of heterogeneous catalysts (ChemReasoner \citep{sprueill2024chemreasoner}), refractory alloys (AtomAgents \citep{ghafarollahi2024atomagents}), and even general ML research automation (AI Scientist \citep{lu2024aiscientist}).

\subsubsection{Thumb-rules for better agentic system}
\label{subsubsec:agent_setup}

Just as in the case of prompting, next we outline some guidelines for setting-up a efficient and accurate agentic system. Its important to understand that an LLM-agent needs a \emph{role} (a system prompt stating its objective and constraints), a set of \emph{tools} it may call, and some \emph{memory}, so that it can plan and act over several steps rather than stopping after a single completion. Further, an agent is not a special type of model---it is a regular LLM that can act as a planner, an extractor, or a verifier, purely based on how it is prompted and which tools it is given access to. A multi-agent system is, therefore, simply a collection of such LLM agents, each assigned a specific role, which are all orchestrated either as a fixed sequence of hand-offs (reproducible) or as a dynamically routed conversation (adaptable) \citep{ghafarollahi2025sciagents}. This orchestration forms the control loop (e.g., ReAct) and determines the degree of autonomy, as stated earlier. Depending on the anticipated complexity and the variability involved in each task, the degree of autonomy should be decided.


For literature extraction, a small set of recurring roles for LLM agents that are often found to be sufficient are: a planner/orchestrator that decomposes the task and routes work to other agents; a retriever that locates relevant passages, tables, or figures, typically using RAG \citep{chiang2024llamp}; extractors specialized for different modalities, including text, tables, and vision-language models for figures \citep{odobesku2025nanominer,fan2024openchemie}; a normalizer that canonicalizes formulae, identifiers, and units; and a verifier or critic that checks extracted records against the source and flags or corrects unsupported values \citep{roy2026comproscanner}. An effective strategy is to separate information extraction from schema formatting, i.e., the task of extraction is further split between two agents where the first identifies all potentially relevant information, and the second agent then maps the output from first agent in desired predefined schema. This helps to improve recall statistics as the first agent is not constrained by the final database structure.

Tools should be used for deterministic operations for which LLMs are either unnecessary or less reliable. Some popular tools include composition parsers that expand variable stoichiometries into explicit compositions (e.g., \(\mathrm{Na_{1-x}Li_xTiO_3}\) at \(x=0.1\) and \(0.3\)), unit converter, name-to-structure resolvers using lookup functions for PubChem, Materials Project or crystal-structure databases, Python interpreters for arithmetic, JSON-schema validators, and vector stores or search indices for retrieval. ChemCrow \citep{bran2024chemcrow}, for example, augmented GPT-4 with eighteen chemistry tools and demonstrated that tool grounding can both improve model capabilities and reduce hallucinations. 
Its best to think of memory in two forms, a temporary working memory that is mainly used to store intermediate results and pass them between agents, and a more persistent, curated and condensed knowledge base that multiple agents consult and update. MatSciKB \citep{zhang2024honeycomb} is a good example demonstration of this. In terms of software infrastructure, frameworks such as CrewAI \citep{crewai2024}, LangGraph \cite{langgraph2024}, and AutoGen \cite{wu2023autogen} are popular libraries for integrating these components (agents, tasks, tools, memory and hand-offs) into complete workflows.

Some other practical guidelines for building an agentic system are as follows. First, the choice of LLM should match the complexity of the task. The planner/orchestrator and verifier should use the strongest available reasoning models, since poor task decomposition or inadequate verification can propagate errors through the entire pipeline. In contrast, high-volume, narrow-scope tasks such as chunk-level extraction or formatting can often be handled by smaller or open-weight models at lower cost. Second, deterministic operations, such as enumeration, arithmetic, canonicalization, and schema validation, should be implemented as tools rather than delegated to LLMs. Third, each agent should have a sufficiently narrow scope that its output can be validated independently. Fourth, it is preferable to start with a small system of two or three agents and introduce additional roles only when they provide a measurable improvement in precision or recall, since each additional hand-off adds latency, cost, and a potential failure mode. Accordingly, a well-defined evaluation metric and a benchmark dataset are therefore essential for determining whether a change in any part of the pipeline genuinely improves its performance. Finally, failures in multi-agent systems are difficult to trace because of the relatively longer extraction pipeline. Thus, it important to log every prompt, tool call and intermediate record to allow traceability.

\newcommand{\Ent}{\mathcal{E}} \newcommand{\Rel}{\mathcal{R}}
\newcommand{\Tri}{\mathcal{T}} \newcommand{\vh}{\mathbf{h}}
\newcommand{\vr}{\mathbf{r}} \newcommand{\vt}{\mathbf{t}}

\subsection{Knowledge graphs: structuring extracted knowledge for reasoning}
\label{subsec:knowledge_graphs}
\begin{figure}[h]
    \centering
    \includegraphics[width=1\linewidth]{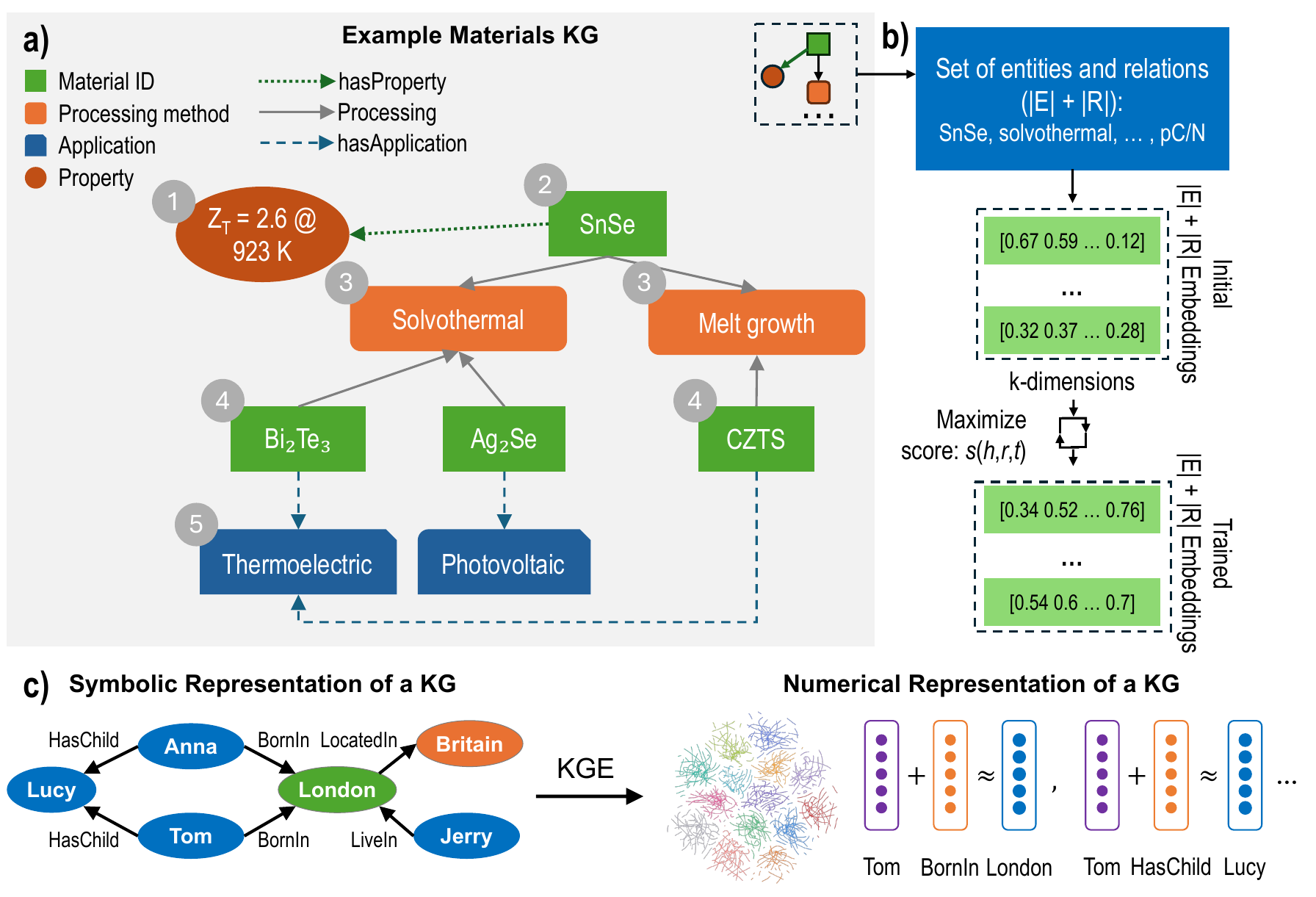}
    \caption{Knowledge graph and graph-embedding approaches. (a) a materials knowledge graph representing entities, material instances, synthesis routes, properties, and applications with an example of multi-hop connectivity; (b) graph representation learning, where entities and relations are encoded using trainable embeddings and score functions in a chosen embedding space; (c) Knowledge Graph (left) to a low-dimensional continuous vector embeddings (right), where relational triples are embedded as vector additions (TransE) \cite{niu2024knowledge, bordes2013transe}.}
    \label{fig:KG}
\end{figure}

The extraction methods discussed so far mostly produce records in the form of flat tuples such as $\langle$\texttt{material}, \texttt{property}, \texttt{value}, \texttt{condition}$\rangle$, which are well suited for downstream property prediction models for materials discovery. A knowledge graph (KG), instead, provides a different representation, in which entities become nodes and relationships become directed edges, each with a specific type (e.g., \emph{material}, \emph{has\_property}). The literature thus becomes a navigable network of \(\langle\texttt{subject}, \texttt{relation}, \texttt{object}\rangle\) triples rather than as rows in a table. Formally, a KG can be represented as \(G=(\Ent,\Rel,\Tri)\), where \(\Ent\) denotes entities (nodes), \(\Rel\) relation types (edges), and $\Tri=\{(h,r,t)\mid h,t\in\Ent,\; r\in\Rel\}$ the observed triples, with head entity $h$ relating to tail entity $t$ via relation $r$ \citep{hogan2021kg}.

Building a KG is not the same as building a tabular database. Given a set of extracted triples $\mathcal{T} \subseteq \mathcal{E} \times \mathcal{R} \times \mathcal{E}$ involving entity types such as \(\texttt{Material}\), \(\texttt{Property}\), \(\texttt{Application}\), \(\texttt{Processing}\), and \(\texttt{DOI}\), and relation types such as \(\texttt{hasProperty}\), \(\texttt{hasApplication}\), \(\texttt{hasProcessing}\), and \texttt{hasDOI}, the knowledge graph is assembled node-by-node and edge-by-edge as each triple is ingested. For an incoming triple \((h,r,t)\), the pipeline first checks whether nodes corresponding to \(h\) and \(t\) already exist under their declared types. If not, they are instantiated. An edge of type \(r\) is then created between the corresponding nodes. For example, \(\texttt{hasProperty}\) links a \(\texttt{Material}\) to a \(\texttt{Property}\), \(\texttt{hasApplication}\) links a \(\texttt{Material}\) to an \(\texttt{Application}\), and \(\texttt{hasProcessing}\) links it to a \(\texttt{Processing}\) entity describing its synthesis or treatment route, while \(\texttt{hasDOI}\) can link the source \(\texttt{DOI}\) node to all other nodes (e.g., \(\texttt{Material}\), \(\texttt{Property}\), etc.). This procedure is repeated over the full triple set, yielding a graph \(\mathcal{G}=(\mathcal{E},\mathcal{R},\mathcal{T})\) formed by the union of the inserted nodes and edges.
Duplicate triples can be collapsed while retaining per-node provenance via the \texttt{hasDOI} edge. The resulting graph supports queries such as retrieving all properties or applications reachable from a given material, or tracing a given property or application claim back to its source record via the attached \texttt{DOI} node.

Another important aspect of a KG is its ontology, i.e., the entity and relation types, and the rules governing how they can be connected. Ontology construction of a KG, similar to relational database schema design, is mostly an expert-driven process since it must encode the relevant domain knowledge. For example, to make property values useful for downstream analysis, the ontology should allow testing conditions and measurement methods to be explicitly associated with property values. To reduce reliance on predefined schemas and expert-driven design, recent works have explored automating ontology construction with LLMs. Zero-shot approaches have been used to first prompt an LLM to propose a schema and then construct the graph directly from raw document text (e.g., ``propose the entity and relationship types needed to represent this passage, then extract triples using them'' \cite{oarga2026scientific}). This has been demonstrated to construct chemical elements and functional groups ontology and to build first ontology for single-atom catalysts, before extracting data and constructing the KG in that domain. \citet{buehler2024graphreasoning} pushes this further by growing the graph iteratively, merging each document's proposed nodes into what already exists, with no schema imposed from outside. Interestingly, the resulting networks exhibited scale-free structure across multiple domains, with hub structures emerging from the extracted data rather than being explicitly designed. Although such automation can reduce the cost of ontology development, it also introduces other challenges, including LLM-induced biases in the resulting ontologies, divergence from established community standards such as EMMO \citep{emmo2023,del2022chameo}, and a continued need for expert validation. Ontology construction therefore remains a largely human-in-the-loop process rather than a fully automated one.

\subsubsection{Advantages of knowledge graphs}
The principal advantage of this representation is its ability to capture and query complex relationships without requiring a rigid, predefined relational schema. Let us understand this using the example in Figure \ref{fig:KG}a. Consider the query \emph{``which materials studied for thermoelectrics share a synthesis route with a known high $Z_T$ compound?''} This query can be answered using the following steps: i) resolve \texttt{Property} nodes with high $Z_T$ values, ii) reverse-traverse the \texttt{hasProperty} edge to find the associated \texttt{Material} nodes (SnSe), iii) from each identified \texttt{Material} node traverse the \texttt{processing} edge to find its \texttt{ProcessingMethod} node (``melt growth" and
``solvothermal"), iv) reverse-traverse the \texttt{processing} edge to find all other \texttt{Material} nodes synthesized via that same processing route, and v) retain only those \texttt{Material} nodes (Bi$_2$Te$_3$, CZTS) connected to an \texttt{Application} node valued ``Thermoelectric" via the \texttt{hasApplication} edge. Such multi-hop traversals are much easier in KGs as compared to relational databases, where the same operation would require multiple table joins. Graph query languages such as Cypher or SPARQL \citep{francis2018cypher,harris2013sparql} allow such multi-hop traversals to be expressed in a human-readable form.

A second advantage is that KGs provide an explicit relational structure for grounding LLMs. Conventional RAG retrieves semantically similar passages from a vector database, but the retrieved passages do not necessarily preserve the relationships among the underlying entities. KG-based RAG, termed GraphRAG, instead identifies entities relevant to the user’s query, retrieves their surrounding subgraphs to a specified number of hops, and converts these subgraphs into context by either serializing them as triples or rendering them as concise natural-language descriptions. This allows the LLM to receive both the supporting facts and the relationships connecting them \citep{edge2024graphrag, peng2024graphrag}, something not possible in the traditional RAG approach. Because nodes and edges can retain links to their source documents, this representation can also provide explicit provenance. The KG therefore serves not only as a structured knowledge store, but also helps to mitigate out-of-distribution hallucinations while offering some degree of interpretation to the generated LLM's response \citep{buehler2024generative}.

A third advantage is that traversal in a graph can be used to derive quantities that may never be explicitly present in the source record. PropNet, for example, represents materials-property relationships as a graph in which edges encode physical relations that compute one property from others \citep{mrdjenovich2020propnet}. Given a subset of measured properties, the graph thus can be used to derive additional quantities by traversing through applicable edges---for instance, elastic modulus and density nodes can yield sound velocity, which in turn provides a Debye-temperature estimate. This provides a natural mechanism for augmenting sparse literature-derived records and, in cases where multiple derivation paths exist, for identifying inconsistencies between alternative models or for uncertainty quantification.

The fourth advantage extends beyond the facts explicitly encoded in the graph. A KG traversal can only retrieve relationships present in \(\Tri\), whereas a literature-derived KG is inherently incomplete since most of the true material-property and material-application pairs have never been written down explicitly. The concept of knowledge graph embeddings (KGE), discussed next, addresses this limitation by learning vector representations of graph's entities and relations such that already observed triples receive high plausibility scores. These learned representations can then be used to score and rank candidate hypothetical triples, thereby identifying top-ranked triples as potentially meaningful relationships \cite{wang2017knowledge, ji2021kgsurvey}.

Despite these advantages, current KG-based methods face the same entity-normalization challenge as relational databases. The same real-world material, property, or other entity may be described inconsistently across papers using synonyms, abbreviations, or minor lexical variants. So naively treating each mention as a distinct node fragments the graph and can obscure cross-paper relationships and trends. Normalization, the process of mapping such variants to a single canonical entity, is therefore essential but nontrivial. For example, ``LiCoO$_2$", ``lithium cobalt oxide", and ``LCO" should map to a single canonical \texttt{Material} node rather than three separate nodes. Some pipelines address this explicitly. MatKG, for example, clusters candidate entity mentions using Levenshtein distance and then uses an LLM to assign a canonical label to each cluster \citep{venugopal2024matkg}. The related MKG pipeline combines chemical-formula extraction, formula/name disambiguation, and an expert-curated dictionary derived from clustering to normalize entities before graph construction \citep{ye2024mkg}. However, even with effective normalization pairwise triples can fail to preserve the complete composition–processing–property context. For example, if a canonical \texttt{Material} node is connected to two different synthesis routes reported in the same study, subsequent \texttt{hasProperty} or \texttt{hasApplication} edges do not necessarily indicate which processing route produced which outcome as all entities in this case will have same \texttt{DOI}.
This is one structural limitation of graphs built on pairwise triples.
Recent work addresses this by moving beyond pairwise triples to hypergraphs, in which a single hyperedge can jointly represent a material, its processing route, and the resulting property or application as one co-occurring fact \citep{stewart2026higherorder}.

\subsubsection{Knowledge graph embeddings}

In a typical KGE model, each entity and relation is represented by a trainable embedding vector of dimension \(k\), together with a predefined scoring function \(s(h,r,t)\) that measures the plausibility of a triple (see Figure \ref{fig:KG}b). Training uses observed triples as positive examples and corrupted triples, generated by randomly replacing the head or tail with another entity, as negative examples. The loss function increases the scores of positive triples and decreases those of negative triples, thereby learning the entity and relation embeddings in this process. During inference, a query with one missing entity, such as (\texttt{MaterialID}, \texttt{hasApplication}, \(?\)) or (\(?\), \texttt{hasApplication}, \texttt{thermoelectrics}), is scored against candidate entities and ranked accordingly, enabling the prediction of new applications for a known material or candidate materials for a target application. However, one issue with the learned KGE is that they are transductive, meaning they cannot provide embeddings of a new entity that was not present during the training. Nonetheless, this can be resolved by deriving entity representations using local graph structure \citep{galkin2022nodepiece} or entity names and textual descriptions \citep{demir2024inferenceunseenentitiesrelations}. 

The choice of scoring function and negative-sampling strategy can substantially affect the relational patterns learned by KGE models \citep{bordes2013transe,trouillon2016complex,dettmers2018conve,lacroix2018canonical,sun2019rotate,mohamed2019loss,ruffinelli2020olddog,ali2021pykeen}. The scoring function determines how relations are represented geometrically and, consequently, which relational patterns can be captured. For example, TransE models a relation as a translation, \(\vh+\vr\approx\vt\) (Fig. 5c), but this has difficulty representing patterns such as one-to-many (a material with many applications) or mutual relations \citep{bordes2013transe,rossi2021knowledge}. Thus, more expressive models, including TransH, TransR, RotatE, and ComplEx, introduce alternative transformations to accommodate different relation patterns \citep{wang2014transh, lin2015transr, trouillon2016complex, sun2019rotate}, as discussed in detail here \cite{cao2024kgesurvey}. The geometry of the embedding space is also important. Euclidean spaces may be inefficient for representing hierarchical relationships, whereas hyperbolic spaces can represent branching hierarchies more compactly at low dimensionality \citep{nickel2017poincare,chami2020atth}. Other approaches represent entities as distributions rather than points, allowing uncertainty or varying reliability to be incorporated into the learned representations \citep{he2015kg2e}.

\subsubsection{Applications enabled by knowledge graphs}
\label{subsubsec:kg_applications}

The applications of materials KGs can be viewed as a progression from organizing and retrieving existing materials knowledge to inferring hidden trends and assisting materials discovery. At the most basic level, KGs provide a structured representation in which complex relationships can be queried and traversed. 
For example, NanoMine linked materials, samples, processing and characterization steps from over 1,700 curated polymer-nanocomposite experiments, enabling composition, processing history, and properties to be queried together with full provenance \citep{mccusker2020nanomine}. At a larger scale, MatKG automatically constructed a materials-science KG containing more than 70,000 entities and 5.4 million triples, linking materials to properties, applications, and synthesis methods, and was released openly in both CSV and RDF formats \citep{venugopal2024matkg}. 

A second application is grounding the LLM response according to the structured materials knowledge present in the graph. Rather than relying solely on the model's parametric knowledge, a KG can be queried to retrieve the relevant entities and relationships, which are then supplied to the LLM as context. For example, \citet{bai2025kgfm} constructed a reticular-materials KG from more than \(10^{5}\) articles on MOFs, COFs, and HOFs, containing approximately 2.53 million nodes and 4.01 million edges. Given a user query, an LLM first translates the question into a Cypher query, retrieves the relevant subgraph from a Neo4j database, and then generates an answer based on the retrieved information, achieving 91.67\% QA accuracy while providing the supporting triples from graph as provenance. KGQA4MAT similarly uses an LLM to translate natural-language questions into Cypher queries over a curated MOF KG, enabling multi-hop question answering without requiring users to formulate graph queries themselves \citep{an2023kgqa4mat}. 

A third application is predicting plausible but previously unreported links between existing entities in the graph. For example, MatKG-2 first constructed an ontology-based KG from LLM-extracted triples and applied TransE, DistMult, and ComplEx KGEs to predict missing links, with TransE achieving a mean reciprocal rank of 0.77 \citep{venugopal2024matkg2}. FMKG similarly combined attribute-based similarity with TransE to predict previously unreported material-application relationships from a graph containing more than 160,000 nodes and 730,000 edges. Importantly, these predictions were evaluated by checking their accuracy against literature from a later date; the authors found that 48.5\% of the predictions made from 2014-era data were subsequently reported within nine years \citep{ye2024mkg}. Other approaches have similarly used KGs to identify new cathode materials (e.g., Li$_2$TiMn$_3$O$_8$) for battery application \cite{nie2022batterykg}.

The most advanced application extends this idea to generating hypotheses involving new relationships or concepts beyond those explicitly encoded in the graph. In SciAgents, for example, paths (triples) sampled from an ontological graph of biologically inspired materials were used as scaffolds to invoke LLM to generate research hypotheses. A selected path, such as ``spider silk'' \(\rightarrow\) ``\(\beta\)-sheet'' \(\rightarrow\) ``toughness'', provides the conceptual structure for a hypothesis that is subsequently expanded, critiqued, and refined by multiple LLM-driven roles (an ontologist, domain scientists, and a critic) \citep{ghafarollahi2025sciagents, buehler2024graphreasoning}. A related approach further extends this concept into a feedback loop \citep{buehler2025agentic}. Rather than reasoning over a static graph, it couples an LLM with a continually evolving graph representation. At each step, the model generates new concepts and relationships, incorporates them into the global graph, and uses the updated structure to formulate subsequent prompts, resulting in the co-evolution of the KG and the inference process. Iterated over many steps, this produces a graph with an emergent scale-free topology characterized by hub formation.

\section{Challenges and Opportunities}

Despite several successful applications of ML models to materials data and knowledge extraction, significant challenges remain. A fundamental difficulty in materials data extraction is the ambiguity in materials terminology, arising from the lack of uniform naming conventions across the field. The same material may be reported using a chemical formula, full name, abbreviation, or trade name, while property, processing, and characterization terms can likewise vary across subfields. Thus, even correctly extracted records may fail to link across papers as different terms are used to refer to the same entity, limiting cross-paper trend analysis. While various ad-hoc post-processing methods using hard-coded rules, regular expressions, or even LLMs, have been used, there is still need for a more general and robust solution. Linking the identified database entity to canonical identifiers, such as PubChem CIDs or materials-specific ontology IDs, offers a promising direction, but more standardized reporting standards will have to developed and adopted to overcome this.

Closely related to this is the need for standardized materials ontology that allow different extraction pipelines to populate compatible or uniform database schemas. An important limitation of current databases is that they often capture only fragments of the materials structure-processing-property-performance tetrahedron.
A material record to be considered complete should actually have all four information present together, however, in most of the extracted databases only two or three of these aspects are present, with others being missed. This limits the value of the database for downstream materials discovery. Thus, more comprehensive extraction pipelines that tracks every samples synthesis, processing, and resulting structure, property and performance details as a single record are needed. 

Another issue that persists is that of extraction from multimodal data. While recent agentic systems such as MaTableGPT \cite{yi2025matablegpt}, ComProScanner \cite{roy2026comproscanner}, MERMaid \cite{leong2025mermaid}, and MatMMExtract \cite{ghosh2026unlocking} have demonstrated the ability to extract information from tables, figures, and other multimodal content, such extraction remains substantially more challenging as well as costly than text extraction. One promising direction is to combine conventional tools such as plot digitization with specialized models for tables and figures. However, the formatting heterogeneity in tables and the compounded nature of figures, where a single caption may cover multiple sub-panels, make reliable extraction of plotted data an open and actively developing problem. Rather than relying on a general-purpose vision-language model to perform the entire process, a more efficient and reliable strategy is to decompose extraction into a series of distinct stages, involving figure or table identification, followed by numerical extraction, semantic interpretation, and, finally, verification. In long term, systemic changes in journal and publisher policies that encourage authors to provide the underlying numerical data for plots and tables as structured supplementary files can help to address this challenge.

Another major limitation is the absence of large, standardized, expert-annotated benchmarks spanning multiple materials classes and data types. Most studies still use small, domain- and task-specific ground-truth datasets, making it difficult to compare performance of different extraction approaches. Thus, a community-wide benchmark constructed from several hundred to thousand papers and encompassing diverse composition, structure, processing, and property records across text, tables, and figures would provide a valuable resource for standardized evaluation. The accessibility and format of the underlying literature is an additional bottleneck. Although many publishers provide programmatic access to their literature databases through APIs,
a substantial fraction of the literature is still inaccessible for automated mining or available only in more difficult to process formats such as PDFs, especially older publications before the 2000s. Thus, broader text- and data-mining licenses from publishers, together with greater availability of machine-readable full text in formats such as HTML or XML, is needed.


Another subtle but an important issue is that of publication bias. Scientific literature overwhelmingly reports successful, ``positive" outcomes, while failed synthesis attempts, negative or null results, and materials with poor performance are rarely published, even though such information is very much needed to learn and predict unsuccessful routes. While some of this information may be present in supplementary files, making them a potentially valuable target for extraction efforts, but addressing this issue will again require more systemic changes. Dedicated venues/journals for negative results or data-sharing policies that encourage the deposition of failed experiments and raw experimental records are some potential solutions.


Finally, the ultimate value of these extracted databases lies in the novelty of the knowledge and discoveries they enable.
Most studies end once the database has been built, but the larger goal should be to use the extracted data to generate new materials knowledge, actionable insights or identify novel materials that can subsequently be synthesized and experimentally validated.
While a few dedicated works \cite{ref5, pan2026diffsyn, ref1, ref22}
have indeed achieved this, more such examples are needed to establish the real value of such materials databases. Using LLM agents grounded in materials KGs for hypothesis generation and validation remains an attractive direction forward. Further, the impact of such databases can also be amplified by coupling such databases with autonomous laboratories, where ML-generated recommendations are directly followed by robotic synthesis and characterization, creating a closed self-driving loop spanning from literature corpus to materials discovery and experimental validation \cite{szymanski2023alab}.


\medskip

\section*{Author Contributions}
A.K.S and R.B. conceived the project. All the authors contributed to the data collection and preparation of the manuscript. All authors read and approved the manuscript. R.B. supervised and directed the overall project.

\section*{Acknowledgments} 
R.B. acknowledges ANRF (File No. SRG/2023/001055), DRDO-DIA and WSAI (SB25261127MMWDWI009014, SB25261508MMWDWI009014) for funding.

\section*{Conflicts of Interest}
The authors declare no competing interests.

\section*{Data Availability Statement}
Not applicable.


\medskip

%

\bibliographystyle{unsrtnat}
\bibliography{references}

@article{olivetti2020data,
  title={Data-driven materials research enabled by natural language processing and information extraction},
  author={Olivetti, Elsa A and Cole, Jacqueline M and Kim, Edward and Kononova, Olga and Ceder, Gerbrand and Han, Thomas Yong-Jin and Hiszpanski, Anna M},
  journal={Applied Physics Reviews},
  volume={7},
  number={4},
  year={2020},
  publisher={AIP Publishing}
}

@article{kononova2021opportunities,
  author  = {Kononova, Olga and He, Tanjin and Huo, Haoyan and
             Trewartha, Amalie and Olivetti, Elsa A. and Ceder, Gerbrand},
  title   = {Opportunities and challenges of text mining in materials research},
  journal = {iScience},
  volume  = {24},
  number  = {3},
  pages   = {102155},
  year    = {2021},
  doi     = {10.1016/j.isci.2021.102155}
}

@article{schilling2025text,
  author  = {Schilling-Wilhelmi, Mara and R{\'i}os-Garc{\'i}a, Marti{\~n}o and
             Shabih, Sherjeel and Gil, Mar{\'i}a Victoria and Miret, Santiago and
             Koch, Christoph T. and M{\'a}rquez, Jos{\'e} A. and
             Jablonka, Kevin Maik},
  title   = {From text to insight: large language models for chemical data
             extraction},
  journal = {Chemical Society Reviews},
  volume  = {54},
  number  = {3},
  pages   = {1125--1150},
  year    = {2025},
  doi     = {10.1039/D4CS00913D}
}

@article{yi2025matablegpt,
  title={MaTableGPT: GPT-Based Table Data Extractor from Materials Science Literature},
  author={Yi, Gyeong Hoon and Choi, Jiwoo and Song, Hyeongyun and Miano, Olivia and Choi, Jaewoong and Bang, Kihoon and Lee, Byungju and Sohn, Seok Su and Buttler, David and Hiszpanski, Anna and others},
  journal={Advanced Science},
  volume={12},
  number={16},
  pages={2408221},
  year={2025},
  publisher={Wiley Online Library}
}

@software{pdfplumber,
author = {Singer-Vine, Jeremy and {The pdfplumber contributors}},
license = {MIT},
month = jun,
title = {{pdfplumber}},
url = {https://github.com/jsvine/pdfplumber},
version = {0.11.10},
year = {2026},
note   = {Accessed: 16 October 2026}
}

@article{swain2016chemdataextractor,
  author  = {Swain, Matthew C. and Cole, Jacqueline M.},
  title   = {{ChemDataExtractor}: A Toolkit for Automated Extraction of Chemical
             Information from the Scientific Literature},
  journal = {Journal of Chemical Information and Modeling},
  volume  = {56},
  number  = {10},
  pages   = {1894--1904},
  year    = {2016},
  doi     = {10.1021/acs.jcim.6b00207}
}

@article{court2018curie,
  author  = {Court, Callum J. and Cole, Jacqueline M.},
  title   = {Auto-generated materials database of {Curie} and {N\'eel}
             temperatures via semi-supervised relationship extraction},
  journal = {Scientific Data},
  volume  = {5},
  pages   = {180111},
  year    = {2018},
  doi     = {10.1038/sdata.2018.111}
}

@article{sierepeklis2022thermoelectric,
  author  = {Sierepeklis, Odysseas and Cole, Jacqueline M.},
  title   = {A thermoelectric materials database auto-generated from the
             scientific literature using {ChemDataExtractor}},
  journal = {Scientific Data},
  volume  = {9},
  number  = {1},
  pages   = {648},
  year    = {2022},
  doi     = {10.1038/s41597-022-01752-1}
}

@article{weston2019matscholar,
  author  = {Weston, Leigh and Tshitoyan, Vahe and Dagdelen, John and
             Kononova, Olga and Trewartha, Amalie and Persson, Kristin A. and
             Ceder, Gerbrand and Jain, Anubhav},
  title   = {Named Entity Recognition and Normalization Applied to Large-Scale
             Information Extraction from the Materials Science Literature},
  journal = {Journal of Chemical Information and Modeling},
  volume  = {59},
  number  = {9},
  pages   = {3692--3702},
  year    = {2019},
  doi     = {10.1021/acs.jcim.9b00470}
}

@article{kononova2019synthesis,
  author  = {Kononova, Olga and Huo, Haoyan and He, Tanjin and Rong, Ziqin and
             Botari, Tiago and Sun, Wenhao and Tshitoyan, Vahe and
             Ceder, Gerbrand},
  title   = {Text-mined dataset of inorganic materials synthesis recipes},
  journal = {Scientific Data},
  volume  = {6},
  number  = {1},
  pages   = {203},
  year    = {2019},
  doi     = {10.1038/s41597-019-0224-1}
}

@article{huo2019semisupervised,
  author  = {Huo, Haoyan and Rong, Ziqin and Kononova, Olga and Sun, Wenhao and
             Botari, Tiago and He, Tanjin and Tshitoyan, Vahe and
             Ceder, Gerbrand},
  title   = {Semi-supervised machine-learning classification of materials
             synthesis procedures},
  journal = {npj Computational Materials},
  volume  = {5},
  pages   = {62},
  year    = {2019},
  doi     = {10.1038/s41524-019-0204-1}
}

@article{gatesrector2019pdf,
  author  = {Gates-Rector, Stacy and Blanton, Thomas},
  title   = {The Powder Diffraction File: a quality materials characterization
             database},
  journal = {Powder Diffraction},
  volume  = {34},
  number  = {4},
  pages   = {352--360},
  year    = {2019},
  doi     ={10.1017/s0885715619000812}
}

@article{kelly2025curieneel,
  title={A dataset of Curie and N{\'e}el temperatures auto-generated with ChemDataExtractor and the Snowball algorithm},
  author={Kelly, Charles R and Cole, Jacqueline M},
  journal={Scientific Data},
  volume={12},
  number={1},
  pages={2000},
  year={2025},
  publisher={Nature Publishing Group UK London}
}

@article{foppiano2024mining,
  author  = {Foppiano, Luca and Lambard, Guillaume and Amagasa, Toshiyuki and
             Ishii, Masashi},
  title   = {Mining experimental data from materials science literature with
             large language models: an evaluation study},
  journal = {Science and Technology of Advanced Materials: Methods},
  volume  = {4},
  number  = {1},
  pages   = {2356506},
  year    = {2024},
  doi     = {10.1080/27660400.2024.2356506}
}

@article{gupta2022matscibert,
  author  = {Gupta, Tanishq and Zaki, Mohd and Krishnan, N. M. Anoop and
             Mausam},
  title   = {{MatSciBERT}: A materials domain language model for text mining and
             information extraction},
  journal = {npj Computational Materials},
  volume  = {8},
  number  = {1},
  pages   = {102},
  year    = {2022},
  doi     = {10.1038/s41524-022-00784-w}
}

@article{jain2013materialsproject,
  author  = {Jain, Anubhav and Ong, Shyue Ping and Hautier, Geoffroy and
             Chen, Wei and Richards, William Davidson and Dacek, Stephen and
             Cholia, Shreyas and Gunter, Dan and Skinner, David and
             Ceder, Gerbrand and Persson, Kristin A.},
  title   = {Commentary: The {Materials Project}: A materials genome approach
             to accelerating materials innovation},
  journal = {APL Materials},
  volume  = {1},
  number  = {1},
  pages   = {011002},
  year    = {2013},
  doi     = {10.1063/1.4812323}
}

@article{curtarolo2012aflow,
  author  = {Curtarolo, Stefano and Setyawan, Wahyu and Hart, Gus L. W. and
             Jahnatek, Michal and Chepulskii, Roman V. and Taylor, Richard H.
             and Wang, Shidong and Xue, Junkai and Yang, Kesong and
             Levy, Ohad and Mehl, Michael J. and Stokes, Harold T. and
             Demchenko, Denis O. and Morgan, Dane},
  title   = {{AFLOW}: An automatic framework for high-throughput materials
             discovery},
  journal = {Computational Materials Science},
  volume  = {58},
  pages   = {218--226},
  year    = {2012},
  doi     = {10.1016/j.commatsci.2012.02.005}
}

@article{kim2017synthesis,
  author  = {Kim, Edward and Huang, Kevin and Saunders, Adam and
             McCallum, Andrew and Ceder, Gerbrand and Olivetti, Elsa},
  title   = {Materials Synthesis Insights from Scientific Literature via Text
             Extraction and Machine Learning},
  journal = {Chemistry of Materials},
  volume  = {29},
  number  = {21},
  pages   = {9436--9444},
  year    = {2017},
  doi     = {10.1021/acs.chemmater.7b03500}
}

@article{jensen2019zeolite,
  author  = {Jensen, Zach and Kim, Edward and Kwon, Soonhyoung and
             Gani, Terry Z. H. and Rom{\'a}n-Leshkov, Yuriy and
             Moliner, Manuel and Corma, Avelino and Olivetti, Elsa},
  title   = {A Machine Learning Approach to Zeolite Synthesis Enabled by
             Automatic Literature Data Extraction},
  journal = {ACS Central Science},
  volume  = {5},
  number  = {5},
  pages   = {892--899},
  year    = {2019},
  doi     = {10.1021/acscentsci.9b00193}
}

@article{tshitoyan2019word2vec,
  author  = {Tshitoyan, Vahe and Dagdelen, John and Weston, Leigh and
             Dunn, Alexander and Rong, Ziqin and Kononova, Olga and
             Persson, Kristin A. and Ceder, Gerbrand and Jain, Anubhav},
  title   = {Unsupervised word embeddings capture latent knowledge from
             materials science literature},
  journal = {Nature},
  volume  = {571},
  number  = {7763},
  pages   = {95--98},
  year    = {2019},
  doi     = {10.1038/s41586-019-1335-8}
}

@article{trewartha2022matbert,
  author  = {Trewartha, Amalie and Walker, Nicholas and Huo, Haoyan and
             Lee, Sanghoon and Cruse, Kevin and Dagdelen, John and
             Dunn, Alexander and Persson, Kristin A. and Ceder, Gerbrand and
             Jain, Anubhav},
  title   = {Quantifying the advantage of domain-specific pre-training on named
             entity recognition tasks in materials science},
  journal = {Patterns},
  volume  = {3},
  number  = {4},
  pages   = {100488},
  year    = {2022},
  doi     = {10.1016/j.patter.2022.100488}
}

@article{dagdelen2024structured,
  author  = {Dagdelen, John and Dunn, Alexander and Lee, Sanghoon and
             Walker, Nicholas and Rosen, Andrew S. and Ceder, Gerbrand and
             Persson, Kristin A. and Jain, Anubhav},
  title   = {Structured information extraction from scientific text with large
             language models},
  journal = {Nature Communications},
  volume  = {15},
  number  = {1},
  pages   = {1418},
  year    = {2024},
  doi     = {10.1038/s41467-024-45563-x}
}

@article{zheng2023chatgptmof,
  author  = {Zheng, Zhiling and Zhang, Oufan and Borgs, Christian and
             Chayes, Jennifer T. and Yaghi, Omar M.},
  title   = {{ChatGPT} Chemistry Assistant for Text Mining and the Prediction
             of {MOF} Synthesis},
  journal = {Journal of the American Chemical Society},
  volume  = {145},
  number  = {32},
  pages   = {18048--18062},
  year    = {2023},
  doi     = {10.1021/jacs.3c05819}
}

@article{jablonka2024predictive,
  author  = {Jablonka, Kevin Maik and Schwaller, Philippe and
             Ortega-Guerrero, Andres and Smit, Berend},
  title   = {Leveraging large language models for predictive chemistry},
  journal = {Nature Machine Intelligence},
  volume  = {6},
  number  = {2},
  pages   = {161--169},
  year    = {2024},
  doi     = {10.1038/s42256-023-00788-1}
}

@misc{khot2023decomposedpromptingmodularapproach,
      title={Decomposed Prompting: A Modular Approach for Solving Complex Tasks}, 
      author={Tushar Khot and Harsh Trivedi and Matthew Finlayson and Yao Fu and Kyle Richardson and Peter Clark and Ashish Sabharwal},
      year={2023},
      eprint={2210.02406},
      archivePrefix={arXiv},
      primaryClass={cs.CL},
      url={https://arxiv.org/abs/2210.02406}, 
}

@article{li2025slm,
  title={SLM-MATRIX: a multi-agent trajectory reasoning and verification framework for enhancing language models in materials data extraction},
  author={Li, Xin and Huang, Zhixuan and Quan, Shu and Peng, Cheng and Ma, Xiaoming},
  journal={npj Computational Materials},
  volume={11},
  number={1},
  pages={241},
  year={2025},
  publisher={Nature Publishing Group UK London}
}

@article{odobesku2025nanominer,
  author  = {Odobesku, Roman and Romanova, Ksenia and Mirzaeva, Sofia and
             Zagorulko, Oleg and Sim, Roman and Khakimullin, Rail and
             Razlivina, Julia and Dmitrenko, Andrei and Vinogradov, Vladimir},
  title   = {Agent-based multimodal information extraction for nanomaterials},
  journal = {npj Computational Materials},
  volume  = {11},
  number  = {1},
  pages   = {194},
  year    = {2025},
  doi     = {10.1038/s41524-025-01674-7}
}

@article{bai2025kgfm,
  title   = {Construction of a knowledge graph for framework material enabled by large language models and its application},
  author  = {Bai, Xuefeng and He, Song and Li, Yi and Xie, Yabo and Zhang, Xin and Du, Wenli and Li, Jian-Rong},
  journal = {npj Computational Materials},
  volume  = {11},
  pages   = {51},
  year    = {2025},
  doi     = {10.1038/s41524-025-01540-6}
}

@article{oarga2026scientific,
  title={Scientific knowledge graph and ontology generation using open large language models},
  author={Oarga, Alexandru and Hart, Matthew and Bran, Andres M and Lederbauer, Magdalena and Schwaller, Philippe},
  journal={Digital Discovery},
  volume={5},
  number={3},
  pages={1269--1279},
  year={2026},
  publisher={The Royal Society of Chemistry}
}

@article{rossi2021knowledge,
  title={Knowledge graph embedding for link prediction: A comparative analysis},
  author={Rossi, Andrea and Barbosa, Denilson and Firmani, Donatella and Matinata, Antonio and Merialdo, Paolo},
  journal={ACM Transactions on Knowledge Discovery from Data (TKDD)},
  volume={15},
  number={2},
  pages={1--49},
  year={2021},
  publisher={ACM New York, NY, USA}
}

@misc{galkin2022nodepiece,
      title={NodePiece: Compositional and Parameter-Efficient Representations of Large Knowledge Graphs}, 
      author={Mikhail Galkin and Etienne Denis and Jiapeng Wu and William L. Hamilton},
      year={2022},
      eprint={2106.12144},
      archivePrefix={arXiv},
      primaryClass={cs.CL},
      url={https://arxiv.org/abs/2106.12144}, 
}

@misc{demir2024inferenceunseenentitiesrelations,
      title={Inference over Unseen Entities, Relations and Literals on Knowledge Graphs}, 
      author={Caglar Demir and N'Dah Jean Kouagou and Arnab Sharma and Axel-Cyrille Ngonga Ngomo},
      year={2024},
      eprint={2410.06742},
      archivePrefix={arXiv},
      primaryClass={cs.LG},
      url={https://arxiv.org/abs/2410.06742}, 
}

@article{glasby2023digimof,
  author  = {Glasby, Lawson T. and Gubsch, Kristian and Bence, Rosalee and
             Oktavian, Rama and Isoko, Kesler and Moosavi, Seyed Mohamad and
             Cordiner, Joan L. and Cole, Jason C. and Moghadam, Peyman Z.},
  title   = {{DigiMOF}: A Database of Metal--Organic Framework Synthesis
             Information Generated via Text Mining},
  journal = {Chemistry of Materials},
  volume  = {35},
  number  = {11},
  pages   = {4510--4524},
  year    = {2023},
  doi     = {10.1021/acs.chemmater.3c00788}
}

@article{huang2020nerbattery,
  author  = {Huang, Shu and Cole, Jacqueline M.},
  title   = {A database of battery materials auto-generated using
             {ChemDataExtractor}},
  journal = {Scientific Data},
  volume  = {7},
  number  = {1},
  pages   = {260},
  year    = {2020},
  doi     = {10.1038/s41597-020-00602-2}
}

@article{del2022chameo,
  title={CHAMEO: An ontology for the harmonisation of materials characterisation methodologies},
  author={Del Nostro, Pierluigi and Goldbeck, Gerhard and Toti, Daniele},
  journal={Applied Ontology},
  volume={17},
  number={3},
  pages={401--421},
  year={2022},
  publisher={SAGE Publications Sage UK: London, England}
}

@misc{emmo2023,
  author       = {Ghedini, Emanuele and Friis, Jesper and Bleken, Francesca L{\o}nstad and Prinz, Manuel and Goldbeck, Gerhard},
  title        = {{EMMO}: Elementary Multiperspective Material Ontology},
  year         = {2023},
  howpublished = {\url{https://github.com/emmo-repo/EMMO}},
  note         = {Version 1.0.1. Accessed: 14 October 2026}
}

@article{groom2016csd,
  author  = {Groom, Colin R. and Bruno, Ian J. and Lightfoot, Matthew P. and
             Ward, Suzanna C.},
  title   = {The {Cambridge Structural Database}},
  journal = {Acta Crystallographica Section B},
  volume  = {72},
  number  = {2},
  pages   = {171--179},
  year    = {2016},
  doi     = {10.1107/S2052520616003954}
}

@article{grazulis2009cod,
  author  = {Gra{\v{z}}ulis, Saulius and Chateigner, Daniel and Downs,
             Robert T. and Yokochi, A. F. T. and Quir{\'o}s, Miguel and
             Lutterotti, Luca and Manakova, Elena and Butkus, Justas and
             Moeck, Peter and Le Bail, Armel},
  title   = {Crystallography {Open Database} -- an open-access collection of
             crystal structures},
  journal = {Journal of Applied Crystallography},
  volume  = {42},
  number  = {4},
  pages   = {726--729},
  year    = {2009},
  doi     = {10.1107/S0021889809016690}
}

@article{belsky2002icsd,
  author  = {Belsky, Alec and Hellenbrandt, Mariette and Karen, Vicky Lynn and
             Luksch, Peter},
  title   = {New developments in the {Inorganic Crystal Structure Database}
             ({ICSD}): accessibility in support of materials research and
             design},
  journal = {Acta Crystallographica Section B},
  volume  = {58},
  number  = {3},
  pages   = {364--369},
  year    = {2002},
  doi     = {10.1107/S0108768102006948}
}

@article{blokhin2018mpds,
  author  = {Blokhin, Evgeny and Villars, Pierre},
  title   = {The {PAULING FILE} Project and {Materials Platform for Data
             Science}: From Big Data Toward Materials Genome},
  journal = {Handbook of Materials Modeling},
  pages   = {1--26},
  year    = {2018},
  doi     = {10.1007/978-3-319-42913-7_62-1}
}

@article{otsuka2011polyinfo,
  author  = {Otsuka, Shingo and Kuwajima, Isao and Hosoya, Junko and
             Xu, Yibin and Yamazaki, Masayoshi},
  title   = {{PoLyInfo}: Polymer Database for Polymeric Materials Design},
  journal = {2011 International Conference on Emerging Intelligent Data and Web
             Technologies},
  pages   = {22--29},
  year    = {2011},
  doi     = {10.1109/EIDWT.2011.13}
}

@article{draxl2018nomad,
  author  = {Draxl, Claudia and Scheffler, Matthias},
  title   = {{NOMAD}: The {FAIR} concept for big data-driven materials science},
  journal = {MRS Bulletin},
  volume  = {43},
  number  = {9},
  pages   = {676--682},
  year    = {2018},
  doi     = {10.1557/mrs.2018.208}
}

@article{rietveld1969,
  author  = {Rietveld, H. M.},
  title   = {A profile refinement method for nuclear and magnetic structures},
  journal = {Journal of Applied Crystallography},
  volume  = {2},
  number  = {2},
  pages   = {65--71},
  year    = {1969},
  doi     = {10.1107/S0021889869006558}
}

@article{hall1991cif,
  author  = {Hall, Sydney R. and Allen, Frank H. and Brown, I. David},
  title   = {The Crystallographic Information File ({CIF}): a new standard
             archive file for crystallography},
  journal = {Acta Crystallographica Section A},
  volume  = {47},
  number  = {6},
  pages   = {655--685},
  year    = {1991},
  doi     = {10.1107/S010876739101067X}
}

@inproceedings{mikolov2013word2vec,
  author    = {Mikolov, Tomas and Sutskever, Ilya and Chen, Kai and
               Corrado, Greg S. and Dean, Jeff},
  title     = {Distributed Representations of Words and Phrases and their
               Compositionality},
  booktitle = {Advances in Neural Information Processing Systems (NeurIPS)},
  volume    = {26},
  pages     = {3111--3119},
  year      = {2013}
}

@inproceedings{pennington2014glove,
  author    = {Pennington, Jeffrey and Socher, Richard and Manning,
               Christopher D.},
  title     = {{GloVe}: Global Vectors for Word Representation},
  booktitle = {Proceedings of the 2014 Conference on Empirical Methods in
               Natural Language Processing (EMNLP)},
  pages     = {1532--1543},
  year      = {2014},
  doi       = {10.3115/v1/D14-1162}
}

@inproceedings{devlin2019bert,
  author    = {Devlin, Jacob and Chang, Ming-Wei and Lee, Kenton and
               Toutanova, Kristina},
  title     = {{BERT}: Pre-training of Deep Bidirectional Transformers for
               Language Understanding},
  booktitle = {Proceedings of the 2019 Conference of the North American Chapter
               of the Association for Computational Linguistics: Human Language
               Technologies (NAACL-HLT)},
  pages     = {4171--4186},
  year      = {2019},
  doi       = {10.18653/v1/N19-1423}
}

@inproceedings{hong2023diminishing,
  title={The diminishing returns of masked language models to science},
  author={Hong, Zhi and Ajith, Aswathy and Pauloski, James and Duede, Eamon and Chard, Kyle and Foster, Ian},
  booktitle={Findings of the Association for Computational Linguistics: ACL 2023},
  pages={1270--1283},
  year={2023}
}

@article{venugopal2024matkg,
  author  = {Venugopal, Vineeth and Olivetti, Elsa},
  title   = {{MatKG}: An autonomously generated knowledge graph in Material
             Science},
  journal = {Scientific Data},
  volume  = {11},
  number  = {1},
  pages   = {217},
  year    = {2024},
  doi     = {10.1038/s41597-024-03039-z}
}

@article{wang2017knowledge,
  title={Knowledge graph embedding: A survey of approaches and applications},
  author={Wang, Quan and Mao, Zhendong and Wang, Bin and Guo, Li},
  journal={IEEE transactions on knowledge and data engineering},
  volume={29},
  number={12},
  pages={2724--2743},
  year={2017},
  publisher={IEEE}
}

@article{mrdjenovich2020propnet,
  author  = {Mrdjenovich, David and Horton, Matthew K. and Montoya, Joseph H.
             and Legaspi, Christian M. and Dwaraknath, Shyam and
             Tshitoyan, Vahe and Jain, Anubhav and Persson, Kristin A.},
  title   = {propnet: A Knowledge Graph for Materials Science},
  journal = {Matter},
  volume  = {2},
  number  = {2},
  pages   = {464--480},
  year    = {2020},
  doi     = {10.1016/j.matt.2019.11.013}
}

@article{kang2024chatmof,
  author  = {Kang, Yeonghun and Kim, Jihan},
  title   = {{ChatMOF}: an artificial intelligence system for predicting and
             generating metal--organic frameworks using large language models},
  journal = {Nature Communications},
  volume  = {15},
  number  = {1},
  pages   = {4705},
  year    = {2024},
  doi     = {10.1038/s41467-024-48998-4}
}

@article{ghafarollahi2025sciagents,
  author  = {Ghafarollahi, Alireza and Buehler, Markus J.},
  title   = {{SciAgents}: Automating Scientific Discovery Through Bioinspired
             Multi-Agent Intelligent Graph Reasoning},
  journal = {Advanced Materials},
  volume  = {37},
  number  = {22},
  pages   = {2413523},
  year    = {2025},
  doi     = {10.1002/adma.202413523}
}

@article{buehler2024generative,
  title={Generative retrieval-augmented ontologic graph and multiagent strategies for interpretive large language model-based materials design},
  author={Buehler, Markus J},
  journal={ACS Engineering Au},
  volume={4},
  number={2},
  pages={241--277},
  year={2024}
}

@article{buehler2024graphreasoning,
  author  = {Buehler, Markus J.},
  title   = {Accelerating scientific discovery with generative knowledge
             extraction, graph-based representation, and multimodal
             intelligent graph reasoning},
  journal = {Machine Learning: Science and Technology},
  volume  = {5},
  number  = {3},
  pages   = {035083},
  year    = {2024},
  doi     = {10.1088/2632-2153/ad7228}
}

@article{lewis2020rag,
  author  = {Lewis, Patrick and Perez, Ethan and Piktus, Aleksandra and
             Petroni, Fabio and Karpukhin, Vladimir and Goyal, Naman and
             K{\"u}ttler, Heinrich and Lewis, Mike and Yih, Wen-tau and
             Rockt{\"a}schel, Tim and Riedel, Sebastian and Kiela, Douwe},
  title   = {Retrieval-Augmented Generation for Knowledge-Intensive {NLP}
             Tasks},
  journal = {Advances in Neural Information Processing Systems (NeurIPS)},
  volume  = {33},
  pages   = {9459--9474},
  year    = {2020}
}

@article{gupta2023discomat,
  author  = {Gupta, Tanishq and Zaki, Mohd and Khatsuriya, Devanshi and
             Hira, Kausik and Krishnan, N. M. Anoop and Mausam},
  title   = {{DiSCoMaT}: Distantly Supervised Composition Extraction from
             Tables in Materials Science Articles},
  journal = {Proceedings of the 61st Annual Meeting of the Association for
             Computational Linguistics (ACL)},
  pages   = {13465--13483},
  year    = {2023},
  doi     = {10.18653/v1/2023.acl-long.753}
}

@article{mccusker2020nanomine,
  author  = {McCusker, Jamie P. and Keshan, Neha and Rashid, Sabbir M. and
             Deagen, Michael and Brinson, L. Cate and McGuinness, Deborah L.},
  title   = {{NanoMine}: A Knowledge Graph for Nanocomposite Materials Science},
  journal = {The Semantic Web -- ISWC 2020},
  series  = {Lecture Notes in Computer Science},
  volume  = {12507},
  pages   = {144--159},
  year    = {2020},
  doi     = {10.1007/978-3-030-62466-8_10}
}

@article{boiko2023coscientist,
  author  = {Boiko, Daniil A. and MacKnight, Robert and Kline, Ben and
             Gomes, Gabe},
  title   = {Autonomous chemical research with large language models},
  journal = {Nature},
  volume  = {624},
  number  = {7992},
  pages   = {570--578},
  year    = {2023},
  doi     = {10.1038/s41586-023-06792-0}
}

@article{bran2024chemcrow,
  author  = {M. Bran, Andres and Cox, Sam and Schilter, Oliver and
             Baldassari, Carlo and White, Andrew D. and Schwaller, Philippe},
  title   = {Augmenting large language models with chemistry tools},
  journal = {Nature Machine Intelligence},
  volume  = {6},
  number  = {5},
  pages   = {525--535},
  year    = {2024},
  doi     = {10.1038/s42256-024-00832-8}
}

@article{szymanski2023alab,
  author  = {Szymanski, Nathan J. and Rendy, Bernardus and Fei, Yuxing and
             Kumar, Rishi E. and He, Tanjin and Milsted, David and
             McDermott, Matthew J. and Gallant, Max and Cubuk, Ekin Dogus and
             Merchant, Amil and Kim, Haegyeom and Jain, Anubhav and
             Bartel, Christopher J. and Persson, Kristin and Zeng, Yan and
             Ceder, Gerbrand},
  title   = {An autonomous laboratory for the accelerated synthesis of novel
             materials},
  journal = {Nature},
  volume  = {624},
  number  = {7990},
  pages   = {86--91},
  year    = {2023},
  doi     = {10.1038/s41586-023-06734-w}
}

@article{gottweis2025coscientist,
  title   = {Accelerating scientific discovery with {Co-Scientist}},
  author  = {Gottweis, Juraj and Weng, Wei-Hung and Daryin, Alexander and Tu, Tao and Sirkovic, Petar and Myaskovsky, Artiom and Glowaty, Grzegorz and Weissenberger, Felix and Orlandi, Alessio and Popovici, Dan and others},
  journal = {Nature},
  year    = {2026},
  volume  = {655},
  pages   = {487–496},  
  doi     = {10.1038/s41586-026-10644-y}
}

@misc{white2023pattern,
      title={A Prompt Pattern Catalog to Enhance Prompt Engineering with ChatGPT}, 
      author={Jules White and Quchen Fu and Sam Hays and Michael Sandborn and Carlos Olea and Henry Gilbert and Ashraf Elnashar and Jesse Spencer-Smith and Douglas C. Schmidt},
      year={2023},
      eprint={2302.11382},
      archivePrefix={arXiv},
      primaryClass={cs.SE},
      url={https://arxiv.org/abs/2302.11382}, 
}

@article{schulhoff2024promptreport,
  title   = {The Prompt Report: A Systematic Survey of Prompt Engineering Techniques},
  author  = {Schulhoff, Sander and Ilie, Michael and Balepur, Nishant and Kahadze, Konstantine and Liu, Amanda and Si, Chenglei and Li, Yinheng and Gupta, Aayush and Han, HyoJung and Schulhoff, Sevien and Dulepet, Pranav Sandeep and Vidyadhara, Saurav and Ki, Dayeon and Agrawal, Sweta and Pham, Chau and Kroiz, Gerson and Li, Feileen and Tao, Hudson and Srivastava, Ashay and Da Costa, Hevander and Gupta, Saloni and Rogers, Megan L. and Goncearenco, Inna and Sarli, Giuseppe and Galynker, Igor and Peskoff, Denis and Carpuat, Marine and White, Jules and Anadkat, Shyamal and Hoyle, Alexander and Resnik, Philip},
  journal = {arXiv preprint arXiv:2406.06608},
  year    = {2024}
}

@article{chen2025unleashing,
  title   = {Unleashing the potential of prompt engineering for large language models},
  author  = {Chen, Banghao and Zhang, Zhaofeng and Langren{\'e}, Nicolas and Zhu, Shengxin},
  journal = {Patterns},
  volume  = {6},
  number  = {6},
  pages   = {101260},
  year    = {2025},
  doi     = {10.1016/j.patter.2025.101260}
}

@article{wei2022cot,
  title   = {Chain-of-Thought Prompting Elicits Reasoning in Large Language Models},
  author  = {Wei, Jason and Wang, Xuezhi and Schuurmans, Dale and Bosma, Maarten and Ichter, Brian and Xia, Fei and Chi, Ed and Le, Quoc V. and Zhou, Denny},
  journal = {arXiv preprint arXiv:2201.11903},
  year    = {2022},
  note    = {Advances in Neural Information Processing Systems 35 (NeurIPS 2022)}
}

@article{gupta2024polymer,
  title   = {Data extraction from polymer literature using large language models},
  author  = {Gupta, Sonakshi and Mahmood, Akhlak and Shetty, Pranav and Adeboye, Aishat and Ramprasad, Rampi},
  journal = {Communications Materials},
  volume  = {5},
  pages   = {269},
  year    = {2024},
  doi     = {10.1038/s43246-024-00708-9}
}

@article{ekuma2025propertyextractor,
  title   = {Dynamic in-context learning with conversational models for data extraction and materials property prediction},
  author  = {Ekuma, Chinedu E.},
  journal = {APL Machine Learning},
  volume  = {3},
  pages   = {016119},
  year    = {2025},
  doi     = {10.1063/5.0254406}
}

@article{kamatchi2026automated,
  title={Automated Extraction of Multicomponent Alloy Data Using Large Language Models for Sustainable Design},
  author={Kamatchi Sundaram, Aravindan and Chakraborty, Mohit and Devathi, Sai Mani Kumar and Prusty, B Pabitramohan and Batra, Rohit},
  journal={Advanced Science},
  pages={e75916},
  year={2026},
  publisher={Wiley Online Library},
  doi = {10.1002/advs.75916}
}

@article{ahlawat2026llamat,
  title   = {A family of large language models for materials research with insights into model adaptability in continued pretraining},
  author  = {Ahlawat, Dhruv and Mishra, Vaibhav and Singh, Somaditya and Zaki, Mohd and Bihani, Vaibhav and Grover, Hargun Singh and Mishra, Biswajit and Miret, Santiago and Mausam and Krishnan, N. M. Anoop},
  journal = {Nature Machine Intelligence},
  volume  = {8},
  pages   = {435--448},
  year    = {2026},
  doi     = {10.1038/s42256-026-01199-8}
}

@article{zheng2024gpt4v,
  title   = {Image and data mining in reticular chemistry powered by GPT-4V},
  author  = {Zheng, Zhiling and He, Zhiguo and Khattab, Omar and Rampal, Nakul and Zaharia, Matei A. and Borgs, Christian and Chayes, Jennifer T. and Yaghi, Omar M.},
  journal = {Digital Discovery},
  volume  = {3},
  number  = {3},
  pages   = {491--501},
  year    = {2024},
  doi     = {10.1039/D3DD00239J}
}

@article{roy2026comproscanner,
  title={ComProScanner: a multi-agent based framework for composition-property structured data extraction from scientific literature},
  author={Roy, Aritra and Grisan, Enrico and Buckeridge, John and Gattinoni, Chiara},
  journal={Digital Discovery},
  volume={5},
  number={4},
  pages={1794--1808},
  year={2026},
  publisher={The Royal Society of Chemistry}
}

@article{lei2024perspective,
  title   = {Materials science in the era of large language models: a perspective},
  author  = {Lei, Ge and Docherty, Ronan and Cooper, Samuel J.},
  journal = {Digital Discovery},
  volume  = {3},
  pages   = {1257--1272},
  year    = {2024},
  doi     = {10.1039/D4DD00074A}
}

@article{liu2025chemistry,
  title   = {Integrating chemistry knowledge in large language models via prompt engineering},
  author  = {Liu, Hongxuan and Yin, Haoyu and Luo, Zhiyao and Wang, Xiaonan},
  journal = {Synthetic and Systems Biotechnology},
  volume  = {10},
  number  = {1},
  pages   = {23--38},
  year    = {2025},
  doi     = {10.1016/j.synbio.2024.07.004}
}

@inproceedings{bordes2013transe,
  author    = {Bordes, Antoine and Usunier, Nicolas and Garcia-Dur{\'a}n,
               Alberto and Weston, Jason and Yakhnenko, Oksana},
  title     = {Translating Embeddings for Modeling Multi-relational Data},
  booktitle = {Advances in Neural Information Processing Systems (NeurIPS)},
  volume    = {26},
  pages     = {2787--2795},
  year      = {2013}
}

@misc{crewai2024,
  author       = {{CrewAI}},
  title        = {{CrewAI}: Framework for orchestrating role-playing,
                  autonomous {AI} agents},
  year         = {2024},
  howpublished = {\url{https://www.crewai.com/}},
  note         = {Open-source multi-agent orchestration framework; accessed 2025}
}

@article{dettmers2023qlora,
  author  = {Dettmers, Tim and Pagnoni, Artidoro and Holtzman, Ari and
             Zettlemoyer, Luke},
  title   = {{QLoRA}: Efficient Finetuning of Quantized {LLMs}},
  journal = {Advances in Neural Information Processing Systems (NeurIPS)},
  volume  = {36},
  year    = {2023},
  note    = {arXiv:2305.14314}
}

@article{foppiano2023automatic,
  title={Automatic extraction of materials and properties from superconductors scientific literature},
  author={Foppiano, Luca and Castro, Pedro Baptista and Ortiz Suarez, Pedro and Terashima, Kensei and Takano, Yoshihiko and Ishii, Masashi},
  journal={Science and Technology of Advanced Materials: Methods},
  volume={3},
  number={1},
  pages={2153633},
  year={2023},
  publisher={Taylor \& Francis},
  doi = {10.1080/27660400.2022.2153633}
}

@inproceedings{hu2022lora,
  author    = {Hu, Edward J. and Shen, Yelong and Wallis, Phillip and
               Allen-Zhu, Zeyuan and Li, Yuanzhi and Wang, Shean and
               Wang, Lu and Chen, Weizhu},
  title     = {{LoRA}: Low-Rank Adaptation of Large Language Models},
  booktitle = {International Conference on Learning Representations (ICLR)},
  year      = {2022},
  note      = {arXiv:2106.09685}
}

@article{jiang2025applications,
  author  = {Jiang, Xue and Wang, Weiren and Tian, Shaohan and Wang, Hao and
             Lookman, Turab and Su, Yanjing},
  title   = {Applications of natural language processing and large language
             models in materials discovery},
  journal = {npj Computational Materials},
  volume  = {11},
  number  = {1},
  pages   = {79},
  year    = {2025},
  doi     = {10.1038/s41524-025-01554-0},
}

@article{nie2022batterykg,
  title={Automating materials exploration with a semantic knowledge graph for Li-ion battery cathodes},
  author={Nie, Zhiwei and Zheng, Shisheng and Liu, Yuanji and Chen, Zhefeng and Li, Shunning and Lei, Kai and Pan, Feng},
  journal={Advanced Functional Materials},
  volume={32},
  number={26},
  pages={2201437},
  year={2022},
  publisher={Wiley Online Library}
}

@article{xie2024solarcell,
  author  = {Xie, Tong and Wan, Yuwei and Zhou, Yufei and Huang, Wei and
             Liu, Yixuan and Linghu, Qingyuan and Wang, Shaozhou and
             Kit, Chunyu and Grazian, Clara and Zhang, Wenjie and Hoex, Bram},
  title   = {Creation of a structured solar cell material dataset and
             performance prediction using large language models},
  journal = {Patterns},
  volume  = {5},
  number  = {5},
  pages   = {100955},
  year    = {2024},
  doi     = {10.1016/j.patter.2024.100955}
}

@article{raccuglia2016nn,
  author  = {Raccuglia, Paul and Elbert, Katherine C. and Adler, Philip D. F. and Falk, Casey and Wenny, Malia B. and Mollo, Aurelio and Zeller, Matthias and Friedler, Sorelle A. and Schrier, Joshua and Norquist, Alexander J.},
  title   = {Machine-learning-assisted materials discovery using failed experiments},
  journal = {Nature},
  volume  = {533},
  pages   = {73--76},
  year    = {2016}
}

@article{hira2024reconstructing,
  title={Reconstructing the materials tetrahedron: challenges in materials information extraction},
  author={Hira, Kausik and Zaki, Mohd and Sheth, Dhruvil and Mausam and Krishnan, NM Anoop},
  journal={Digital Discovery},
  volume={3},
  number={5},
  pages={1021--1037},
  year={2024},
  publisher={The Royal Society of Chemistry},
  doi = {10.1039/D4DD00032C}
}

@article{hogan2021kg,
  author  = {Hogan, Aidan and Blomqvist, Eva and Cochez, Michael and d'Amato, Claudia
             and de Melo, Gerard and Gutierrez, Claudio and Kirrane, Sabrina and
             Gayo, Jos{\'e} Emilio Labra and Navigli, Roberto and Neumaier, Sebastian
             and Ngonga Ngomo, Axel-Cyrille and Polleres, Axel and Rashid, Sabbir M.
             and Rula, Anisa and Schmelzeisen, Lukas and Sequeda, Juan and Staab, Steffen
             and Zimmermann, Antoine},
  title   = {Knowledge Graphs},
  journal = {ACM Computing Surveys},
  volume  = {54}, number = {4}, pages = {1--37}, year = {2021},
  doi     = {10.1145/3447772}
}

@article{cao2024kgesurvey,
  author  = {Cao, Jiahang and Fang, Jinyuan and Meng, Zaiqiao and Liang, Shangsong},
  title   = {Knowledge Graph Embedding: A Survey from the Perspective of Representation Spaces},
  journal = {ACM Computing Surveys},
  volume  = {56}, number = {6}, pages = {1--42}, year = {2024},
  doi     = {10.1145/3643806}
}

@article{ji2021kgsurvey,
  author  = {Ji, Shaoxiong and Pan, Shirui and Cambria, Erik and Marttinen, Pekka and Yu, Philip S.},
  title   = {A Survey on Knowledge Graphs: Representation, Acquisition, and Applications},
  journal = {IEEE Transactions on Neural Networks and Learning Systems},
  volume  = {33}, number = {2}, pages = {494--514}, year = {2022},
  doi     = {10.1109/TNNLS.2021.3070843}
}

@inproceedings{wang2014transh,
  author    = {Wang, Zhen and Zhang, Jianwen and Feng, Jianlin and Chen, Zheng},
  title     = {Knowledge Graph Embedding by Translating on Hyperplanes},
  booktitle = {Proceedings of the AAAI Conference on Artificial Intelligence},
  pages     = {1112--1119}, year = {2014}
}

@inproceedings{lin2015transr,
  author    = {Lin, Yankai and Liu, Zhiyuan and Sun, Maosong and Liu, Yang and Zhu, Xuan},
  title     = {Learning Entity and Relation Embeddings for Knowledge Graph Completion},
  booktitle = {Proceedings of the AAAI Conference on Artificial Intelligence},
  pages     = {2181--2187}, year = {2015}
}

@inproceedings{trouillon2016complex,
  author    = {Trouillon, Th{\'e}o and Welbl, Johannes and Riedel, Sebastian and
               Gaussier, {\'E}ric and Bouchard, Guillaume},
  title     = {Complex Embeddings for Simple Link Prediction},
  booktitle = {Proceedings of the 33rd International Conference on Machine Learning (ICML)},
  pages     = {2071--2080}, year = {2016}
}

@inproceedings{sun2019rotate,
  author    = {Sun, Zhiqing and Deng, Zhi-Hong and Nie, Jian-Yun and Tang, Jian},
  title     = {RotatE: Knowledge Graph Embedding by Relational Rotation in Complex Space},
  booktitle = {International Conference on Learning Representations (ICLR)},
  year      = {2019},
  url       = {https://openreview.net/forum?id=HkgEQnRqYQ}
}

@inproceedings{he2015kg2e,
  author    = {He, Shizhu and Liu, Kang and Ji, Guoliang and Zhao, Jun},
  title     = {Learning to Represent Knowledge Graphs with Gaussian Embedding},
  booktitle = {Proceedings of the 24th ACM International Conference on Information
               and Knowledge Management (CIKM)},
  pages     = {623--632}, year = {2015},
  doi       = {10.1145/2806416.2806502}
}

@inproceedings{nickel2017poincare,
  author    = {Nickel, Maximilian and Kiela, Douwe},
  title     = {Poincar{\'e} Embeddings for Learning Hierarchical Representations},
  booktitle = {Advances in Neural Information Processing Systems (NeurIPS)},
  volume    = {30}, year = {2017}
}

@inproceedings{chami2020atth,
  author    = {Chami, Ines and Wolf, Adva and Juan, Da-Cheng and Sala, Frederic and
               Ravi, Sujith and R{\'e}, Christopher},
  title     = {Low-Dimensional Hyperbolic Knowledge Graph Embeddings},
  booktitle = {Proceedings of the 58th Annual Meeting of the Association for
               Computational Linguistics (ACL)},
  pages     = {6901--6914}, year = {2020},
  doi       = {10.18653/v1/2020.acl-main.617}
}

@article{edge2024graphrag,
  author  = {Edge, Darren and Trinh, Ha and Cheng, Newman and Bradley, Joshua and
             Chao, Alex and Mody, Apurva and Truitt, Steven and Larson, Jonathan},
  title   = {From Local to Global: A Graph RAG Approach to Query-Focused Summarization},
  journal = {arXiv preprint arXiv:2404.16130},
  year    = {2024}
}

@article{peng2024graphrag,
  author  = {Peng, Boci and Zhu, Yun and Liu, Yongchao and Bo, Xiaohe and Shi, Haizhou
             and Hong, Chuntao and Zhang, Yan and Tang, Siliang},
  title   = {Graph Retrieval-Augmented Generation: A Survey},
  journal = {arXiv preprint arXiv:2408.08921},
  year    = {2024}
}

@inproceedings{francis2018cypher,
  author    = {Francis, Nadime and Green, Alastair and Guagliardo, Paolo and
               Libkin, Leonid and Lindaaker, Tobias and Marsault, Victor and
               Plantikow, Stefan and Rydberg, Mats and Selmer, Petra and Taylor, Andr{\'e}s},
  title     = {Cypher: An Evolving Query Language for Property Graphs},
  booktitle = {Proceedings of the 2018 International Conference on Management of Data (SIGMOD)},
  pages     = {1433--1445}, year = {2018},
  doi       = {10.1145/3183713.3190657}
}

@misc{harris2013sparql,
  author       = {Harris, Steve and Seaborne, Andy},
  title        = {SPARQL 1.1 Query Language},
  howpublished = {W3C Recommendation},
  year         = {2013},
  url          = {https://www.w3.org/TR/sparql11-query/}
}

@inproceedings{mohamed2019loss,
  author    = {Mohamed, Sameh K. and Nov{\'a}{\v{c}}ek, V{\'i}t and
               Vandenbussche, Pierre-Yves and Mu{\~n}oz, Emir},
  title     = {Loss Functions in Knowledge Graph Embedding Models},
  booktitle = {Proceedings of the Workshop on Deep Learning for Knowledge Graphs
               (DL4KG@ESWC)},
  series    = {CEUR Workshop Proceedings},
  volume    = {2377}, pages = {1--10}, year = {2019},
  url       = {https://ceur-ws.org/Vol-2377/paper_1.pdf}
}

@inproceedings{ruffinelli2020olddog,
  author    = {Ruffinelli, Daniel and Broscheit, Samuel and Gemulla, Rainer},
  title     = {You {CAN} Teach an Old Dog New Tricks! On Training Knowledge Graph
               Embeddings},
  booktitle = {International Conference on Learning Representations (ICLR)},
  year      = {2020}
}

@article{ali2021pykeen,
  author  = {Ali, Mehdi and Berrendorf, Max and Hoyt, Charles Tapley and
             Vermue, Laurent and Galkin, Mikhail and Sharifzadeh, Sahand and
             Fischer, Asja and Tresp, Volker and Lehmann, Jens},
  title   = {Bringing Light Into the Dark: A Large-Scale Evaluation of Knowledge
             Graph Embedding Models Under a Unified Framework},
  journal = {IEEE Transactions on Pattern Analysis and Machine Intelligence},
  volume  = {44}, number = {12}, pages = {8825--8845}, year = {2022},
  doi     = {10.1109/TPAMI.2021.3124805}
}

@inproceedings{dettmers2018conve,
  author    = {Dettmers, Tim and Minervini, Pasquale and Stenetorp, Pontus and
               Riedel, Sebastian},
  title     = {Convolutional 2D Knowledge Graph Embeddings},
  booktitle = {Proceedings of the AAAI Conference on Artificial Intelligence},
  pages     = {1811--1818}, year = {2018}
}

@inproceedings{lacroix2018canonical,
  author    = {Lacroix, Timoth{\'e}e and Usunier, Nicolas and Obozinski, Guillaume},
  title     = {Canonical Tensor Decomposition for Knowledge Base Completion},
  booktitle = {Proceedings of the 35th International Conference on Machine Learning (ICML)},
  pages     = {2863--2872}, year = {2018}
}

@article{an2023kgqa4mat,
  author  = {An, Yuan and Greenberg, Jane and Kalinowski, Alex and Zhao, Xintong
             and Hu, Xiaohua and Uribe-Romo, Fernando J. and Langlois, Kyle and
             Furst, Jacob and Gomez-Gualdron, Diego A.},
  title   = {Knowledge Graph Question Answering for Materials Science (KGQA4MAT):
             Developing Natural Language Interface for Metal--Organic Frameworks
             Knowledge Graph (MOF-KG)},
  journal = {arXiv preprint arXiv:2309.11361},
  year    = {2023}
}

@article{hendrycks2021mmlu,
  title={Measuring Massive Multitask Language Understanding},
  author={Hendrycks, Dan and Burns, Collin and Basart, Steven and Zou, Andy and Mazeika, Mantas and Song, Dawn and Steinhardt, Jacob},
  journal={arXiv preprint arXiv:2009.03300},
  year={2021},
  note={Published as a conference paper at ICLR 2021}
}

@article{itani2025thermoelectric,
  title   = {Large Language Model-Driven Database for Thermoelectric Materials},
  author  = {Itani, Suman and Zhang, Yibo and Zang, Jiadong},
  journal = {arXiv preprint arXiv:2501.00564},
  year    = {2025}
}

@article{itani2025nemad,
  title   = {The Northeast Materials Database for magnetic materials},
  author  = {Itani, Suman and Zhang, Yibo and Zang, Jiadong},
  journal = {Nature Communications},
  volume  = {16},
  pages   = {9415},
  year    = {2025}
}

@article{walls2025catminer,
  title   = {Use of Large Language Models for Extracting and Analyzing Data from Heterogeneous Catalysis Literature},
  author  = {Walls, Benjamin W. and Linic, Suljo},
  journal = {ACS Catalysis},
  volume  = {15},
  number  = {17},
  pages   = {14751--14763},
  year    = {2025}
}

@article{farris2025accelerating,
  title   = {Accelerating Catalysis Understanding via Large Language Model Data Extraction and Shallow Machine Learning Techniques},
  author  = {Farris, Brianna R. and Leonard, Kevin C.},
  journal = {JACS Au},
  volume  = {5},
  number  = {11},
  pages   = {5578--5589},
  year    = {2025}
}

@article{mashhadimoslem2026co2,
  title   = {Toward smart CO$_2$ capture by the synthesis of metal organic frameworks using large language models},
  author  = {Mashhadimoslem, Hossein and Abdol, Mohammad Ali and Zanganeh, Kourosh and Shafeen, Ahmed and Liu, Encheng and Zendehboudi, Sohrab and Elkamel, Ali and Yu, Aiping},
  journal = {Digital Discovery},
  volume  = {5},
  number  = {1},
  pages   = {384--396},
  year    = {2026}
}

@article{chizhevskiy2026hea,
  title={High Entropy Alloys Database generated with Large Language Model},
  author={Chizhevskiy, Vladimir and Markovi{\'c}, Gordana and Benrazzouq, Salah-eddine and Cvelbar, Uro{\v{s}} and Nomin{\'e}, Alexandre and Zava{\v{s}}nik, Janez},
  journal={Scientific data},
  volume={13},
  number={1},
  pages={612},
  year={2026},
  publisher={Nature Publishing Group UK London}
}

@article{desai2025perovskite,
  title   = {Bridging the Synthesizability Gap in Perovskites by Combining Computations, Literature Data, and PU Learning},
  author  = {Desai, Rushik and Ahn, Junyeong and Strachan, Alejandro and Mannodi-Kanakkithodi, Arun},
  journal = {arXiv preprint arXiv:2510.06166},
  year    = {2025}
}

@article{pan2026diffsyn,
  title={DiffSyn: a generative diffusion approach to materials synthesis planning},
  author={Pan, Elton and Kwon, Soonhyoung and Liu, Sulin and Xie, Mingrou and Hoffman, Alexander J and Duan, Yifei and Prein, Thorben and Sheriff, Killian and Roman-Leshkov, Yuriy and Moliner, Manuel and others},
  journal={Nature Computational Science},
  volume={6},
  number={4},
  pages={404--416},
  year={2026},
  publisher={Nature Publishing Group US New York}
}

@article{polak2024extracting,
  title={Extracting accurate materials data from research papers with conversational language models and prompt engineering},
  author={Polak, Maciej P and Morgan, Dane},
  journal={Nature Communications},
  volume={15},
  number={1},
  pages={1569},
  year={2024},
  publisher={Nature Publishing Group UK London}
}

@article{circi2024tables,
  title   = {How Well Do Large Language Models Understand Tables in Materials Science?},
  author  = {Circi, Defne and Khalighinejad, Ghazal and Chen, Anlan and Dhingra, Bhuwan and Brinson, L. Catherine},
  journal = {Integrating Materials and Manufacturing Innovation},
  volume  = {13},
  number  = {3},
  pages   = {669--687},
  year    = {2024}
}

@article{duan2025substitution,
  title={LLM-empowered literature mining for material substitution studies in sustainable concrete},
  author={Duan, Yifei and Tian, Yixi and Ghosh, Soumya and Venugopal, Vineeth and Chen, Jie and Olivetti, Elsa A},
  journal={Resources, Conservation and Recycling},
  volume={221},
  pages={108379},
  year={2025},
  publisher={Elsevier}
}

@inproceedings{Friedrich2020SOFCExp,
  author    = {Friedrich, Annemarie and Adel, Heike and Tomazic, Federico and
               Hingerl, Johannes and Benteau, Renou and Marusczyk, Anika and
               Lange, Lukas},
  title     = {The {SOFC-Exp} Corpus and Neural Approaches to Information
               Extraction in the Materials Science Domain},
  booktitle = {Proceedings of the 58th Annual Meeting of the Association for
               Computational Linguistics ({ACL})},
  pages     = {1255--1268},
  year      = {2020},
  publisher = {Association for Computational Linguistics},
  doi       = {10.18653/v1/2020.acl-main.116}
}

@inproceedings{vaswani2017attention,
  author    = {Vaswani, Ashish and Shazeer, Noam and Parmar, Niki and
               Uszkoreit, Jakob and Jones, Llion and Gomez, Aidan N. and
               Kaiser, {\L}ukasz and Polosukhin, Illia},
  title     = {Attention Is All You Need},
  booktitle = {Advances in Neural Information Processing Systems ({NeurIPS})},
  volume    = {30},
  pages     = {5998--6008},
  year      = {2017}
}

@article{bommasani2021foundation,
  title={On the opportunities and risks of foundation models},
  author={Bommasani, Rishi and Hudson, Drew A and Adeli, Ehsan and Altman, Russ and Arora, Simran and von Arx, Sydney and Bernstein, Michael S and Bohg, Jeannette and Bosselut, Antoine and Brunskill, Emma and others},
  journal={arXiv preprint arXiv:2108.07258},
  year={2021}
}

@inproceedings{howard2018ulmfit,
  author    = {Howard, Jeremy and Ruder, Sebastian},
  title     = {Universal Language Model Fine-tuning for Text Classification},
  booktitle = {Proceedings of the 56th Annual Meeting of the Association for
               Computational Linguistics ({ACL})},
  pages     = {328--339},
  year      = {2018},
  publisher = {Association for Computational Linguistics},
  doi       = {10.18653/v1/P18-1031}
}

@article{brown2020language,
  title={Language models are few-shot learners},
  author={Brown, Tom and Mann, Benjamin and Ryder, Nick and Subbiah, Melanie and Kaplan, Jared D and Dhariwal, Prafulla and Neelakantan, Arvind and Shyam, Pranav and Sastry, Girish and Askell, Amanda and others},
  journal={Advances in neural information processing systems},
  volume={33},
  pages={1877--1901},
  year={2020},
}

@article{min2023recent,
  author  = {Min, Bonan and Ross, Hayley and Sulem, Elior and
             Ben Veyseh, Amir Pouran and Nguyen, Thien Huu and
             Sainz, Oscar and Agirre, Eneko and Heintz, Ilana and Roth, Dan},
  title   = {Recent Advances in Natural Language Processing via Large Pre-trained
             Language Models: A Survey},
  journal = {ACM Computing Surveys},
  volume  = {56},
  number  = {2},
  pages   = {1--40},
  year    = {2023},
  doi     = {10.1145/3605943}
}

@article{polak2024flexible,
   title={Flexible, model-agnostic method for materials data extraction from text using general purpose language models},
   volume={3},
   ISSN={2635-098X},
   url={http://dx.doi.org/10.1039/D4DD00016A},
   DOI={10.1039/d4dd00016a},
   number={6},
   journal={Digital Discovery},
   publisher={Royal Society of Chemistry (RSC)},
   author={Polak, Maciej P. and Modi, Shrey and Latosinska, Anna and Zhang, Jinming and Wang, Ching-Wen and Wang, Shaonan and Hazra, Ayan Deep and Morgan, Dane},
   year={2024},
   pages={1221–1235} }

@article{zhao2023surveyllm,
  author  = {Zhao, Wayne Xin and Zhou, Kun and Li, Junyi and Tang, Tianyi and
             Wang, Xiaolei and Hou, Yupeng and Min, Yingqian and others},
  title   = {A Survey of Large Language Models},
  journal = {arXiv preprint arXiv:2303.18223},
  year    = {2023},
  doi     = {10.48550/arXiv.2303.18223}
}

@inproceedings{agarwal2024manyshot,
  title     = {Many-Shot In-Context Learning},
  author    = {Agarwal, Rishabh and Singh, Avi and Zhang, Lei M. and Bohnet, Bernd
               and Rosias, Luis and Chan, Stephanie C. Y. and Zhang, Biao and
               Anand, Ankesh and Abbas, Zaheer and Nova, Azade and Co-Reyes,
               John D. and Chu, Eric and Behbahani, Feryal and Faust, Aleksandra
               and Larochelle, Hugo},
  booktitle = {Advances in Neural Information Processing Systems (NeurIPS)},
  year      = {2024},
  eprint    = {2404.11018},
  archivePrefix = {arXiv},
  primaryClass  = {cs.LG},
  url       = {https://openreview.net/forum?id=AB6XpMzvqH}
}

@article{liu2024lostmiddle,
  title   = {Lost in the Middle: How Language Models Use Long Contexts},
  author  = {Liu, Nelson F. and Lin, Kevin and Hewitt, John and Paranjape, Ashwin
             and Bevilacqua, Michele and Petroni, Fabio and Liang, Percy},
  journal = {Transactions of the Association for Computational Linguistics},
  volume  = {12},
  pages   = {157--173},
  year    = {2024},
  doi     = {10.1162/tacl_a_00638}
}

@article{ghosh2026agentic,
  title={LLM-based AI agents for automated extraction of material properties and structural features},
  author={Ghosh, Subham and Tewari, Abhishek},
  journal={Computational Materials Science},
  volume={265},
  pages={114521},
  year={2026},
  publisher={Elsevier}
}

@inproceedings{yao2023react,
  author    = {Yao, Shunyu and Zhao, Jeffrey and Yu, Dian and Du, Nan and Shafran, Izhak and Narasimhan, Karthik and Cao, Yuan},
  title     = {{ReAct}: Synergizing reasoning and acting in language models},
  booktitle = {International Conference on Learning Representations (ICLR)},
  year      = {2023}
}

@misc{chandrasekhar2026materialdatabaseagentmultimodal,
      title={Material Database Agent: A Multimodal Agentic Framework for Scientific Literature Mining}, 
      author={Achuth Chandrasekhar and Omid Barati Farimani and Radheesh Sharma Meda and Amir Barati Farimani},
      year={2026},
      eprint={2605.04278},
      archivePrefix={arXiv},
      primaryClass={cs.CL},
      url={https://arxiv.org/abs/2605.04278}, 
}

@article{pham2026chemgraph,
  author  = {Pham, Thang D. and Tanikanti, Aditya and Ke{\c{c}}eli, Murat},
  title   = {{ChemGraph} as an agentic framework for computational chemistry workflows},
  journal = {Communications Chemistry},
  volume  = {9},
  pages   = {33},
  year    = {2026},
  doi     = {10.1038/s42004-025-01776-9}
}

@article{ritchhart2026cicero,
  author  = {Ritchhart, Andrew and Allec, Sarah I. and Butreddy, Pravalika and Kulesa, Krista and Wang, Qingpu and Nguyen, Dan Thien and Ziatdinov, Maxim and Nakouzi, Elias},
  title   = {Agentic workflow enables the recovery of critical materials from complex feedstocks via selective precipitation},
  journal = {Materials Horizons},
  volume  = {13},
  number  = {14},
  pages   = {7008},
  year    = {2026},
  doi     = {10.1039/d6mh00475j}
}

@article{chen2025chemeagle,
  author  = {Chen, Yufan and Leung, Ching Ting and Yu, Bowen and Sun, Jianwei and Huang, Yong and Li, Linyan and Chen, Hao and Gao, Hanyu},
  title   = {A multi-agent system enables versatile information extraction from the chemical literature},
  journal = {arXiv preprint arXiv:2507.20230},
  year    = {2025},
  doi     = {10.48550/arXiv.2507.20230}
}

@misc{yang2026agentcatllmagentextracting,
      title={AgentCAT: An LLM Agent for Extracting and Analyzing Catalytic Reaction Data from Chemical Engineering Literature}, 
      author={Wei Yang and Zihao Liu and Tao Tan and Xiao Hu and Hong Xie and Lulu Li Xin Li and Jianyu Han and Defu Lian and Mao Ye},
      year={2026},
      eprint={2602.18479},
      archivePrefix={arXiv},
      primaryClass={physics.chem-ph},
      url={https://arxiv.org/abs/2602.18479}, 
}

@article{mcnaughton2024cactus,
  title={Cactus: Chemistry agent connecting tool usage to science},
  author={McNaughton, Andrew D and Sankar Ramalaxmi, Gautham Krishna and Kruel, Agustin and Knutson, Carter R and Varikoti, Rohith A and Kumar, Neeraj},
  journal={ACS omega},
  volume={9},
  number={46},
  pages={46563--46573},
  year={2024},
  publisher={ACS Publications}
}

@inproceedings{wang2023planandsolve,
  author    = {Wang, Lei and Xu, Wanyu and Lan, Yihuai and Hu, Zhiqiang and Lan, Yunshi and Lee, Roy Ka-Wei and Lim, Ee-Peng},
  title     = {Plan-and-Solve prompting: improving zero-shot chain-of-thought reasoning by large language models},
  booktitle = {Proceedings of the 61st Annual Meeting of the Association for Computational Linguistics (ACL)},
  pages     = {2609--2634},
  year      = {2023}
}

@article{xu2023rewoo,
  author  = {Xu, Binfeng and Peng, Zhiyuan and Lei, Bowen and Mukherjee, Subhabrata and Liu, Yuchen and Xu, Dongkuan},
  title   = {{ReWOO}: decoupling reasoning from observations for efficient augmented language models},
  journal = {arXiv preprint arXiv:2305.18323},
  year    = {2023}
}

@inproceedings{shinn2023reflexion,
  author    = {Shinn, Noah and Cassano, Federico and Gopinath, Ashwin and Narasimhan, Karthik and Yao, Shunyu},
  title     = {Reflexion: language agents with verbal reinforcement learning},
  booktitle = {Advances in Neural Information Processing Systems (NeurIPS)},
  year      = {2023}
}

@article{madaan2023selfrefine,
  title={Self-refine: Iterative refinement with self-feedback},
  author={Madaan, Aman and Tandon, Niket and Gupta, Prakhar and Hallinan, Skyler and Gao, Luyu and Wiegreffe, Sarah and Alon, Uri and Dziri, Nouha and Prabhumoye, Shrimai and Yang, Yiming and others},
  journal={Advances in neural information processing systems},
  volume={36},
  pages={46534--46594},
  year={2023}
}

@inproceedings{yao2023tot,
  author    = {Yao, Shunyu and Yu, Dian and Zhao, Jeffrey and Shafran, Izhak and Griffiths, Thomas L. and Cao, Yuan and Narasimhan, Karthik},
  title     = {Tree of Thoughts: deliberate problem solving with large language models},
  booktitle = {Advances in Neural Information Processing Systems (NeurIPS)},
  year      = {2023}
}

@article{karpas2022mrkl,
  title={Mrkl systems: A modular, neuro-symbolic architecture that combines large language models, external knowledge sources and discrete reasoning},
  author={Karpas, Ehud and Abend, Omri and Belinkov, Yonatan and Lenz, Barak and Lieber, Opher and Ratner, Nir and Shoham, Yoav and Bata, Hofit and Levine, Yoav and Leyton-Brown, Kevin and others},
  journal={arXiv preprint arXiv:2205.00445},
  year={2022}
}

@article{abou2025agentic,
  title={Agentic AI: a comprehensive survey of architectures, applications, and future directions},
  author={Abou Ali, Mohamad and Dornaika, Fadi and Charafeddine, Jinan},
  journal={Artificial Intelligence Review},
  volume={59},
  number={1},
  pages={11},
  year={2025},
  publisher={Springer}
}

@article{jia2024llmatdesign,
  title   = {{LLMatDesign}: Autonomous Materials Discovery with Large Language Models},
  author  = {Jia, Shuyi and Zhang, Chao and Fung, Victor},
  journal = {arXiv preprint arXiv:2406.13163},
  year    = {2024},
  doi     = {10.48550/arXiv.2406.13163}
}

@article{ghafarollahi2024atomagents,
  title   = {{AtomAgents}: Alloy design and discovery through physics-aware
             multi-modal multi-agent artificial intelligence},
  author  = {Ghafarollahi, Alireza and Buehler, Markus J.},
  journal = {arXiv preprint arXiv:2407.10022},
  year    = {2024},
  doi     = {10.48550/arXiv.2407.10022}
}

@article{swanson2024virtuallab,
  title   = {The Virtual Lab: {AI} Agents Design New {SARS-CoV-2} Nanobodies with
             Experimental Validation},
  author  = {Swanson, Kyle and Wu, Wesley and Bulaong, Nash L. and Pak, John E. and
             Zou, James},
  journal = {bioRxiv},
  year    = {2024},
  doi     = {10.1101/2024.11.11.623004}
}

@article{ni2024matpilot,
  title   = {{MatPilot}: an {LLM}-enabled {AI} Materials Scientist under the Framework
             of Human-Machine Collaboration},
  author  = {Ni, Ziqi and Li, Yahao and Hu, Kaijia and Han, Kunyuan and Xu, Ming and
             Chen, Xingyu and Liu, Fengqi and Ye, Yicong and Bai, Shuxin},
  journal = {arXiv preprint arXiv:2411.08063},
  year    = {2024},
  doi     = {10.48550/arXiv.2411.08063}
}

@article{darvish2024organa,
  title   = {{ORGANA}: A Robotic Assistant for Automated Chemistry Experimentation and
             Characterization},
  author  = {Darvish, Kourosh and Skreta, Marta and Zhao, Yuchi and Yoshikawa, Naruki and
             Som, Sagnik and Bogdanovic, Miroslav and Cao, Yang and Hao, Han and
             Xu, Haoping and Aspuru-Guzik, Al{\'a}n and Garg, Animesh and
             Shkurti, Florian},
  journal = {arXiv preprint arXiv:2401.06949},
  year    = {2024},
  doi     = {10.48550/arXiv.2401.06949}
}

@article{lu2024aiscientist,
  title   = {The {AI} Scientist: Towards Fully Automated Open-Ended Scientific Discovery},
  author  = {Lu, Chris and Lu, Cong and Lange, Robert Tjarko and Foerster, Jakob and
             Clune, Jeff and Ha, David},
  journal = {arXiv preprint arXiv:2408.06292},
  year    = {2024},
  doi     = {10.48550/arXiv.2408.06292}
}

@article{ref50,
  author  = {Itani, Suman and Zhang, Yibo and Itani, Ranjit and Zang, Jiadong},
  title   = {Large Language Models for Superconductor Discovery},  
  journal = {arXiv preprint arXiv:2512.10847},
  year    = {2025},
  doi     = {10.48550/arXiv.2512.10847}
}

@article{ref66,
  author  = {Shang, Lijun and Yu, Yadong and Kang, Wenqiang and Zhou, Jian and Gao, Dongyue and Xiang, Pan and Liu, Zhe and Dai, Mengyan and Guo, Zhonglu and Sun, Zhimei},
  title   = {LLMs-Powered Accurate Extraction, Querying and Intelligent Management of Literature-derived 2D Materials Data},
  journal = {arXiv preprint arXiv:2511.20691},
  year    = {2025},
  doi     = {10.48550/arXiv.2511.20691}
}

@article{ref4,
  author  = {Rameshbabu, Koushik and Luo, Jing and Shargh, Ali and El-Awady, Khalid A. and El-Awady, Jaafar A.},
  title   = {From Papers to Property Tables: A Priority-Based LLM Workflow for Materials Data Extraction},
  journal = {arXiv preprint arXiv:2604.07584},
  year    = {2026},
  doi     = {10.48550/arXiv.2604.07584}
}

@article{ref51,
  author  = {Cheng, Mouyang and He, Wenhao and Jin, Zhuotao and Yu, Bowen and Li, Ju and Kozinsky, Boris and Wang, Yao and Volkov, Pavel and Deng, Liangzi and Chu, Ching-Wu and Wen, Xiao-Gang and Li, Mingda},
  title   = {Can LLMs extract scientific consensus? A case study in high-temperature superconductivity},
  journal = {arXiv preprint arXiv:2606.07570},
  year    = {2026},
  doi     = {10.48550/arXiv.2606.07570}
}

@article{ref34,
  author  = {Chong, Curtis and Colindres, Jorge},
  title   = {LitXBench: A Benchmark for Extracting Experiments from Scientific Literature},
  journal = {arXiv preprint arXiv:2604.07649},
  year    = {2026},
  doi     = {10.48550/arXiv.2604.07649}
}

@inproceedings{ref37,
  author    = {Cheung, Jerry Junyang and Zhuang, Yuchen and Li, Yinghao and Shetty, Pranav and Zhao, Wantian and Grampurohit, Sanjeev and Ramprasad, Rampi and Zhang, Chao},
  title     = {PolyIE: A Dataset of Information Extraction from Polymer Material Scientific Literature},
  booktitle = {Proceedings of the 2024 Conference of the North American Chapter of the Association for Computational Linguistics: Human Language Technologies (Volume 1: Long Papers)},
  pages     = {2370--2385},
  year      = {2024},
  doi       = {10.18653/v1/2024.naacl-long.131}
}

@article{ref41,
  author  = {Htet, Hein and Ibrahim, Amgad Ahmed Ali and Sasaki, Yutaka and Asahi, Ryoji},
  title   = {Extracting ORR Catalyst Information for Fuel Cell from Scientific Literature},
  journal = {arXiv preprint arXiv:2507.07499},
  year    = {2025},
  doi     = {10.48550/arXiv.2507.07499}
}

@article{mullick2024matscire,
  title={Matscire: Leveraging pointer networks to automate entity and relation extraction for material science knowledge-base construction},
  author={Mullick, Ankan and Ghosh, Akash and Chaitanya, G Sai and Ghui, Samir and Nayak, Tapas and Lee, Seung-Cheol and Bhattacharjee, Satadeep and Goyal, Pawan},
  journal={Computational Materials Science},
  volume={233},
  pages={112659},
  year={2024},
  publisher={Elsevier}
}

@misc{ref63,
  title         = {L\textsuperscript{2}M\textsuperscript{3}OF: A Large Language Multimodal Model for Metal-Organic Frameworks},
  author        = {Cui, Jiyu and Wu, Fang and Zhao, Haokai and Feng, Minggao and Evangelopoulos, Xenophon and Cooper, Andrew I. and Choi, Yejin},
  year          = {2025},
  eprint        = {2510.20976},
  archivePrefix = {arXiv},
  primaryClass  = {cs.CL},
  url           = {https://arxiv.org/abs/2510.20976}
}

@article{ye2024mkg,
  title   = {Construction and Application of Materials Knowledge Graph in Multidisciplinary Materials Science via Large Language Model},
  author  = {Ye, Yanpeng and Ren, Jie and Wang, Shaozhou and Wan, Yuwei and Wang, Haofen and Razzak, Imran and Hoex, Bram and Xie, Tong and Zhang, Wenjie},
  journal = {arXiv preprint arXiv:2404.03080},
  year    = {2024}
}

@article{stewart2026higherorder,
  title   = {Higher-Order Knowledge Representations for Agentic Scientific Reasoning},
  author  = {Stewart, Isabella A. and Buehler, Markus J.},
  journal = {arXiv preprint arXiv:2601.04878},
  year    = {2026}
}

@article{wu2023autogen,
  title={Autogen: Enabling next-gen llm applications via multi-agent conversation},
  author={Wu, Qingyun and Bansal, Gagan and Zhang, Jieyu and Wu, Yiran and Li, Beibin and Zhu, Erkang and Jiang, Li and Zhang, Xiaoyun and Zhang, Shaokun and Liu, Jiale and others},
  journal={arXiv preprint arXiv:2308.08155},
  year={2023}
}

@article{sumers2023coala,
  title={Cognitive Architectures for Language Agents},
  author={Sumers, Theodore R and Yao, Shunyu and Narasimhan, Karthik and Griffiths, Thomas L},
  journal={arXiv preprint arXiv:2309.02427},
  year={2023}
}

@article{liu2023agentbench,
  title={AgentBench: Evaluating LLMs as Agents},
  author={Liu, Xiao and Yu, Hao and Zhang, Hanyu and Xu, Yifan and Lei, Xuanyu and Lai, Hanyu and Gu, Yu and Ding, Hangwei and Men, Kaiwen and Yang, Kejuan and others},
  journal={arXiv preprint arXiv:2308.03688},
  year={2023}
}

@article{niu2024knowledge,
  title={Knowledge graph embeddings: A comprehensive survey on capturing relation properties},
  author={Niu, Guanglin},
  journal={arXiv preprint arXiv:2410.14733},
  year={2024}
}

@article{buehler2025agentic,
  title={Agentic deep graph reasoning yields self-organizing knowledge networks},
  author={Buehler, Markus J},
  journal={Journal of Materials Research},
  volume={40},
  number={15},
  pages={2204--2242},
  year={2025},
  publisher={Springer}
}

@article{lafuente2015power,
author = {Lafuente, Barbara and Downs, R.T. and Yang, H. and Stone, Nathan},
year = {2015},
month = {11},
pages = {1-30},
title = {The power of databases: The RRUFF project},
isbn = {9783110417104},
journal = {Highlights in Mineralogical Crystallography},
doi = {10.1515/9783110417104-003}
}

@article{ward2014making,
  title={Making materials science and engineering data more valuable research products},
  author={Ward, Charles H and Warren, James A and Hanisch, Robert J},
  journal={Integrating Materials and Manufacturing Innovation},
  volume={3},
  number={1},
  pages={292--308},
  year={2014},
  publisher={Springer}
}

@article{hall1951deformation,
  author  = {Hall, E. O.},
  title   = {The deformation and ageing of mild steel: III. Discussion of results},
  journal = {Proceedings of the Physical Society. Section B},
  volume  = {64},
  pages   = {747--753},
  year    = {1951},
  doi     = {10.1088/0370-1301/64/9/303}
}

@article{HumeRothery1961,
  author     = {Hume-Rothery, W.},
  title     = {Atomic theory for students of metallurgy},
  journal   = {Journal of the Less Common Metals},
  volume    = {3},
  number    = {3},
  pages     = {264},
  year      = {1961},
  publisher = {Elsevier},
  doi       = {10.1016/0022-5088(61)90071-6}
}

@article{vegard1921konstitution,
  author  = {Vegard, L.},
  title   = {Die Konstitution der Mischkristalle und die Raumf{\"u}llung der Atome},
  journal = {Zeitschrift f{\"u}r Physik},
  volume  = {5},
  pages   = {17--26},
  year    = {1921},
  doi     = {10.1007/BF01349680}
}

@article{saal2013oqmd,
  author  = {Saal, James E. and Kirklin, Scott and Aykol, Muratahan
             and Meredig, Bryce and Wolverton, Chris},
  title   = {Materials design and discovery with high-throughput density
             functional theory: The Open Quantum Materials Database (OQMD)},
  journal = {JOM},
  volume  = {65},
  pages   = {1501--1509},
  year    = {2013},
  doi     = {10.1007/s11837-013-0755-4}
}

@article{cavignac2026alexandria,
  author  = {Cavignac, Th{\'e}o and Schmidt, Jonathan and De Breuck, Pierre-Paul
             and Loew, Antoine and Cerqueira, Tiago F. T.
             and Wang, Hai-Chen and Bochkarev, Anton and Lysogorskiy, Yury
             and Romero, Aldo H. and Drautz, Ralf and Botti, Silvana
             and Marques, Miguel A. L.},
  title   = {AI-Driven expansion and application of the Alexandria database},
  journal = {Journal of Physics: Materials},
  volume  = {9},
  number  = {2},
  pages   = {025014},
  year    = {2026},
  doi     = {10.1088/2515-7639/ae6620}
}

@misc{langgraph2024,
  author       = {{LangChain AI}},
  title        = {{LangGraph}: Building language agents as graphs},
  year         = {2024},
  howpublished = {\url{https://github.com/langchain-ai/langgraph}},
  note         = {Accessed 2026}
}

@article{achiam2023gpt,
  title={Gpt-4 technical report},
  author={Achiam, Josh and Adler, Steven and Agarwal, Sandhini and Ahmad, Lama and Akkaya, Ilge and Aleman, Florencia Leoni and Almeida, Diogo and Altenschmidt, Janko and Altman, Sam and Anadkat, Shyamal and others},
  journal={arXiv preprint arXiv:2303.08774},
  year={2023}
}

@misc{anthropic2024claude3,
  author       = {{Anthropic}},
  title        = {The {Claude 3} Model Family: {Opus}, {Sonnet}, {Haiku}},
  year         = {2024},
  howpublished = {Model card},
  url          = {https://www.anthropic.com/claude-3-model-card}
}

@article{bojanowski2017enriching,
  title={Enriching word vectors with subword information},
  author={Bojanowski, Piotr and Grave, Edouard and Joulin, Armand and Mikolov, Tomas},
  journal={Transactions of the association for computational linguistics},
  volume={5},
  pages={135--146},
  year={2017},
  publisher={MIT Press One Rogers Street, Cambridge, MA 02142-1209, USA journals-info~…}
}

@article{choudhary2024atomgpt,
  title={Atomgpt: Atomistic generative pretrained transformer for forward and inverse materials design},
  author={Choudhary, Kamal},
  journal={The Journal of Physical Chemistry Letters},
  volume={15},
  number={27},
  pages={6909--6917},
  year={2024},
  publisher={ACS Publications}
}

@article{choudhary2025diffractgpt,
  title={DiffractGPT: Atomic structure determination from x-ray diffraction patterns using a generative pretrained transformer},
  author={Choudhary, Kamal},
  journal={The Journal of Physical Chemistry Letters},
  volume={16},
  number={8},
  pages={2110},
  year={2025}
}

@article{lee2026agapi,
  title={AGAPI-Agents: an open-access agentic AI platform for accelerated materials design on AtomGPT. org},
  author={Lee, Jaehyung and Ely, Justin and Zhang, Kent and Ajith, Akshaya and Campbell, Charles Rhys and Choudhary, Kamal},
  journal={The Journal of Physical Chemistry Letters},
  volume={17},
  number={26},
  pages={7221--7231},
  year={2026},
  publisher={ACS Publications}
}

@article{lupulescu2015asm,
  title={ASM micrograph database™},
  author={Lupulescu, Afina and Flowers, Tyler and Vermillion, Linda and Henry, Scott},
  journal={Metallography, Microstructure, and Analysis},
  volume={4},
  number={4},
  pages={322--327},
  year={2015},
  publisher={Springer}
}

@article{mostafa2026eraf4xrd,
  title={ERAF4XRD: A multimodal agentic framework for constructing validated experimental X-ray diffraction databases from scientific literature},
  author={Mostafa, Afnan and Ratcliff, William and Billinge, Simon JL and Abdolrahim, Niaz},
  journal={arXiv preprint arXiv:2609.18583},
  year={2026}
}

@article{dartsi2026automating,
  title={Automating Chemical Reasoning in High-Throughput Phase Identification With a Probabilistic, LLM-Guided Framework},
  author={Dartsi, Olympia and Walters, Lauren N and Trewartha, Amalie E and Torrisi, Steven B and Volk, Amanda and Joyce, Liam and Ceder, Gerbrand and Jain, Anubhav},
  journal={Advanced Science},
  pages={e76450},
  year={2026},
  publisher={Wiley Online Library}
}

@article{hira2025matskraft,
  title={MatSKRAFT: A framework for large-scale materials knowledge extraction from scientific tables},
  author={Hira, Kausik and Zaki, Mohd and Krishnan, NM and others},
  journal={arXiv preprint arXiv:2509.10448},
  year={2025}
}

@misc{openai2022chatgpt,
  author       = {{OpenAI}},
  title        = {Introducing {ChatGPT}},
  year         = {2022},
  howpublished = {Blog post},
  url          = {https://openai.com/blog/chatgpt}
}

@misc{alma991024648149705706,
language = {eng},
publisher = {Springer},
address = {Berlin},
title = {SpringerMaterials : the Landolt-Börnstein database.},
url = {http://materials.springer.com},
year = {1900},
}

@inproceedings{
venugopal2024matkg2,
title={Mat{KG}-2: Unveiling precise material science ontology through autonomous committees of {LLM}s},
author={Vineeth Venugopal and Elsa Olivetti},
booktitle={AI for Accelerated Materials Design - NeurIPS 2023 Workshop},
year={2023},
url={https://openreview.net/forum?id=cR1iE6MQ1y}
}

@article{zagorac2019recent,
author = "Zagorac, D. and M{\"{u}}ller, H. and Ruehl, S. and Zagorac, J. and Rehme, S.",
title = "{Recent developments in the Inorganic Crystal Structure Database: theoretical crystal structure data and related features}",
journal = "Journal of Applied Crystallography",
year = "2019",
volume = "52",
number = "5",
pages = "918--925",
month = "Oct",
doi = {10.1107/S160057671900997X},
url = {https://doi.org/10.1107/S160057671900997X},
}

@article{duan2026information,
title = {Information extraction for materials science using large language models: A survey of recent advances and new frontiers},
journal = {Materials Today Communications},
volume = {54},
pages = {115551},
year = {2026},
issn = {2352-4928},
doi = {https://doi.org/10.1016/j.mtcomm.2026.115551},
author = {Yutong Duan and Ying Zhang and Runjia Yu and Dongyuan Li and Satoshi Kosugi and Manabu Okumura and Kotaro Funakoshi},
}

@article{dong2022deepxrd,
  title={DeepXRD, a deep learning model for predicting XRD spectrum from material composition},
  author={Dong, Rongzhi and Zhao, Yong and Song, Yuqi and Fu, Nihang and Omee, Sadman Sadeed and Dey, Sourin and Li, Qinyang and Wei, Lai and Hu, Jianjun},
  journal={ACS applied materials \& interfaces},
  volume={14},
  number={35},
  pages={40102--40115},
  year={2022},
  publisher={ACS Publications},
  doi= {10.1021/acsami.2c05812}
}

@article{johansen2026decifer,
  title={decifer: Crystal structure prediction from powder diffraction data using autoregressive language models},
  author={Johansen, Frederik Lizak and Friis-Jensen, Ulrik and Dam, Erik Bj{\o}rnager and Jensen, Kirsten Marie {\O}rnsbjerg and Mercado, Roc{\'\i}o and Selvan, Raghavendra},
  journal={arXiv preprint arXiv:2502.02189},
  year={2025}
}

@misc{grobid2008,
    title = {GROBID},
    howpublished = {\url{https://github.com/grobidOrg/grobid}},
    publisher = {GitHub},
    year = {2008--2026},
    archivePrefix = {swh},
    eprint = {1:dir:dab86b296e3c3216e2241968f0d63b68e8209d3c}
}

@article{tkaczyk2015cermine,
  title={CERMINE: automatic extraction of structured metadata from scientific literature},
  author={Tkaczyk, Dominika and Szostek, Pawe{\l} and Fedoryszak, Mateusz and Dendek, Piotr Jan and Bolikowski, {\L}ukasz},
  journal={International Journal on Document Analysis and Recognition (IJDAR)},
  volume={18},
  number={4},
  pages={317--335},
  year={2015},
  publisher={Springer}
}

@misc{scienceparse,
  author       = {{Allen Institute for Artificial Intelligence}},
  title        = {{Science Parse: Parse scientific papers in PDF form into structured text}},
  year         = {2016},
  howpublished = {\url{https://github.com/allenai/science-parse}},
  note         = {Accessed: 16 October 2026}
}

@article{touvron2023llama2,
  title={Llama 2: Open foundation and fine-tuned chat models},
  author={Touvron, Hugo and Martin, Louis and Stone, Kevin and Albert, Peter and Almahairi, Amjad and Babaei, Yasmine and Bashlykov, Nikolay and Batra, Soumya and Bhargava, Prajjwal and Bhosale, Shruti and others},
  journal={arXiv preprint arXiv:2307.09288},
  year={2023}
}

@article{ref53,
  title={Auto-generating a database on the fabrication details of perovskite solar devices},
  author={Valencia, Agnes and Liu, Fei and Zhang, Xiangyang and Bo, Xiangkun and Li, Weilu and Daoud, Walid A},
  journal={Scientific data},
  volume={12},
  number={1},
  pages={270},
  year={2025},
  publisher={Nature Publishing Group UK London}
}

@article{ref22,
  title={Inverse design of materials that exhibit the magnetocaloric effect by text-mining of the scientific literature and generative deep learning},
  author={Court, Callum J and Jain, Apoorv and Cole, Jacqueline M},
  journal={Chemistry of Materials},
  volume={33},
  number={18},
  pages={7217--7231},
  year={2021},
  publisher={ACS Publications}
}

@article{ref5,
  title={ZeoSyn: A comprehensive zeolite synthesis dataset enabling machine-learning rationalization of hydrothermal parameters},
  author={Pan, Elton and Kwon, Soonhyoung and Jensen, Zach and Xie, Mingrou and G{\'o}mez-Bombarelli, Rafael and Moliner, Manuel and Rom{\'a}n-Leshkov, Yuriy and Olivetti, Elsa},
  journal={ACS Central Science},
  volume={10},
  number={3},
  pages={729},
  year={2024}
}

@article{ref36,
  title={Automated knowledge extraction from polymer literature using natural language processing},
  author={Shetty, Pranav and Ramprasad, Rampi},
  journal={Iscience},
  volume={24},
  number={1},
  year={2021},
  publisher={Elsevier}
}

@inproceedings{beltagy2019scibert,
  title={SciBERT: A pretrained language model for scientific text},
  author={Beltagy, Iz and Lo, Kyle and Cohan, Arman},
  booktitle={Proceedings of the 2019 conference on empirical methods in natural language processing and the 9th international joint conference on natural language processing (EMNLP-IJCNLP)},
  pages={3615--3620},
  year={2019}
}

@inproceedings{wadden-etal-2019-entity,
  title={Entity, relation, and event extraction with contextualized span representations},
  author={Wadden, David and Wennberg, Ulme and Luan, Yi and Hajishirzi, Hannaneh},
  booktitle={Proceedings of the 2019 conference on empirical methods in natural language processing and the 9th international joint conference on natural language processing (EMNLP-IJCNLP)},
  pages={5784--5789},
  year={2019}
}

@article{ref23,
  title={Magbert: magnetics knowledge aware language model coupled with a question answering pipeline for Curie temperature extraction task},
  author={Zhumabayeva, Aigerim and Ranjan, Nikhil and Takac, Martin and Sanvito, Stefano and Ucar, Huseyin},
  journal={The Journal of Physical Chemistry C},
  volume={128},
  number={31},
  pages={13217--13229},
  year={2024},
  publisher={ACS Publications}
}

@article{huang2022batterybert,
  title={BatteryBERT: A pretrained language model for battery database enhancement},
  author={Huang, Shu and Cole, Jacqueline M},
  journal={Journal of chemical information and modeling},
  volume={62},
  number={24},
  pages={6365},
  year={2022}
}

@article{ref1,
  title={Accelerating materials discovery for polymer solar cells: Data-driven insights enabled by natural language processing},
  author={Shetty, Pranav and Adeboye, Aishat and Gupta, Sonakshi and Zhang, Chao and Ramprasad, Rampi},
  journal={Chemistry of Materials},
  volume={36},
  number={16},
  pages={7676--7689},
  year={2024},
  publisher={ACS Publications}
}

@article{shetty2023polymer,
  author  = {Shetty, Pranav and Rajan, Arunkumar Chitteth and
             Kuenneth, Christopher and Gupta, Sonakshi and
             Panchumarti, Lakshmi Prerana and Holm, Lauren and Zhang, Chao and
             Ramprasad, Rampi},
  title   = {A general-purpose material property data extraction pipeline from
             large polymer corpora using natural language processing},
  journal = {npj Computational Materials},
  volume  = {9},
  number  = {1},
  pages   = {52},
  year    = {2023},
  doi     = {10.1038/s41524-023-01003-w}
}

@article{ref9,
  title={A database of stress-strain properties auto-generated from the scientific literature using ChemDataExtractor},
  author={Kumar, Pankaj and Kabra, Saurabh and Cole, Jacqueline M},
  journal={Scientific Data},
  volume={11},
  number={1},
  pages={1273},
  year={2024},
  publisher={Nature Publishing Group UK London}
}

@article{dieb2021matscie,
  title={MatScIE: An automated tool for the generation of databases of methods and parameters used in the computational materials science literature},
  author={Guha, Souradip and Mullick, Ankan and Agrawal, Jatin and Ram, Swetarekha and Ghui, Samir and Lee, Seung-Cheol and Bhattacharjee, Satadeep and Goyal, Pawan},
  journal={Computational Materials Science},
  volume={192},
  pages={110325},
  year={2021},
  publisher={Elsevier}
}

@article{ref65,
  title={Text-mined dataset of gold nanoparticle synthesis procedures, morphologies, and size entities},
  author={Cruse, Kevin and Trewartha, Amalie and Lee, Sanghoon and Wang, Zheren and Huo, Haoyan and He, Tanjin and Kononova, Olga and Jain, Anubhav and Ceder, Gerbrand},
  journal={Scientific data},
  volume={9},
  number={1},
  pages={234},
  year={2022},
  publisher={Nature Publishing Group UK London}
}

@article{wu2026empowering,
  title={Empowering Chemistry Experts With Large Language Models for Literature Interpretation in Single-Atom Catalysis toward Advanced Oxidation},
  author={Wu, Jing-Hang and Shi, Ran and Zhou, Xiao and Zhang, Liang and Chen, Kong and Yu, Han-Qing and Wu, Yuen},
  journal={Angewandte Chemie},
  volume={138},
  number={2},
  pages={e202520525},
  year={2026},
  publisher={Wiley Online Library}
}

@article{ref28,
  title={Advanced scientific information mining using LLM-driven approaches in layered cathode materials for sodium-ion batteries},
  author={Na, Youwan and Kim, Jeffrey J and Park, Chanhyoung and Hwang, Jaewon and Kim, Changgi and Lee, Hokyung and Lee, Jehoon},
  journal={Materials Advances},
  volume={6},
  number={8},
  pages={2543--2548},
  year={2025},
  publisher={The Royal Society of Chemistry}
}

@article{ref62,
  title={Mof-chemunity: Literature-informed large language models for metal--organic framework research},
  author={Pruyn, Thomas Michael and Aswad, Amro and Khan, Sartaaj Takrim and Huang, Ju and Black, Robert and Moosavi, Seyed Mohamad},
  journal={Journal of the American Chemical Society},
  volume={147},
  number={47},
  pages={43474--43486},
  year={2025},
  publisher={ACS Publications}
}

@article{chiang2024llamp,
  title={LLaMP: Large language model made powerful for high-fidelity materials knowledge retrieval and distillation},
  author={Chiang, Yuan and Hsieh, Elvis and Chou, Chia-Hong and Riebesell, Janosh},
  journal={arXiv preprint arXiv:2401.17244},
  year={2024}
}

@InProceedings{zhou2024lats,
  title = 	 {Language Agent Tree Search Unifies Reasoning, Acting, and Planning in Language Models},
  author =       {Zhou, Andy and Yan, Kai and Shlapentokh-Rothman, Michal and Wang, Haohan and Wang, Yu-Xiong},
  booktitle = 	 {Proceedings of the 41st International Conference on Machine Learning},
  pages = 	 {62138--62160},
  year = 	 {2024},
  editor = 	 {Salakhutdinov, Ruslan and Kolter, Zico and Heller, Katherine and Weller, Adrian and Oliver, Nuria and Scarlett, Jonathan and Berkenkamp, Felix},
  volume = 	 {235},
  series = 	 {Proceedings of Machine Learning Research},
  month = 	 {21--27 Jul},
  publisher =    {PMLR},
  url = 	 {https://proceedings.mlr.press/v235/zhou24r.html},
}

@article{sadruddin2026scikgextract,
  title={Designing an agentic AI workflow for structured information extraction from scientific text: A case study on atomic layer deposition of ZnO and IGZO},
  author={Sadruddin, Sameer and Poupaki, Eleni and D’Souza, Jennifer and Auer, S{\"o}ren and Watkins, Alex and Karasulu, Bora and Mackus, Adriaan JM and Kessels, Erwin},
  journal={Journal of Vacuum Science \& Technology A},
    volume = {44},
    number = {3},
    pages = {032413},
    year = {2026},
  publisher={AIP Publishing}
}

@article{ref32,
  title={A general LLM-powered text mining framework: Applied to extract high entropy alloys},
  author={Yuan, Haolun and Zeng, Jun and Zuo, Jie and Wang, Xin and Xu, Dingguo},
  journal={Computational Materials Science},
  volume={264},
  pages={114476},
  year={2026},
  publisher={Elsevier}
}

@article{ref30,
  title={LLMB: AI Agent for Lithium Metal Battery Research Using Large Language Model},
  author={Lee, Jaewoong and Woo, Junhee and Kim, Younghun and Kim, Sejin and Paulina, Cinthya and Park, Hyunmin and Kim, Hee-Tak and Park, Steve and Kim, Jihan},
  journal={ACS Central Science},
  volume={12},
  number={4},
  pages={484},
  year={2026}
}

@article{fan2024openchemie,
  title={OpenChemIE: an information extraction toolkit for chemistry literature},
  author={Fan, Vincent and Qian, Yujie and Wang, Alex and Wang, Amber and Coley, Connor W and Barzilay, Regina},
  journal={Journal of Chemical Information and Modeling},
  volume={64},
  number={14},
  pages={5521--5534},
  year={2024},
  publisher={ACS Publications}
}

@article{ansari2024dziner,
  title={dziner: Rational inverse design of materials with ai agents},
  author={Ansari, Mehrad and Watchorn, Jeffrey and Brown, Carla E and Brown, Joseph S},
  journal={arXiv preprint arXiv:2410.03963},
  year={2024}
}

@inproceedings{sprueill2024chemreasoner,
author = {Sprueill, Henry W. and Edwards, Carl and Agarwal, Khushbu and Olarte, Mariefel V. and Sanyal, Udishnu and Johnston, Conrad and Liu, Hongbin and Ji, Heng and Choudhury, Sutanay},
title = {CHEMREASONER: heuristic search over a large language model's knowledge space using quantum-chemical feedback},
year = {2024},
publisher = {JMLR.org},
booktitle = {Proceedings of the 41st International Conference on Machine Learning},
articleno = {1885},
numpages = {24},
location = {Vienna, Austria},
series = {ICML'24}
}

@inproceedings{zhang2024honeycomb,
    title = "{H}oney{C}omb: A Flexible {LLM}-Based Agent System for Materials Science",
    author = "Zhang, Huan  and
      Song, Yu  and
      Hou, Ziyu  and
      Miret, Santiago  and
      Liu, Bang",
    editor = "Al-Onaizan, Yaser  and
      Bansal, Mohit  and
      Chen, Yun-Nung",
    booktitle = "Findings of the {A}ssociation for {C}omputational {L}inguistics: {EMNLP} 2024",
    month = nov,
    year = "2024",
    address = "Miami, Florida, USA",
    publisher = "Association for Computational Linguistics",
    url = "https://aclanthology.org/2024.findings-emnlp.192/",
    doi = "10.18653/v1/2024.findings-emnlp.192",
    pages = "3369--3382",
}

@article{leong2025mermaid,
  title={MERMaid: Universal multimodal mining of chemical reactions from PDFs using vision-language models},
  author={Leong, Shi Xuan and Pablo-Garc{\'\i}a, Sergio and Wong, Brandon and Aspuru-Guzik, Al{\'a}n},
  journal={Matter},
  volume={8},
  number={12},
  year={2025},
  pages={102331},
  publisher={Elsevier}
}

@misc{ghosh2026unlocking,
      title={MatMMExtract: An Open-Source Pipeline for Panel-Level Extraction of Grounded Image-Text Pairs from Materials Science Literature}, 
      author={Subham Ghosh and Shubham Tiwari and Mohammad Ibrahim and Abhishek Tewari},
      year={2026},
      eprint={2606.29667},
      archivePrefix={arXiv},
      primaryClass={cs.CV},
      url={https://arxiv.org/abs/2606.29667}, 
}

@article{roy2026beyond,
  title={Beyond Text and Tables: Vision-Language Model Integration in ComProScanner for Extracting Materials Data from Scientific Figures with High Accuracy},
  author={Roy, Aritra and Grisan, Enrico and Gattinoni, Chiara and Buckeridge, John},
  journal={arXiv preprint arXiv:2606.00065},
  year={2026}
}

@article{agrawal2024pixtral,
  title={Pixtral 12B},
  author={Agrawal, Pravesh and Antoniak, Szymon and Hanna, Emma Bou and Bout, Baptiste and Chaplot, Devendra and Chudnovsky, Jessica and Costa, Diogo and De Monicault, Baudouin and Garg, Saurabh and Gervet, Theophile and others},
  journal={arXiv preprint arXiv:2410.07073},
  year={2024}
}

@article{yang2025qwen3,
  title={Qwen3 technical report},
  author={Yang, An and Li, Anfeng and Yang, Baosong and Zhang, Beichen and Hui, Binyuan and Zheng, Bo and Yu, Bowen and Gao, Chang and Huang, Chengen and Lv, Chenxu and others},
  journal={arXiv preprint arXiv:2505.09388},
  year={2025}
}

\end{document}